\documentclass[aps,prb,twocolumn,amsmath,amssymb,superscriptaddress,floatfix,scrartcl,eqsecnum,longbibliography]{revtex4-1}

\usepackage{extarrows}
\usepackage{slashed}
\usepackage{amsmath,amsfonts,amssymb,graphics,graphicx,epsfig,color,times,indentfirst,layout}
\usepackage{lipsum}
\usepackage{mathrsfs}
\usepackage[unicode=true,pdfusetitle,bookmarks=false,colorlinks=true,citecolor=black,urlcolor=black,linkcolor=black]{hyperref}
\usepackage{tikz}
\usepackage{tikz-cd}
\usetikzlibrary{arrows}
\usetikzlibrary{intersections}
\usetikzlibrary{shapes.geometric}

\begin{document}

\title{Holographic Entanglement Renormalization of Topological Insulators}

\author{Xueda Wen}
\affiliation{Institute for Condensed Matter Theory and Department of Physics,
University of Illinois at Urbana-Champaign, 1110 West Green St, Urbana IL 61801}

\author{Gil Young Cho}
\affiliation{Institute for Condensed Matter Theory and Department of Physics,
University of Illinois at Urbana-Champaign, 1110 West Green St, Urbana IL 61801}
\affiliation{Department of Physics, Korea Advanced Institute of Science and Technology, Daejeon 305-701, Korea}

\author{Pedro L. S. Lopes}
\affiliation{Instituto de Fisica Gleb Wataghin, Universidade Estadual de Campinas, Campinas, SP 13083-970, Brazil}

\author{Yingfei Gu}
\affiliation{Department of Physics, Stanford University, Stanford, California 94305, USA}

\author{Xiao-Liang Qi}
\affiliation{Department of Physics, Stanford University, Stanford, California 94305, USA}

\author{Shinsei Ryu}
\affiliation{Institute for Condensed Matter Theory and Department of Physics,
University of Illinois at Urbana-Champaign, 1110 West Green St, Urbana IL 61801}

\date{\today}

\begin{abstract}
We study the real-space entanglement renormalization group flows of topological band insulators in
(2+1) dimensions by using the continuum multi-scale entanglement renormalization ansatz (cMERA). Given
the ground state of a Chern insulator, we construct and study its cMERA by paying attention, in particular, to
how the bulk holographic geometry and the Berry curvature depend on the topological properties of the ground
state. It is found that each state defined at different energy scale of cMERA carries a nonzero Berry flux, which
is emanated from the UV layer of cMERA, and flows towards the IR. Hence, a topologically nontrivial UV state
flows under the RG to an IR state, which is also topologically nontrivial. On the other hand, we found that there
is an obstruction to construct the exact ground state of a topological insulator with a topologically trivial IR
state. I.e., if we try to construct a cMERA for the ground state of a Chern insulator by taking a topologically
trivial IR state, the resulting cMERA does not faithfully reproduce the exact ground state at all length scales.
\end{abstract}

\maketitle

\tableofcontents

\section{Introduction}

Entanglement renormalization\cite{Vidal2005}, as a real space renormalization group (RG),
has received substantial attention recently because of the following two main reasons:
firstly, its efficiency in numerically finding ground states of quantum many-body systems; secondly, on the conceptual side,
its close connection with the Anti-de Sitter space/conformal field theory (AdS/CFT) correspondence.

An entanglement renormalization method addresses the computational obstacle (`entanglement')
of finding a highly entangled many-body ground state. One defines a set of unitary transformations, 
which recognizes the degrees of freedom and efficiently removes the spurious entanglement, the obstacle to finding the ground state.
Combining such transformations with the coarse-graining procedure of the real space RG, the multi-scale
entanglement renormalization ansatz (MERA) enforces that the quantum entanglement at different length
scales is removed under successive applications of the RG transformation, allowing one to study highly entangled quantum states.
As a powerful variational ansatz, the lattice MERA has been demonstrated to accurately approximate ground states of various quantum many body systems,
including symmetry broken phases\cite{VidalPRL2008,VidalPRB2008,Rams2008,VidalPRL2009,VidalPRA2009}
and topologically ordered phases\cite{Vidal2008top,Vidal2009top} in (1+1) and (2+1) dimensions.
In addition, to apply entanglement renormalization to quantum field theories (which are defined in an
inherently continuous spacetime), a continuum version of MERA, namely continuum MERA (cMERA), was recently developed\cite{Haegeman,Ryu2012}.

It is conjectured that the lattice MERA may be understood as a discrete
`realization' of the AdS/CFT correspondence\cite{Swingle2012}, where it is suggested that the
MERA may capture the key geometric properties of AdS spacetime. Some recent developments along this idea can be found in
Refs. \onlinecite{Van2010,Van2010b,Vilaplana2011,Swingle2,Vilaplana2,Matsueda,Okunishi,Bao,Miyaji2015A,Miyaji1503,Miyaji2015B}.
See also Refs. \onlinecite{EHM1,EHM2} where a  similar construction was proposed under the name of ``Exact holographic mapping''.

\begin{figure}
\includegraphics[width=3.75in]{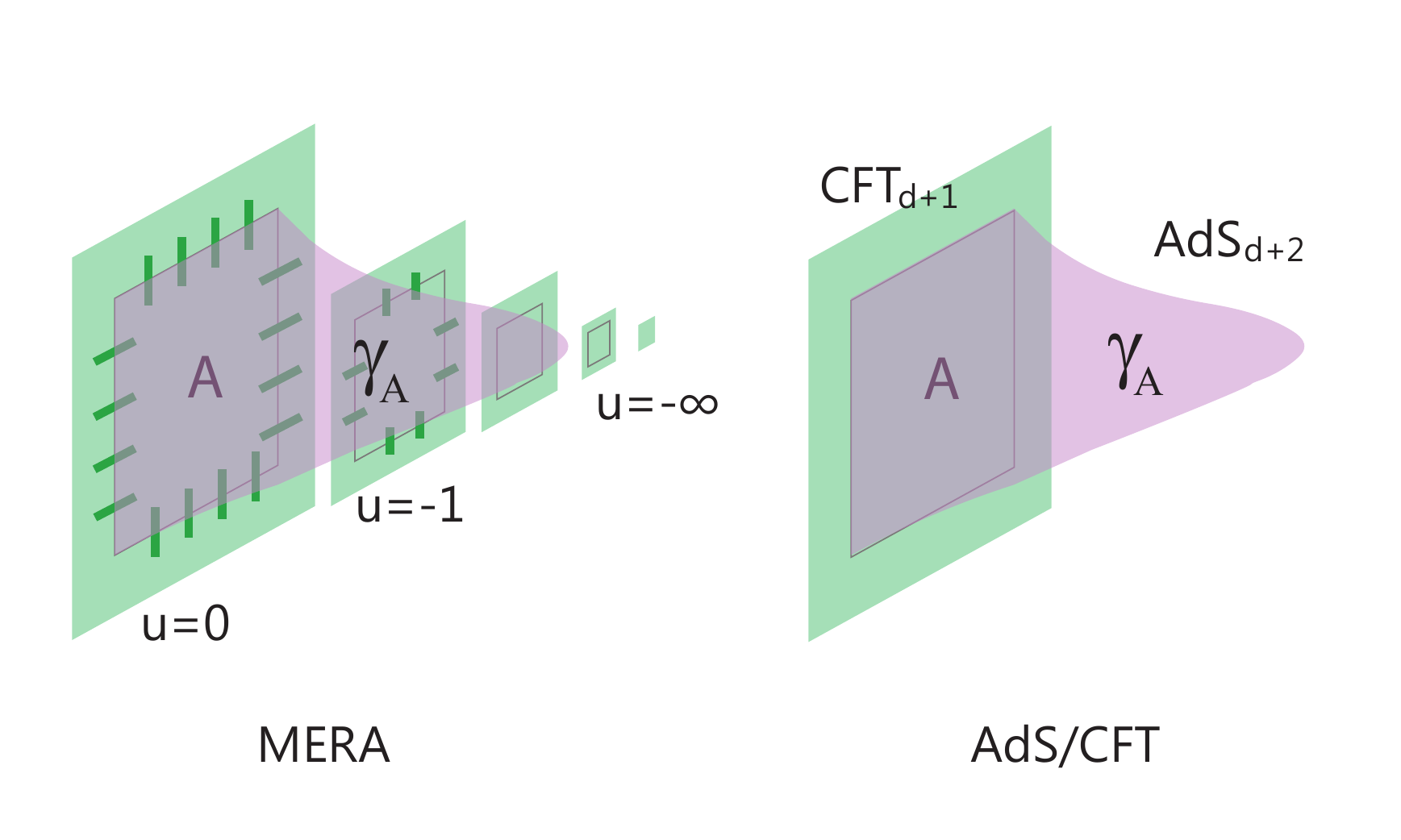}
\caption{Comparison between the calculations of entanglement entropy in the lattice MERA and AdS/CFT frameworks, respectively. The purple surface represents the minimal surfaces $\gamma_A$. The green solid bonds in the lattice MERA represent disentanglers. For the lattice MERA, the entanglement entropy of a subsystem $A$ can be expressed as $S_A\propto\text{min}[\#\text{Bonds}(\gamma_A)]$, while for AdS/CFT one has $S_A\propto\text{min}[\text{Area}]$.
 }\label{MeraAdS}
\end{figure}

The connection between the lattice MERA and AdS/CFT
may also be understood based on the observation that the entanglement entropy in the lattice
MERA can be estimated in a way similar to the holographic formula of the entanglement entropy in
AdS/CFT. In the classical limit of AdS/CFT, {\it i.e.,} when the gravity is described by the Einstein
equation, the entanglement entropy $S_A$ of a subsystem $A$ can be obtained by calculating
the minimal area surface\cite{RyuPRL2006}
\begin{equation}\label{AdSEntropy}
S_A=\frac{\text{Area}(\gamma_A)}{4G_N},
\end{equation}
where $\gamma_A$ is the minimal area surface embedded in a \emph{higher dimensional} AdS spacetime
whose boundary is $A$, and $G_N$ is the Newton constant of gravity in the AdS space. In the lattice 
MERA, the entanglement entropy of a subsystem $A$ is estimated by partitioning the MERA tensor network
into two parts, one which includes the subsystem $A$ and its complement. It should be noted that there
is no unique way to bipartition the network, and we label a set of partitions at different levels of the RG flow by $\gamma_A$. $S_A$
is then bounded by
\begin{equation}\label{MERAEntropy}
S_A\le \text{Min}_{\gamma_A}\text{Bonds}(\gamma_A)\cdot \log J.
\end{equation}
where $\text{Bonds}(\gamma_A)$ represents, for a given choice of the partitioning $\gamma_A$, 
the number of bonds connecting the two parts of the MERA network,
and $J$ is the dimension of bonds of the disentangler (see below). In particular, if each
bond is maximally entangled, then the entanglement entropy $S_A$ will be determined by the
minimal area of $\gamma_A$, in a fashion similar to the AdS case as shown in Fig.\ \ref{MeraAdS}. By identifying
$\gamma_A$ in the lattice MERA and the area $\gamma_A$ in AdS space, we can find
\begin{equation}\label{BondArea}
\text{Bonds}(\gamma_A)\simeq \frac{\gamma_A}{4G_N}
\end{equation}
up to a constant. 

The requirement of maximally entangled bonds is crucial for this identification; as discussed in
Ref. \onlinecite{Ryu2012}, if the bonds are not maximally entangled, the estimation of
entanglement entropy $S_A$ becomes more complicated, as one needs to consider the
bonds which are far from the minimal area. Equivalently, the calculation in terms of
tensor network is expected to become `non-local'. At the same time, it is known that the bulk
gravity in AdS space is non-local if one does not take the 't Hooft limit. The case of non-maximal entangled bonds in the lattice MERA may, therefore,
correspond to the quantum gravity limit in AdS space. The strong parallelism between MERA and AdS/CFT in calculating entanglement entropy suggests that
the `emergent' geometry appearing in the tensor network representation of the lattice MERA might be the dual AdS space of the quantum states at the boundary (see Fig. \ref{MeraAdS}).

Although a complete understanding of the connection between AdS/CFT and the lattice MERA
is still lacking, some progress has been made recently. In Ref.\ \onlinecite{Ryu2012}, the
expression for the holographic metric in the extra dimension (which is parametrized by the RG step) has been
proposed based on quantum field theory data in the continuous version MERA.
Furthermore, in a following work\cite{Ryu2013}, the holographic metric
after a quantum quench is also studied. It is found that the quenched holographic metric
qualitatively agrees with its gravity dual given by a half of the AdS black hole spacetime.
From the point view of cMERA, it has also been shown that the conformally invariant
boundary states are dual to trivial spacetimes of zero volume\cite{Miyaji2015A}, and
the bulk local states and corresponding operators in the
three-dimensional AdS space can be constructed using Ishibashi states in two-dimensional CFTs\cite{Miyaji1503,Miyaji2015B,Verlinde,Nakayama2015}.
In a different approach, MERA has been proposed to be related to the kinematic space, \textit{i.e.} 
the space of geodesic surfaces in AdS space \cite{Czech1505,Czech1512}.

In the previously mentioned references, the cMERA study is mainly focused on free boson or free fermion systems with trivial
topological properties. In the context of non-trivial topology, the AdS/CFT correspondence of Chern-Simons (CS) theories has been
studied recently\cite{Ryu_CS}. It is now therefore desirable to construct the cMERA dual to such AdS/CS
correspondence -- the main aim of our work. To achieve this goal, we develop the cMERA analysis of
Chern band insulators in (2+1) dimensions. 

Besides the cMERA dual of AdS/CS, there are other motivations for our study as follows:

(i)
Recently, tensor network methods have been applied extensively to topological phases in two dimensions
\cite{Read13,Cirac13,Cirac13b}. In these works, the exact projected entangled pair states
(PEPS) representations of chiral topological states are obtained, although the correlations
decay as an inverse power law. On the other hand, MERA has been constructed for exactly
solvable lattice models with topological order including the Kitaev toric code, the Levin-Wen string-net 
models and the AKLT model \cite{Vidal2008top,Vidal2009top,Vidal2013}. It is noted, however, that topological insulators\cite{Kane,QiRMP,QiPRB},
\emph{e.g.} Chern band insulators, have not been explicitly constructed with the lattice MERA,
despite some related recent studies\cite{Swingle14}. In Ref.\onlinecite{Swingle14},
it is shown that the lattice MERA representation of a gapped topological phase, including Chern band
insulators, should exist. By taking a bond dimension of order polynomial $L$, where $L$ is the system size,
the corresponding lattice MERA should be able to achieve high overlap with the true ground state in the
thermodynamic limit. Finding a concrete MERA network fulfilling this construction has, nonetheless, proved to be 
a hard task. In the present work, we find that the ground state of Chern band insulators may be
straightforwardly constructed in the framework of cMERA, which may shed light on our understanding of
the lattice MERA structure of topological insulators.

(ii)
In the previous studies, the IR state of cMERA is usually chosen as a topologically trivial state with
no entanglement whatsoever\cite{Haegeman,Ryu2012,Ryu2013}. In contrast, the ground state of a Chern
insulator in (2+1)D  carries a nonzero quantized momentum-space Berry-flux (in units of $2\pi$). It is thus interesting to ask what 
happens to such Berry flux if one performs entanglement renormalization procedures. Before the calculation, 
one may guess that there are mainly \emph{three} possibilities, depending on the choice of IR states as follows. (i) 
If the IR state carries a zero Berry-flux, then there must be a drain for the Berry curvature in the (3+1)D bulk of
cMERA towards the IR, corresponding to a magnetic-monopole-like structure. It is expected that a phase transition may happen
through the renormalization procedure in this case.  (ii) If the IR state carries a nonzero Berry flux whose amount does not equal
to the Berry flux at the UV layer, \emph{i.e.}, $\Phi(\text{IR})\neq \Phi(\text{UV})$, we expect
that part of the Berry curvature flows to the IR layer, and the other part is absorbed by the
magnetic-monopole in the bulk of cMERA. Again, there may be a phase transition in this case. (iii)
If the IR state carries the same amount of Berry flux as the UV state, \emph{i.e.},
$\Phi(\text{IR})= \Phi(\text{UV})$, we expect that all the Berry curvature emanated from the
UV layer flows to the IR layer, and there is no magnetic monopole in the bulk of cMERA.
In this case, no phase transition happens. It is thus necessary to obtain quantitative and exact picture on the pattern of Berry curvature
flow in the bulk of cMERA.

(iii)
Besides the aforementioned topological properties, it is also interesting to
study the geometric properties of cMERA. From the AdS/CFT correspondence point of view, different
phases at the boundary correspond to different bulk space geometries in a higher dimension.
In the prior studies on topological insulators, it is known that the momentum-metric can
capture new aspects of topological phases\cite{Shunji}. Now in cMERA, we have an emergent
holographic metric in the renormalization direction\cite{Ryu2012}. It is interesting to ask
if this holographic metric can display novel information about topological insulators, and how
the topological properties and geometric properties affect each other in the bulk of cMERA.

In this paper, we set up towards answering some of these questions.
The paper is organized as follows. In Sec.\ \ref{SecII}, we give a short review of cMERA in various versions. 
In Sec.\ \ref{SecIII}, starting
from a topologically trivial IR state, we construct cMERA for four different fermionic systems
in (2+1) dimensions, {\it i.e.}, non-relativistic Chern insulators, non-relativistic trivial
insulators, relativistic insulators with positive mass and relativistic insulators with negative mass, respectively.
Then we study the holographic geometry, band inversion, and Berry curvature flow in the bulk of cMERA for
different phases. In Sec.\ \ref{SecIV}, we construct the cMERA for Chern insulators with a
topologically nontrivial IR state, and study the corresponding holographic geometry,
band inversion as well as Berry curvature flow in the bulk of cMERA.
In Sec.\ \ref{conclusion}, we summarize our work and mention some future directions.

\section{Entanglement renormalization}
\label{SecII}

In this section, we give a brief introduction and review of cMERA.
For the completeness of this work, we also give a short review of the lattice MERA in Appendix \ref{LatticeMERA}.
Both the lattice MERA and cMERA are developed in order to find the ground state of many-body 
systems by making use of the variational
principle. As an implementation of real space renormalization group, they are different from
the conventional method developed by Migdal, Kadanoff and Wilson\cite{Kadanoff,Wilson,Fisher}.
For the lattice MERA and cMERA, short-ranged entanglement is removed during the process of
coarse graining, instead of removing {\it high-energy} degrees of freedom as in the conventional RG
formalism.

\subsection{Brief review of cMERA}

The continuum version of MERA (cMERA) was proposed
to understand quantum field theories within the lattice MERA scheme\cite{Haegeman}.
The formulation of cMERA is very helpful for making an explicit connection between the
entanglement renormalization and the AdS/CFT duality. In particular, it is found that an emergent metric
in the extra holographic direction can be defined in cMERA, where the holographic metric
shows properties expected from AdS/CFT\cite{Ryu2012}.To avoid confusions in later discussions, 
in the following parts, we discuss three different pictures of cMERA respectively.

We start from an IR state
\begin{equation}
|\Psi(u_{\text{IR}}=-\infty)\rangle \equiv |\Omega\rangle,
\end{equation}
which may be either entangled or unentangled, and aim to find a UV target state
\begin{equation}
\begin{split}
|\Psi(u_{\text{UV}}=0)\rangle &\equiv |\Psi\rangle,
\end{split}
\end{equation}
where $|\Psi\rangle$ represents the ground state of a given Hamiltonian at the UV length scale,
{\it e.g.,} a lattice in the condensed matter systems. With the same spirit as in the lattice
MERA, at each layer $u$, we use the disentangling operators to remove short-ranged entanglement
and perform isometry operationts to coarse-grain  (see Appendix \ref{LatticeMERA}). Compared to the lattice MERA, we
can formally replace the disentanglers and isometries as follows
\begin{equation}
\begin{split}
V_u\to V(u)=&e^{-iK(u)du},\\
W_u\to W(u)=&e^{-iL(u)du},
\end{split}
\end{equation}
where $du$ represents the {\it infinitesimal} RG step and $K(u)$ represents a local interaction of the form
\begin{equation}
K(u)=\int k(\mathbf{r},u)d^d\mathbf{r},
\end{equation}
with $k(\mathbf{r},u)$ being a local combination of local field operators $\psi(\mathbf{r})$ and
$\partial^n_{r}\psi(\mathbf{r}),~r = |\mathbf{r}|$  (in which we have assumed rotational
symmetry of the wavefunctions so that the disentangler is also rotationally symmetric),
and their adjoints. Here one may make the Gaussian ansatz\cite{Haegeman}
\begin{equation}
K(u)=\sum_{n=0}^{\infty}\int a_n(u)\psi^{\dag}(\mathbf{r})\partial^n_{r}\psi(\mathbf{r})
+\bar{a}_n(u) \partial^n_{r}\psi^{\dag}(\mathbf{r})\psi(\mathbf{r}) d\mathbf{r},
\end{equation}
where $a_n(u)$($\bar{a}_n(u)$) is a complex function which depends on the layer $u$.
$L$ is the generator of scale transformations
\begin{equation}
L=-\frac{i}{2}\int \psi^{\dag}(\mathbf{r})\mathbf{r}\cdot\mathbf{\nabla}_{\mathbf{r}}\psi(\mathbf{r})-\mathbf{r}\cdot \mathbf{\nabla}_{\mathbf{r}}\psi^{\dag}(\mathbf{r})\psi(\mathbf{r})d\mathbf{r},
\end{equation}
based on which one can find
\begin{equation}\label{scale}
\begin{split}
e^{-iuL}\psi(\mathbf{r})e^{iuL}=&e^{\frac{d}{2}u}\psi(e^u \mathbf{r}),\\
e^{-iuL}\psi(\mathbf{k})e^{iuL}=&e^{-\frac{d}{2}u}\psi(e^{-u} \mathbf{k}).\\
\end{split}
\end{equation}

\subsubsection{Schr$\ddot{o}$dinger picture}

\begin{figure}
\includegraphics[width=3.05in]{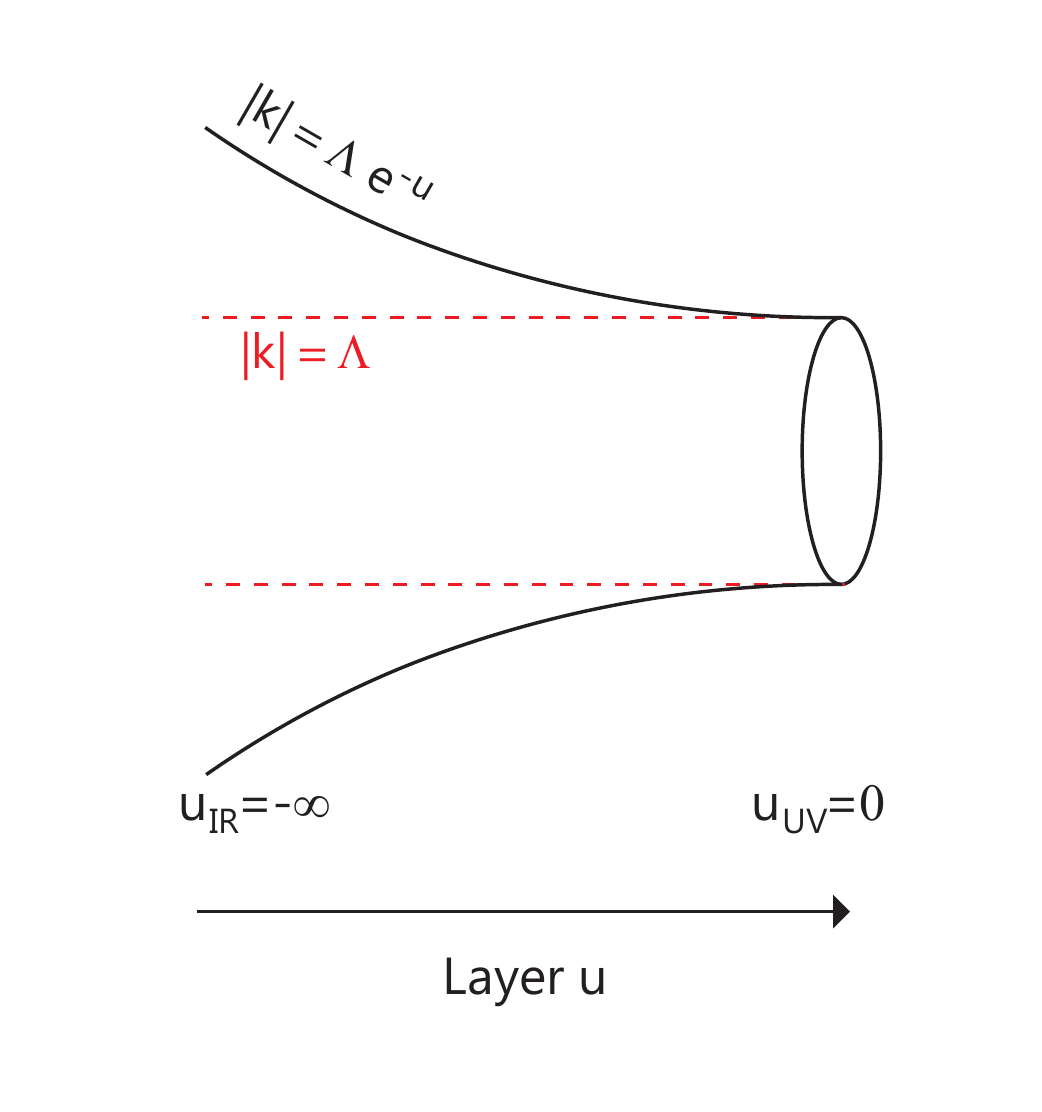}
\caption{Scheme of the momentum region where the Hilbert space in the `Schr$\ddot{o}$dinger' picture is defined. At layer $u$, the Hilbert space is defined within $0\le |k|\le \Lambda e^{-u}$, as indicated by the solid line. The disentangler $K(u)$ creates/removes entanglement with a constant cut-off $\Lambda$, as indicated by the red dotted line. Effectively, as $u$ goes deeper towards $u_{\text{IR}}$, the disentangler creates/removes entanglement for smaller $|k|$ (\textit{i.e.}, $|k|e^{\mu}$) in the original system at $u_{\text{UV}}$, which corresponds to a larger length scale in real space.
}\label{Fig3}
\end{figure}

The continuum version of many-body wavefunction in layer $u$ (see Eq.(\ref{Psi0}) for the lattice version) may be written as
\begin{equation}\label{PsiH}
|\Psi^S(u=0)\rangle=\mathcal{P}\exp\left[-i\int_{u_{\text{IR}}}^{0}\big(K(u)+L\big)du\right]|\Omega^S\rangle,
\end{equation}
where $\mathcal{P}$ is the path ordering operator as in Eq.\ (\ref{Psi0}), and $S$ represents the
`Schr$\ddot{o}$dinger' picture. The physical interpretation of cMERA is similar to that of the lattice
MERA, {\it i.e.}, the UV target state $|\Psi^S(u=0)\rangle=|\Psi\rangle$ can be constructed from an
IR state $|\Omega^S\rangle$ by adding short-ranged entanglement with $K(u)$ and doing scale
transformations by $L$ repeatedly. The opposite way from the UV limit to the IR limit may also be
interpreted straightforwardly. As $u$ varies from $u_{\text{UV}}=0$ to $u_{\text{IR}}=-\infty$, by
removing entanglement and doing scale transformations repeatedly, we end up with a state
$|\Omega^S\rangle$ which may be unentangled.  Eq.\ (\ref{PsiH}) can be generalized to an
arbitrary layer $u$ as
\begin{equation}\label{PsiLayerU}
\begin{split}
|\Psi^S(u=0)\rangle=&\mathcal{P}\exp\left[-i\int_{u}^{0}\big(K(u')+L\big)du'\right]|\Psi^S(u)\rangle.\\
\end{split}
\end{equation}
Based on this, we obtain `Schr$\ddot{o}$dinger's equation'
\begin{equation}\label{Sequation}
i\frac{\partial}{\partial u}|\Psi^S(u)\rangle=\left[K(u)+L\right]|\Psi^S(u)\rangle.
\end{equation}
It is beneficial to check the Hilbert space in which $|\Psi^S(u)\rangle$ is defined. First, one
 notes that the disentangler $K(u)$ only creates (or removes) entanglement and will not
change the Hilbert space. Therefore, one only needs to check the effect of $L$.
Now we consider a single particle state in momentum space in $(d+1)$ dimensions. At layer $u_{\text{UV}}=0$,
the single particle state can be written as
\begin{equation}
|\phi^S(u=0)\rangle=\int_{|\mathbf{k}|\le \Lambda} d^d\mathbf{k} \phi(\mathbf{k})\psi^{\dag}(\mathbf{k})|\text{vac}\rangle,
\end{equation}
where $\Lambda$ is a UV cut-off in momentum space.
Then the single particle state at layer $u$ according to Eq.\ (\ref{Sequation}) reads
\begin{equation}
\begin{split}
|\phi^S(u)\rangle=&e^{-iLu}|\phi(u=0)\rangle\\
=&\int_{|\mathbf{k}|\le \Lambda} d^d\mathbf{k} \phi(\mathbf{k})   e^{-iLu}\psi^{\dag}(\mathbf{k})e^{iLu}|\text{vac}\rangle,
\end{split}
\end{equation}
where we have used $e^{-iLu}|\text{vac}\rangle=|\text{vac}\rangle$. By using the
formula in Eq.\ (\ref{scale}) one can find
\begin{equation}
|\phi^S(u)\rangle=
\int_{|\mathbf{k}|\le \Lambda e^{-u}} d^d\mathbf{k} e^{-\frac{d}{2}u}\phi(\mathbf{k}e^u)  \psi^{\dag}(\mathbf{k})|\text{vac}\rangle,
\end{equation}
which means $|\phi^S(u)\rangle$ is now defined in the region $0\le |\mathbf{k}|\le \Lambda e^{-u}$.
At the same time, $K(u)$ is assumed to create or remove entanglement with a constant
cut-off $|\mathbf{k}|\le \Lambda$, which is independent of the layer $u$.\cite{Haegeman} As shown
in Fig.\ \ref{Fig3}, we plot schematically the region where the Hilbert space at layer $u$ is
defined, as well as the region within which entanglement is created or removed (on which
the disentangler operates). It can be found that as $u$ goes deeper towards $u_{\text{IR}}$,
the disentangler $K(u)$ effectively creates/removes entanglement for smaller $|\mathbf{k}|$ in the
layer $u=u_{\text{UV}}$. This is as expected because in the lattice MERA as $u$ goes deeper
towards $u_{\text{IR}}$, entanglement is created/removed in a larger length scale,
which corresponds to a smaller momentum scale.

\subsubsection{Heisenberg picture}
For convenience, we define the unitary operator
\begin{equation}
U(0,u)=\mathcal{P}\exp\left[-i\int_{u}^{0}\big(K(u')+L\big)du'\right].
\end{equation}
Suppose $O$ is some local operator defined in the layer $u=u_{\text{UV}}=0$, then by
moving to the `Heisenberg picture', one can define
$O(u)$ at layer $u$ as the following
\begin{equation}
O(u)=U(0,u)^{-1}\cdot O\cdot U(0,u),
\end{equation}
based on which one can get `Heisenberg's equation of motion'
\begin{equation}
\frac{dO(u)}{du}=-i[K(u)+L, O(u)].
\end{equation}
It is noted that the `Heisenberg' picture is used in Ref.\ \onlinecite{Haegeman}.

\subsubsection{Interaction picture}

As will be seen later, it is useful to move to the `interaction' picture, {\it i.e.},
\begin{equation}
|\Psi^I(u)\rangle=e^{iuL}|\Psi^S(u)\rangle.
\end{equation}
Combining with the Schr$\ddot{o}$dinger's equation in Eq.\ (\ref{Sequation}), one can obtain
\begin{equation}\label{EOM}
i\frac{\partial}{\partial u}|\Psi^I(u)\rangle=\hat{K}(u)|\Psi^{I}(u)\rangle,
\end{equation}
where
\begin{equation}\label{Kueffective}
\hat{K}(u)=e^{iuL}K(u)e^{-iuL}.
\end{equation}
Then the wavefunction $|\Psi^I(u)\rangle$ at layer $u$ can be expressed as
\begin{equation}\label{UnitaryInteraction1}
|\Psi^I(u)\rangle=\mathcal{P}\exp\left(-i\int_{u_{\text{IR}}}^{u}\hat{K}(u')du'\right)|\Omega^I\rangle,\\
\end{equation}
or
\begin{equation}\label{UnitaryInteraction2}
|\Psi^I(u)\rangle=\widetilde{\mathcal{P}}\exp\left(i\int_{u}^{0}\hat{K}(u')du'\right)|\Psi^I(u=0)\rangle,\\
\end{equation}
where $\tilde{\mathcal{P}}$ represents the path ordering operator which orders
operators in an {\it opposite order} relative to $\mathcal{P}$.
In this way, at each layer $u$, $|\Psi^I(u)\rangle$ is defined in the same Hilbert space with
$0\le |k|\le \Lambda$. The unitary operation defined in Eqs.\ (\ref{UnitaryInteraction1}) or
(\ref{UnitaryInteraction2}), after factoring out scale transformation, creates/removes
entanglement within $|\mathbf{k}|\le \Lambda e^u$ (See also Eq.\ (\ref{Khat})
in the next section for example.). Note that in the language of AdS/CFT, the factor
$e^{-iuL}$ corresponds to the warp factor of the AdS metric.

\begin{figure}[htb]
\includegraphics[width=3.25in]{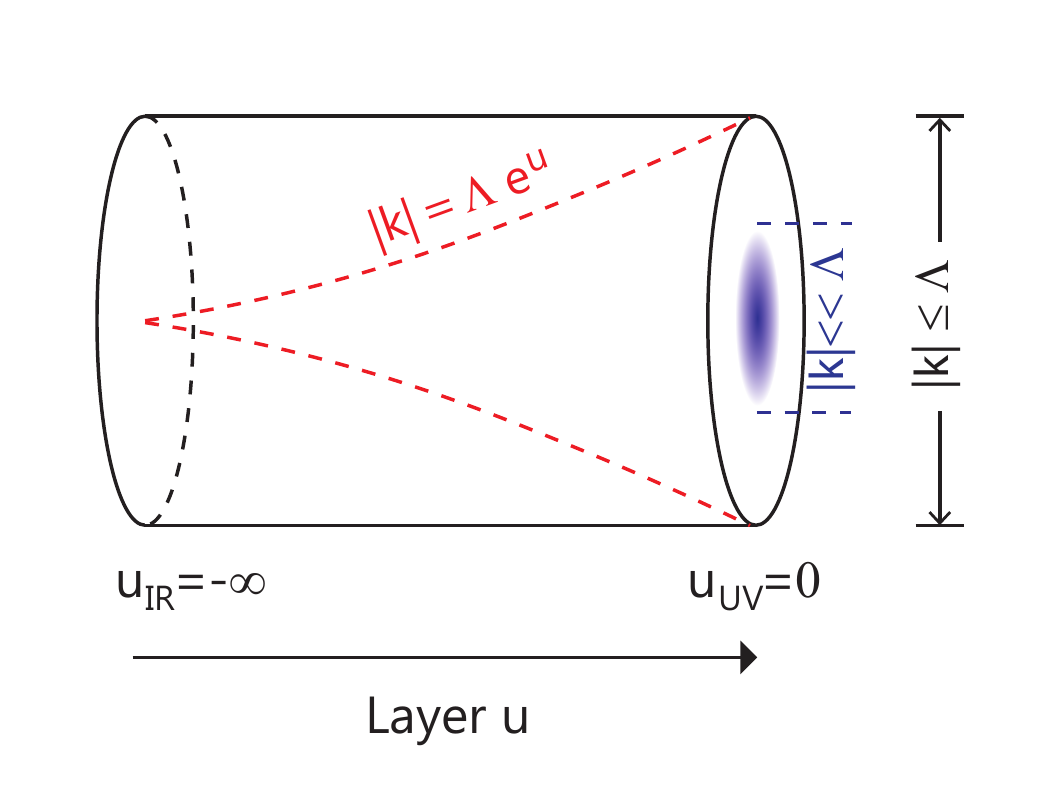}
\caption{Scheme of the momentum region where the Hilbert space in the `interaction' picture is defined. The Hilbert space at each layer $u$ is the same, in correspondence with the `interaction' picture of MERA in Fig.\ref{InMERA}. The boundary $|\mathbf{k}|=\Lambda e^u$ defines a cone within which quantum entanglement can be created/removed. The region defined by $|\mathbf{k}|\ll \Lambda$ is the low energy physics region.
}\label{Fig5}
\end{figure}

The merit of the `interaction' picture is that at each layer $u$ of the cMERA, we have the
same Hilbert space defined in $0\le |\mathbf{k}|\le \Lambda$ in momentum space.
This allows us to define and calculate the overlap $\langle \Psi(u)|\Psi(u+du)\rangle$,
from which we extract the emergent metric $g_{uu}(u)$ in the holographic direction,
as will be discussed later. To have an intuitive picture, in Fig.\ \ref{Fig5} we show
schematically the momentum region where the Hilbert space is defined in the interaction picture. The boundary
$|\mathbf{k}|=\Lambda e^u$ defines a cone in which entanglement can be created/removed.
Given an IR state $|\Omega\rangle$ at $u=u_{\text{IR}}$, the region outside the
cone is trivial because no entanglement is added. This is in analogy with the dangling
unentangled $|0\rangle$s in Fig.\ \ref{InMERA}.

In the rest of this paper, we will work
in the `interaction' picture, and for convenience we will simply write $|\Psi^I(u)\rangle$
as $|\Psi(u)\rangle$.

\section{cMERA of topological insulators with a topologically trivial IR state}
\label{SecIII}

Here we refer the IR state to be topologically trivial (nontrivial) if the
corresponding cMERA constructed wavefunction
$|\Psi(u)\rangle$ at each layer $u$ carries a zero (nonzero) Berry flux.

\subsection{cMERA of Chern insulators in (2+1)D}

In Refs.\ \onlinecite{Haegeman,Ryu2012,Ryu2013}, cMERA of a relativistic free fermion
system has been studied. This method can be generalized to various gapped phases
in a straightforward way.
Here we focus on a two-band free fermion system in (2+1) dimensions,
defined by the Hamiltonian
\begin{equation}\label{H0}
\begin{split}
H=&\int d^2\mathbf{k} \psi^{\dag}(\mathbf{k}) \left[\mathbf{R}(\mathbf{k})\cdot\mathbf{\sigma}\right] \psi(\mathbf{k}),\\
\end{split}
\end{equation}
where $\psi(\mathbf{k})=[\psi_1(\mathbf{k})\ \ \psi_2(\mathbf{k})]^T$, and $\psi_{1,2}(\mathbf{k})$
are fermion operators satisfying the canonical anti-commutation relation
$
\{\psi_1(\mathbf{k}),\psi^\dag_1(\mathbf{k}')\}=\{\psi_2(\mathbf{k}),\psi^\dag_2(\mathbf{k'})\}=\delta(\mathbf{k}-\mathbf{k}').
$
As a comparison, besides the Chern insulators, we will also consider
 non-relativistic trivial insulators
and relativistic insulators with $m>0$ and $m<0$, respectively. For convenience of labeling,
we will use $a,b,c,d$ to represent non-relativistic Chern insulators, non-relativistic trivial insulators,
relativistic insulators with $m>0$, and relativistic insulators with $m<0$ respectively
\footnote{Here we use the terminology 
`non-relativistic' (`relativistic') simply because the dispersion relation is $\sim k^2$ ($\sim k$) at UV
limit $k\to\infty$. Alternatively, one can refer to these phases as insulators with (without) regularization at UV limit.}, with
\begin{equation}\label{Ri}
\left\{
\begin{split}
&\mathbf{R}^a(\mathbf{k})=(k_x,k_y, m-k^2), \ \ \ &m>0\\
&\mathbf{R}^b(\mathbf{k})=(k_x,k_y, m-k^2), \ \ \ &m<0\\
&\mathbf{R}^c(\mathbf{k})=(k_x,k_y, m), \ \ \ &m>0\\
&\mathbf{R}^d(\mathbf{k})=(k_x,k_y, m), &m<0
\end{split}
\right.
\end{equation}
where $k=|\mathbf{k}|$.
The ground state corresponding to the Hamiltonian in Eq. (\ref{H0}) can be expressed as 
\begin{equation}\label{target}
|\Psi\rangle=
\prod_{|\mathbf{k}|\le \Lambda}
\left(
u_{\mathbf{k}}\psi_2^{\dag}(\mathbf{k})-v_{\mathbf{k}}\psi_1^{\dag}(\mathbf{k})
\right)|\text{vac}\rangle,
\end{equation}
where $u_{\mathbf{k}}$ and $v_{\mathbf{k}}$ are expressed in terms of $\mathbf{R}(\mathbf{k})$
(See Appendix \ref{cMERA4} for details.).
Our aim is to find a proper IR state $|\Omega\rangle$ and the associated
disentanglers which generate $|\Psi \rangle$ as the UV state.
 Next, we will derive the expression for the
wavefunction $|\Psi(u)\rangle$  at each layer $u$ of cMERA.
The wavefunction $|\Psi(u)\rangle$ is supposed to interpolate
$|\Omega \rangle$ to $|\Psi \rangle$ as $u$ sweeps over
$(u_{\text{IR}}, u_{\text{UV}}] = (-\infty, 0]$.

In the `interaction' picture of cMERA, the fermion operator
$\widetilde{\psi}(\mathbf{k},u)$ in layer $u$ is related with $\psi(\mathbf{k})$ as
\begin{equation}\label{unitary001}
\widetilde{\psi}(\mathbf{k},u)=\mathcal{P}e^{-i\int_{u_{\text{IR}}}^{u} \hat{K}(s)ds    }
\psi(\mathbf{k})
\tilde{\mathcal{P}}e^{i\int_{u_{\text{IR}}}^{u} \hat{K}(s)ds    }.
\end{equation}
Since the free fermion model is gaussian, one may make the gaussian ansatz
for the disentangler
\begin{equation}\label{Khat}
\begin{split}
\hat{K}(u)=&
i\int d^2\mathbf{k}\big[g_{\mathbf{k}}(u)\psi_1^\dag(\mathbf{k})\psi_2(\mathbf{k})
+g_{\mathbf{k}}^{\ast}(u)
\psi_1(\mathbf{k})\psi_2^\dag(\mathbf{k}) \big],
\end{split}
\end{equation}
where $g_{\mathbf{k}}(u)$ is chosen of the form
\begin{equation}\label{gku001}
g_{\mathbf{k}}(u)=g(u)\Gamma\left(\frac{|\mathbf{k}|}{\Lambda e^{u}}\right)\frac{|\mathbf{k}|}{\Lambda e^u}e^{-i\theta_{\mathbf{k}}}
=: g_{\mathbf{k}}^r(u)e^{-i\theta_{\mathbf{k}}},
\end{equation}
where $\Gamma(x)=\Theta(1-|x|)$ is the hard cut-off function, $g(u)$
is a complex function that we need to solve for, and $\theta_{\mathbf{k}}$ is defined through $k\cos\theta_{\mathbf{k}}=k_x$ and
$k\sin\theta_{\mathbf{k}}=k_y$. The disentangler in Eq.\ (\ref{Khat})
indicates that at each
layer $u$, the quantum entanglement can be created/removed only within the region
$|\mathbf{k}|\le\Lambda e^u$, as schematically shown in Fig.\ \ref{Fig5}. In fact, based on the expression of
$g_{\mathbf{k}}(u)$ in Eq.\ (\ref{gku001}), one can find that the disentangler
adds/removes entanglement mainly in the region $|\mathbf{k}|\simeq \Lambda e^u$.

It is noted that Eq.\ (\ref{unitary001}) can be considered as a unitary transformation, {\it i.e.},
\begin{equation}
\widetilde{\psi}(\mathbf{k},u)=M_{\mathbf{k}}(u)\psi(\mathbf{k}),
\end{equation}
where we have introduced the matrix $M_{\mathbf{k}}(u)$ as
\begin{small}
\begin{equation}
\begin{split}
M_{\mathbf{k}}(u):=
\begin{split}
\left(
\begin{array}{ll}
P_{\mathbf{k}}(u)\ \ &Q_{\mathbf{k}}(u)\\
-Q^{\ast}_{\mathbf{k}}(u)\ \ &P^{\ast}_{\mathbf{k}}(u)
\end{array}
\right)\end{split}
=\widetilde{P}\exp\left(\int_{u_{\text{IR}}}^u G_{\mathbf{k}}(u')du'\right),
\end{split}
\end{equation}
\end{small}
with $|P_{\mathbf{k}}(u)|^2+|Q_{\mathbf{k}}(u)|^2=1$ and
\begin{equation}
G_{\mathbf{k}}(u)=\left(
\begin{array}{llll}
&0 \ \ \ &-g_{\mathbf{k}}(u)\\
&g^{\ast}_{\mathbf{k}}(u)\ \ \ &0
\end{array}
\right).
\end{equation}
Equivalently, one has
\begin{equation}\label{diff001}
\frac{dM_{\mathbf{k}}(u)}{du}=G_{\mathbf{k}}(u)M_{\mathbf{k}}(u).
\end{equation}
By solving the differential equation above, one can obtain the general solution as
\begin{equation}\label{QP}
\left\{
\begin{split}
Q_{\mathbf{k}}(u)=&-ie^{-i\theta_{\mathbf{k}}}\left(
A^{\ast}e^{-i\int^u g^r_{\mathbf{k}}(u')du'}-B^{\ast}e^{i\int^u g^r_{\mathbf{k}}(u')du'}
\right),\\
P_{\mathbf{k}}(u)=&Ae^{i\int^u g^r_{\mathbf{k}}(u')du'}+Be^{-i\int^u g^r_{\mathbf{k}}(u')du'}.
\end{split}
\right.
\end{equation}
The wavefunction $|\Psi(u)\rangle$ at layer $u$ constructed from cMERA is defined by
\begin{equation}
\widetilde{\psi}_1(\mathbf{k},u)|\Psi(u)\rangle=0,\ \ \widetilde{\psi}^\dag_2(\mathbf{k},u)|\Psi(u)\rangle=0,
\end{equation}
with the explicit expression
\begin{equation}\label{GeneralForm001}
\begin{split}
|\Psi(u)\rangle=&\prod_{|\mathbf{k}|\le \Lambda}\widetilde{\psi}^\dag_2(\mathbf{k},u)|\text{vac}\rangle\\
=&
\prod_{|\mathbf{k}|\le \Lambda}\left(
P_{\mathbf{k}}(u)\psi^{\dag}_2(\mathbf{k})-Q_{\mathbf{k}}(u)\psi^{\dag}_1(\mathbf{k})
\right)|\text{vac}\rangle.
\end{split}
\end{equation}
Therefore, now our task is reduced to solving differential equations in Eq.\ (\ref{diff001})
under the boundary conditions
\begin{equation}
|\Psi(u=u_{\text{IR}})\rangle=|\Omega\rangle, \ \ \ |\Psi(u=u_{\text{UV}})\rangle=|\Psi\rangle.
\end{equation}
It is noted that there may be many choices of $|\Omega\rangle$. In the prior study on free fermion
systems\cite{Haegeman,Ryu2012,Ryu2013}, $|\Omega\rangle$ is chosen as an unentangled state, \textit{e.g.},
$|\Omega\rangle=\prod_{|\mathbf{k}|\le \Lambda}\psi_2^{\dag}(\mathbf{k})|\text{vac}\rangle$.

For the non-relativistic Chern insulator,
by comparing $|\Psi(u)\rangle$ with the boundary condition at the UV limit (see Eq.\ (\ref{targetA0})),
one can simply set $A$ and $B$ in Eq.\ (\ref{QP}) to be real, so that $A=B=1/2$. Then one can obtain
\begin{equation}\label{WFchern}
\left\{
\begin{split}
Q_{\mathbf{k}}(u)=&-e^{-i\theta_{\mathbf{k}}}\sin \int^u_{u_{\text{IR}}} g^r_{\mathbf{k}}(u')du'\\
P_{\mathbf{k}}(u)=&\cos \int^u_{u_{\text{IR}}} g^r_{\mathbf{k}}(u')du',
\end{split}
\right.
\end{equation}
based on which one can find that in the IR limit,
\begin{equation}
|\Psi(u\to u_{\text{IR}})\rangle=\prod_{|\mathbf{k}|\le \Lambda}\psi^\dag_2(\mathbf{k},u)|\text{vac}\rangle,
\end{equation}
which is the unentangled IR state used in the prior studies on free fermion systems.\cite{Haegeman,Ryu2012,Ryu2013}

The same procedure applies to the other three phases in Eq.\ (\ref{Ri}).
In particular, for the relativistic insulators with $m>0$ (\textit{i.e.}, $i=c$), $Q^b_{\mathbf{k}}(u)$ and $P^b_{\mathbf{k}}(u)$
have the same expressions as those in Eq.\ (\ref{WFchern}). On the other hand, for the two phases with $m<0$ (\textit{i.e.}, $i=b,d$), one has
\begin{equation}\label{NegativeM}
\left\{
\begin{split}
Q_{\mathbf{k}}^{b(d)}(u)=&-\cos \int^u_{u_{\text{IR}}} g^r_{\mathbf{k}}(u')du',\\
P_{\mathbf{k}}^{b(d)}(u)=&-e^{-i\theta_{\mathbf{k}}}\sin \int^u_{u_{\text{IR}}} g^r_{\mathbf{k}}(u')du'.\\
\end{split}
\right.
\end{equation}
It is straightforward to check that the IR states for the four phases are $|\Omega^{a(c)}\rangle=\prod_{|\mathbf{k}|\le \Lambda}\psi_2^{\dag}(\mathbf{k})|\text{vac}\rangle$ and $|\Omega^{b(d)}\rangle=\prod_{|\mathbf{k}|\le \Lambda}\psi_1^{\dag}(\mathbf{k})|\text{vac}\rangle$, all of which are un-entangled states.
The difference between $|\Omega^{a(c)}\rangle$
and $|\Omega^{b(d)}\rangle$ are simply caused by the sign change of the mass term $m$.

Next, by considering the boundary condition at the UV limit, \textit{i.e.},
\begin{equation}
|\Psi^i(u=u_{\text{UV}})\rangle=|\Psi^i\rangle,
\end{equation}
where $i=a,b,c,d$,
one can fix the form of $g(u)$ (and thus those of $g^r_{\mathbf{k}}(u)$ and $g_{\mathbf{k}}(u)$) in the disentangler
(see the Appendix for details of calculation) as follows
\begin{equation}\label{gua}
\begin{split}
g^{a}(u)
&=\frac{1}{2}\frac{\Lambda e^u(m+\Lambda^2 e^{2u})}{(m-\Lambda^2 e^{2u})^2+\Lambda^2 e^{2u}}\\
-&\text{arctan}\frac{\Lambda e^u}{\sqrt{(m-\Lambda^2 e^{2u})^2+\Lambda^2 e^{2u}}+(m-\Lambda^2 e^{2u})},
\end{split}
\end{equation}
\begin{equation}\label{gub}
\begin{split}
g^{b}(u)
&=\frac{1}{2}\frac{\Lambda e^u(m+\Lambda^2 e^{2u})}{(m-\Lambda^2 e^{2u})^2+\Lambda^2 e^{2u}}\\
+&\text{arctan}\frac{\Lambda e^u}{\sqrt{(m-\Lambda^2 e^{2u})^2+\Lambda^2 e^{2u}}-(m-\Lambda^2 e^{2u})},
\end{split}
\end{equation}
\begin{equation}\label{guc}
\begin{split}
g^{c }(u)=&
\frac{1}{2}\frac{m\Lambda e^u}{m^2+\Lambda^2 e^{2u}}-\text{arctan}\frac{\Lambda e^u}{\sqrt{m^2+\Lambda^2 e^{2u}}+m},
\end{split}
\end{equation}
and
\begin{equation}\label{gud}
\begin{split}
g^{d}(u)
 =&
\frac{1}{2}\frac{m\Lambda e^u}{m^2+\Lambda^2 e^{2u}}+\text{arctan}\frac{\Lambda e^u}{\sqrt{m^2+\Lambda^2 e^{2u}}-m}.
\end{split}
\end{equation}
\begin{figure}
\includegraphics[width=3.30in]{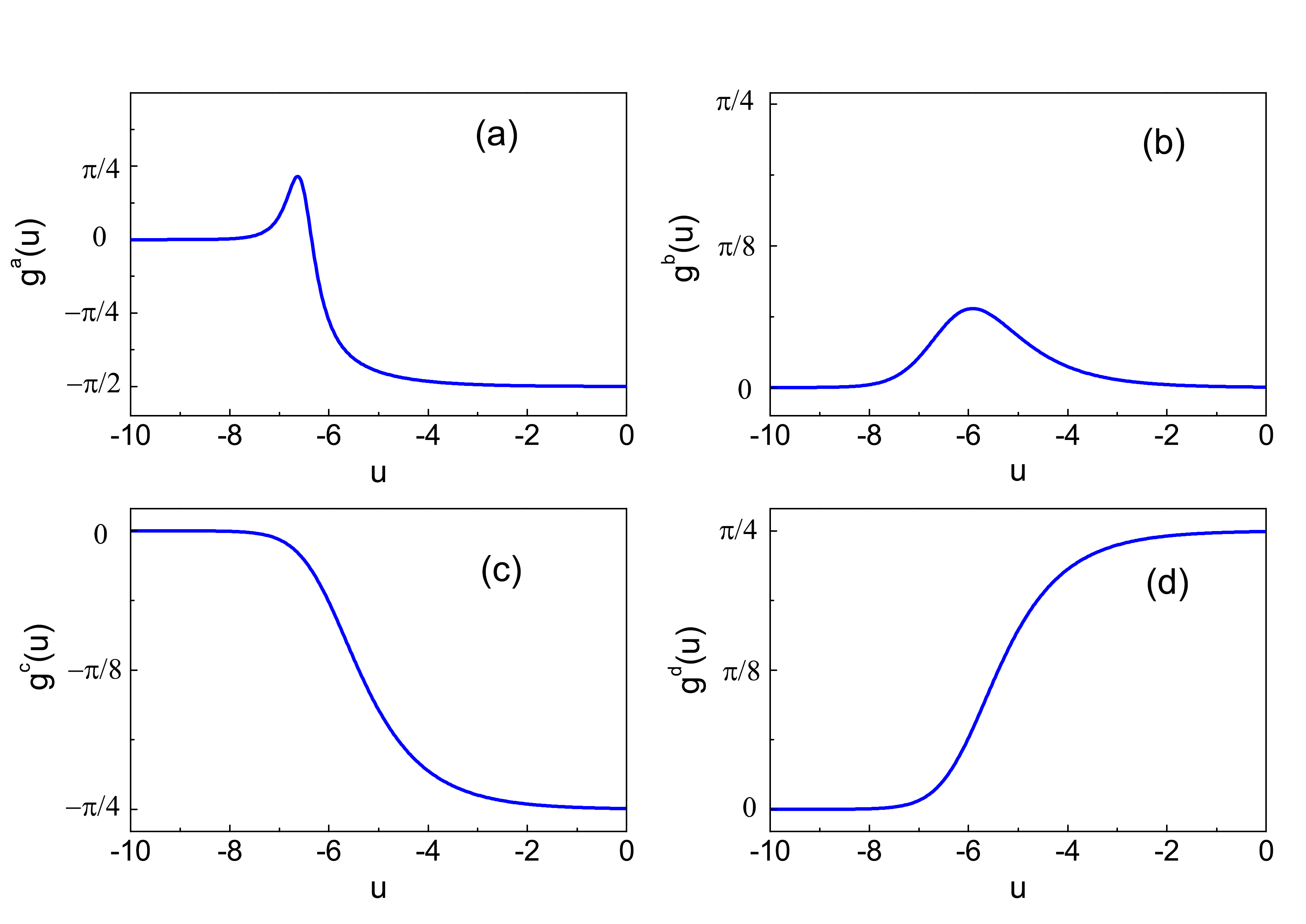}
\caption{$g^i(u)$ for (a) a non-relativistic Chern insulator (b) a non-relativistic trivial insulator
(c) a relativistic insulator with $m>0$ and (d) a relativistic insulator with $m<0$. The parameters we use are (a) $m=2$ (b) $ m=-2$ (c) $m=2$ and (d) $m=-2$. And $\Lambda=1000$ is used for all cases. }
\label{gu}
\end{figure}
As shown in Fig.\ \ref{gu}, it is noted that in the IR limit, one has $g(u)=0$ for all the four phases, which indicates that no
entanglement is added/removed in the IR layers. In the UV limit, one has
\begin{equation}
\left\{
\begin{split}
g^{a}(u_{\text{UV}})=&-\pi/2,\\
g^{b}(u_{\text{UV}})=&0,\\
g^{c}(u_{\text{UV}})=&-\pi/4,\\
g^{d}(u_{\text{UV}})=&+\pi/4.\\
\end{split}
\right.
\end{equation}
Furthermore, there are some more interesting
features in $g^i(u)$. For example, one can observe a
peak as well as a sign change in $g^a(u)$.  We will see how these features play an
important role in determining the Berry curvature flow in the bulk
of cMERA later.
In addition,  by considering the limit $|m|\to 0$ in Eqs. (\ref{guc}) and (\ref{gud}), one can obtain
\begin{equation}
g^c(u,m\to 0^+)=-\frac{\pi}{4}
\end{equation}
and
\begin{equation}\label{Nmassless}
g^d(u,m\to 0^-)=+\frac{\pi}{4},
\end{equation}
which reproduce the result in Ref.\ \onlinecite{Ryu2012}.
$g^{c(d)}(u)$ is independent of layer $u$
because the $|m|\to 0$ limit in relativistic insulators corresponds to a critical point,
and therefore the corresponding bulk theory in cMERA is scale invariant.
This is similar to the scale invariant lattice MERA\cite{Vidalreview01,Vidalreview02},
where both the tensor network structure and disentanglers do not change
as one goes deeper towards $u_{\text{IR}}$.

\subsection{Emergent Holographic Metric in cMERA}
The definition of holographic metric $g_{uu}(u)$ for a general quantum field theory in cMERA
was discussed in Ref.\onlinecite{Ryu2013}. By comparing with the classical gravity limit of AdS/CFT,
the authors find the metric $g_{uu}(u)$ should measure the density of the strength of the
disentanglers. One natural choice is the quantum metric defined through the overlap between wavefunctions $|\Psi(u)\rangle$ and $|\Psi(u+du)\rangle$ in the following
\begin{equation}
g_{uu}(u)=\frac{1}{N}\int d^2\mathbf{k} g_{uu}(u,\mathbf{k}),
\end{equation}
where
\begin{equation}\label{guu001}
g_{uu}(\mathbf{k},u)du^2=1-|\langle \Psi(\mathbf{k},u)|\Psi(\mathbf{k},u+du)\rangle|^2,
\end{equation}
and $N$ is the normalization factor with the concrete form $N=\int d^d\mathbf{k}$.
$|\Psi(\mathbf{k},u)\rangle$ is a single-particle wavefunction defined as
\begin{equation}\label{SingleWave}
|\Psi(\mathbf{k},u)\rangle
=
\left(
P_{\mathbf{k}}(u)\psi^{\dag}_2(\mathbf{k})-Q_{\mathbf{k}}(u)\psi^{\dag}_1(\mathbf{k})
\right)|\text{vac}\rangle.
\end{equation}

To have a better understanding of this definition, one may consider the limit that
no entanglement is added at layer $u$, which means
$\langle \Psi(\mathbf{k},u)|\Psi(\mathbf{k},u+du)\rangle=1$, and therefore one ends
with $g_{uu}(u)=0$. On the other hand, if more entanglement is added at layer $u$,
then the overlap $|\langle \Psi(\mathbf{k},u)|\Psi(\mathbf{k},u+du)\rangle|$ becomes
smaller, and therefore one has a larger $g_{uu}(u)$. This means
$g_{uu}(u)$ can indeed measure the density of the strength of disentanglers.
To see clearly the relation between $g_{uu}(u)$ and disentanglers $\hat{K}(\mathbf{k},u)$,
one notes that Eq.\ (\ref{guu001}) can be rewritten as
\begin{equation}\label{guudefinition}
\begin{split}
g_{uu}(\mathbf{k},u)=&\text{Re}\langle \partial_u\Psi(\mathbf{k},u)|\partial_u \Psi(\mathbf{k},u)\rangle\\
&-\langle \partial_u\Psi(\mathbf{k},u)|\Psi(\mathbf{k},u)\rangle\langle \Psi(\mathbf{k},u)|\partial_u\Psi(\mathbf{k},u)\rangle.
\end{split}
\end{equation}
Then by using Eq.\ (\ref{EOM}), one can immediately obtain
\begin{equation}
\begin{split}
g_{uu}(\mathbf{k},u)=&\langle \Psi(\mathbf{k},u)|\hat{K}^2(\mathbf{k},u)|\Psi(\mathbf{k},u)\rangle\\
&-\left|\Psi(\mathbf{k},u)|\hat{K}(\mathbf{k},u)|\Psi(\mathbf{k},u)\rangle\right|^2.
\end{split}
\end{equation}

Next we will apply the definition of $g_{uu}(u)$ to concrete systems, \textit{e.g.}, Chern insulators in (2+1) dimensions.
The cMERA constructed wavefunction for Chern insulators at layer $u$ has been obtained in Eqs.\ (\ref{GeneralForm001})
and (\ref{WFchern}). Based on Eq.\ (\ref{guudefinition}), one can find
\begin{equation}
\begin{split}
g_{uu}(\mathbf{k},u)=&\left(g^r_{\mathbf{k}}(u)\right)^2=g^2(u)\frac{k^2}{\Lambda^2 e^{2u}}\Gamma\left(\frac{k}{\Lambda e^u}\right).
\end{split}
\end{equation}
Therefore, for Chern insulators in (2+1) dimensions, one can get
\begin{equation}
g_{uu}(u)=\frac{\int_{|\mathbf{k}|\le \Lambda e^u}d^2\mathbf{k} g^2(u)\frac{k^2}{\Lambda^2 e^{2u}} }{\int_{|\mathbf{k}|\le \Lambda e^u} d^2 \mathbf{k}}=\frac{1}{2}g(u)^2.
\end{equation}
In addition, by checking the other three phases, it is straightforward to obtain
\begin{equation}
g_{uu}^i(u)=\frac{1}{2}g^i(u)^2,
\end{equation}
where $i=a,b,c,d$, and the explicit expression of $g^i(u)$ has been obtained in Eqs.\ (\ref{gua})$\sim$(\ref{gud}).
Note that for all the four cases, $g_{uu}^i$ vanishes in the IR layers.

For other components of the metric, one can find their general expressions in Appendices \ref{MetricOther}.

\subsection{Band inversion in cMERA of Chern insulators}

To study the Chern band insulator, it is helpful to check the behavior of pseudo
spin configuration ${\vec d}(\mathbf{k},u):=\langle\Psi(\mathbf{k},u)|\vec{\sigma}|\Psi(\mathbf{k},u)\rangle$,
the $z$ component of which can be used to track the band inversion of the
corresponding Hamiltonian.
For convenience, we denote
\begin{equation}\label{DvarphiK}
\varphi_{\mathbf{k}}(u)=\int^u_{u_{\text{IR}}} g^r_{\mathbf{k}}(u')du'.
\end{equation}
Then based on Eqs.\ (\ref{WFchern}) and (\ref{NegativeM}),
one has
\begin{equation}\label{Pspin}
\begin{split}
d^{a(c)}_z(\mathbf{k},u)=&-\cos 2\varphi^{a(c)}_k(u),\\
d^{b(d)}_z(\mathbf{k},u)=&\cos 2\varphi^{b(d)}_k(u),
\end{split}
\end{equation}
where the minus sign difference results from $m>0$ for phase $a(c)$ and $m<0$ for phase $b(d)$.

\begin{figure}
\includegraphics[width=3.50in]{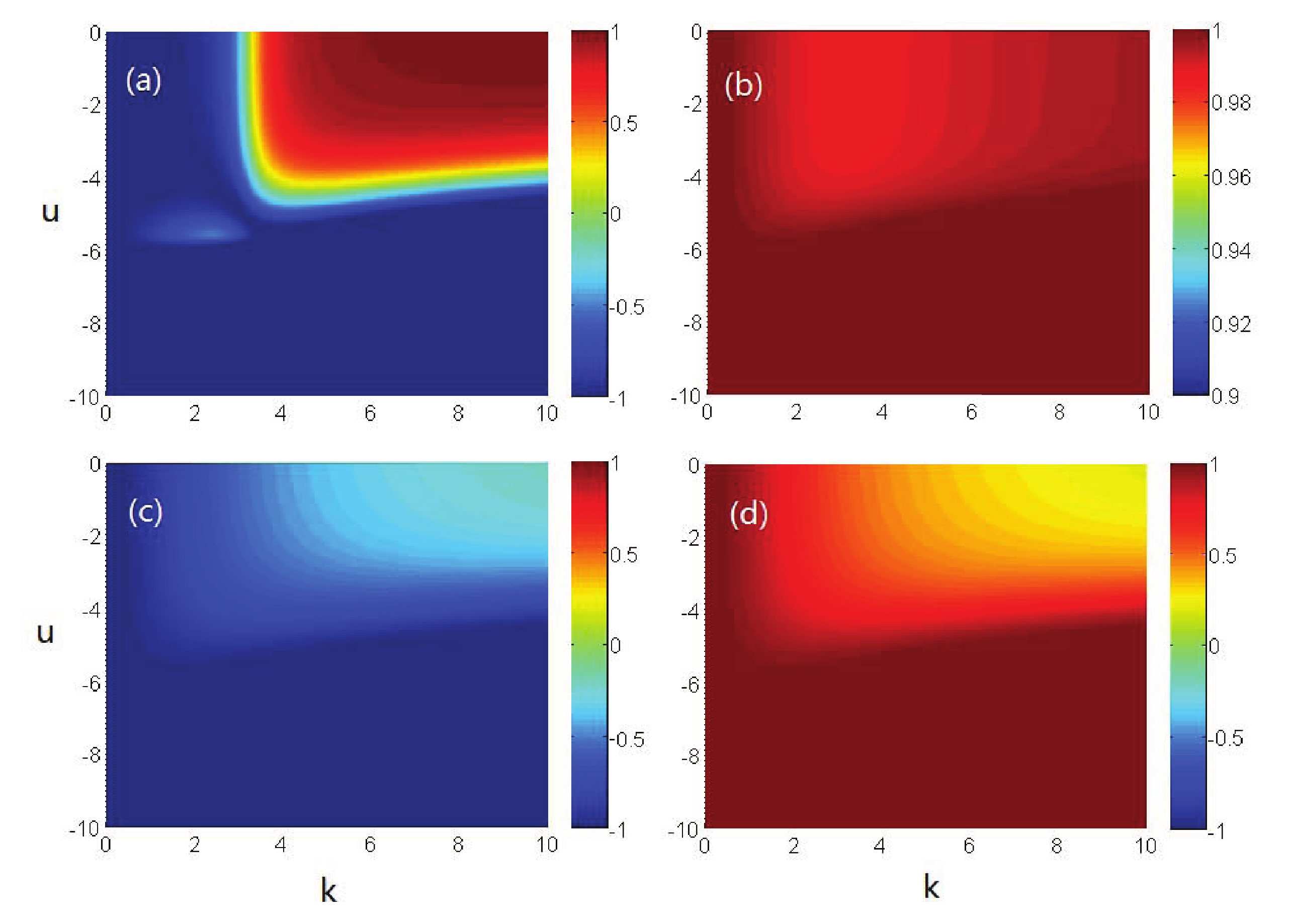}
\caption{ $d^{i}_z(\mathbf{k},u)$  for (a) a non-relativistic Chern insulator (b) a non-relativistic trivial insulator (c) a relativistic insulator with $m>0$ and (d) a relativistic insulator with $m<0$. Band inversion happens only for case (a), which is related with the UV behavior of $g^i(u)$. The parameters we choose are $m_a=m_c=10$, $m_b=m_d=-10$ and $\Lambda=1000$, based on which one has $u^{\ast}\simeq -5.76$ by using Eq.\ (\ref{ustar}). }\label{inversion}
\end{figure}

As shown in Fig.\ \ref{inversion}, we plot $\langle\Psi(\mathbf{k},u)|\sigma^z|\Psi(\mathbf{k},u)\rangle$
as a function of layer $u$ and momentum $|\mathbf{k}|$ in the region $|\mathbf{k}|\ll \Lambda$.
One can find that the band inversion happens only for the non-relativistic Chern insulator,
which agrees with our knowledge in the UV limit.
As shown in Fig.\ \ref{inversion}(a), for cMERA of Chern insulators, band inversion happens
in the UV layer $u=u_{\text{UV}}$.
As $u$ goes deeper towards $u_{\text{IR}}$, the band inversion insists until $u$ is near $u^{\ast}$,
which is defined by
\begin{equation}\label{ustar}
|m|=\Lambda^2 e^{2u^{\ast}}.
\end{equation}
It is noted that for relativistic insulators, $u^{\ast}$ is defined by $|m|=\Lambda e^{u^{\ast}}$.
Next we will discuss how the band inversion in region $|\mathbf{k}|\ll \Lambda$
is related with the behavior of $g^i(u)$. For convenience, we divide each plot in Fig.\ \ref{inversion}
into three regions as follows.
\begin{equation}\nonumber
\left\{
\begin{split}
&\text{Region I}: \ \ \ u_{\text{IR}}<u<u^{\ast},\\
&\text{Region II}: \ \ \ u^{\ast}<u\le u_{\text{UV}}, \ \ k<k^{\ast},\\
&\text{Region III}: \ \ \ u^{\ast}<u\le u_{\text{UV}}, \ \ k^{\ast}<k\ll \Lambda,
\end{split}
\right.
\end{equation}
where $k^{\ast}$ is defined by
$|m|=(k^{\ast})^2$ for non-relativistic insulators, and $|m|=k^{\ast}$ for relativistic insulators.
The relation between $k^{\ast}$ and $u^{\ast}$ is
\begin{equation}\label{kstar}
k^{\ast}=\Lambda e^{u^{\ast}}.
\end{equation}

Then the behavior of $d^i_z(\mathbf{k},u)$ can be analyzed as follows.

(i) \text{Region I}:
From the behavior of  $g^i(u)$ in Fig.\ \ref{gu}, we have $g^i(u<u^{\ast})\simeq 0$.
Therefore, by using the definition in Eq.\ (\ref{DvarphiK}), one has $\varphi_{\mathbf{k}}^i\simeq 0$.
Then one can immediately obtain
\begin{equation}\label{DregionsI}
\left\{
\begin{split}
d^{a(c)}_z(\mathbf{k},u)\simeq &-1,\\
d^{b(d)}_z(\mathbf{k},u)\simeq&+1.
\end{split}
\right.
\end{equation}
Note that the result is independent of momentum $\mathbf{k}$. In other words,
for $u_{\text{IR}}<u<u^{\ast}$, as we change momentum $k$, there is no band inversion happening.

(ii)
\text{Region II}:
In this region, $g^i(u)$ has a finite value for $u^{\ast}<u<u_{\text{UV}}$.
However, the factor $|\mathbf{k}|/\Lambda e^u$ in Eq.\ (\ref{gku001}) goes to zero as we increase
$u$ from $u=u^{\ast}$. Then, again, one has $\varphi_{\mathbf{k}}^i(u)\simeq 0$ and $d^i_z(\mathbf{k},u)$
shows the same feature as that in Eq.\ (\ref{DregionsI}).

(iii) \text{Region III}:
In this region, one can replace $g^{i}(u)$ with $g^{i}(u_{\text{UV}})$ as an approximation.
Then $\varphi_{\mathbf{k}}^{i}(u)$ in Eq.\ (\ref{DvarphiK}) can be expressed as
\begin{equation}\nonumber
\varphi_{\mathbf{k}}^{i}(u> u^{\ast})=g^{i}(u_{\text{UV}})\int_{\log k/\Lambda}^{u}du'\frac{k}{\Lambda e^{u'}},
\end{equation}
which can be simplified as
\begin{equation}\nonumber
\varphi_{\mathbf{k}}^{i}(u> u^{\ast})=g^{i}(u_{\text{UV}}) \left(1-\frac{k}{\Lambda e^u}\right).
\end{equation}
By considering the limit $u\to u_{\text{UV}}$ and $k\ll \Lambda e^{u_{\text{UV}}}$, it is straightforward to obtain
\begin{equation}\nonumber
\left\{
\begin{split}
d^{a}_z(\mathbf{k},u\to u_{\text{UV}})&\to 1\\
d^{b}_z(\mathbf{k},u\to u_{\text{UV}})&\to 1\\
d^{c}_z(\mathbf{k},u\to u_{\text{UV}})&\to 0^{-}\\
d^{d}_z(\mathbf{k},u\to u_{\text{UV}})&\to 0^{+}\\
\end{split}
\right.
\end{equation}
which can be observed in the upper right of each plot in Fig.\ \ref{inversion}.

In short sum, with appropriate approximation, we show that (i) For $u_{\text{IR}}<u<u^{\ast}$,
there is no band inversion for all the four phases as we change momentum $k$. (ii) For $u^{\ast}<u<u_{\text{UV}}$,
as we increase the momentum $k$ across $k=k^{\ast}$, the value of $d^{i}_z(\mathbf{k},u)$
changes as follows
\begin{equation}\nonumber
\left\{
\begin{split}
d^{a}_z(\mathbf{k},u):\ \ \ -1\to +1,\\
d^{b}_z(\mathbf{k},u):\ \ \ +1\to +1,\\
d^{c}_z(\mathbf{k},u):\ \ \ -1\to 0^-,\\
d^{d}_z(\mathbf{k},u):\ \ \ +1\to 0^+.\\
\end{split}
\right.
\end{equation}
One can find that only the non-relativistic Chern insulator shows the band inversion behavior
for $u>u^{\ast}$,
which indicates that the system is in a topologically nontrivial phase. On the other hand,
for $u<u^{\ast}$, there is no band inversion happening as we change $k$, which
indicates the system is in a topologically trivial phase. Therefore, as $u$ goes across
$u^{\ast}$ from the IR side to the UV side, it seems that we have a phase transition
from a topologicaly trivial phase to a topologically nontrvial phase.
Therefore, it may be viewed as a `topological phase transition' in the direction of entanglement
renormalization.

Before we end this part, we emphasize that the discussion above is based on the assumption
$k\ll \Lambda$, \emph{i.e.}, we focus on the low energy physics region. In the following parts,
we will study the topological property of the four systems in the whole region $0\le k\le\Lambda$.

\subsection{Berry curvature flow in cMERA of Chern insulators }

To further understand the `topological phase transition' in the previous part,
we study the Berry curvature flow in the bulk of cMERA for a Chern insulator.
It is known that Chern insulators are distinguished from trivial insulators by a nonzero
quantized Chern number, which can be viewed as a Berry flux in momentum space.
Therefore, there must be some Berry curvature emanated from the UV layer of cMERA
for a Chern insulator.
On the other hand, we know that the IR state is unentangled and there is no Berry curvature.
One may ask where does the Berry curvature flow? We will study this problem
in this part.

\begin{figure}
\includegraphics[width=3.50in]{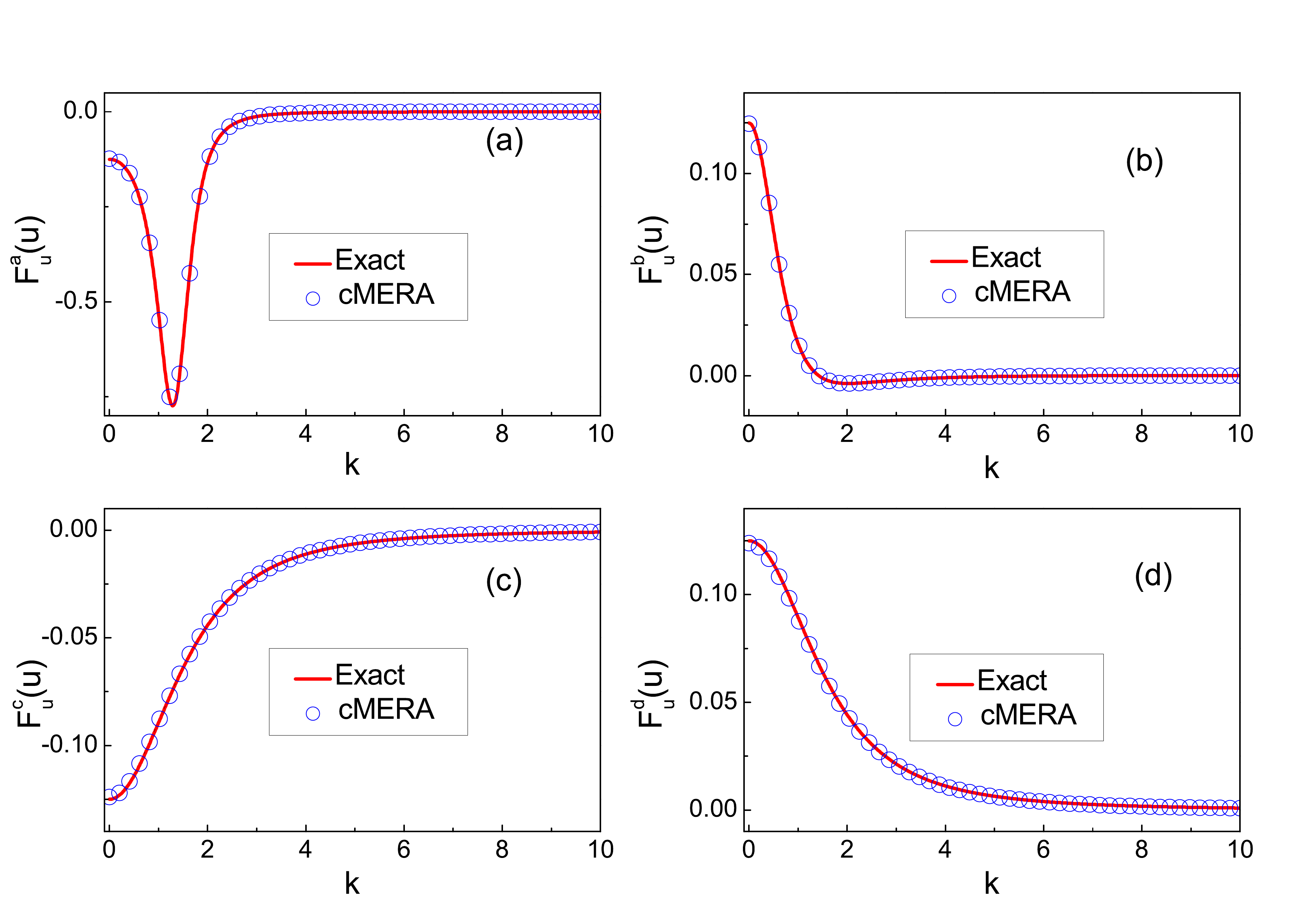}
\caption{Comparison of the cMERA constructed Berry curvature $\mathcal{F}^i_{u}(k,\theta_{\mathbf{k}};u=u_{\text{UV}})$ and
the exact results for (a) a non-relativistic Chern insulator (b) a non-relativistic trivial insulator (c) a relativistic  insulator with $m>0$ and (d) a relativistic insulator with $m<0$. The parameters we used are (a) $m=2$ (b) $m=-2$ (c) $m=2$ and (d) $m=-2$. We use $\Lambda=1000$ for all cases. }\label{BerryCompare}
\end{figure}

Based on the cMERA constructed single particle wavefunction $|\Psi(\mathbf{k},u)\rangle$
in Eq.\ (\ref{SingleWave}), one can obtain the Berry connection for a Chern insulator as follows
\begin{small}
\begin{equation}\label{connection}
\left\{
\begin{split}
\mathcal{A}_k(k,\theta_{\mathbf{k}};u)=&-i\langle \Psi(\mathbf{k},u)|\partial_{k} |\Psi(\mathbf{k},u)\rangle=0,\\
\mathcal{A}_{\theta_{\mathbf{k}}}(k,\theta_{\mathbf{k}};u)=&-\frac{i}{k} \langle \Psi(\mathbf{k},u)|\partial_{\theta_{\mathbf{k}}} |\Psi(\mathbf{k},u)\rangle
=-\frac{1}{k}\sin^2\varphi_{\mathbf{k}}(u),\\
\mathcal{A}_u(k,\theta_{\mathbf{k}};u)=&-i\langle \Psi(\mathbf{k},u)|\partial_{u} |\Psi(\mathbf{k},u)\rangle=0.\\
\end{split}
\right.
\end{equation}
\end{small}
The Berry curvature can be obtained by calculating
\begin{equation}\nonumber
\vec{\mathcal{F}}=\mathbf{\nabla}\times \vec{\mathcal{A}},
\end{equation}
which can be explicitly expressed as
\begin{equation}\label{vector}
\begin{split}
\vec{\mathcal{F}}^a(k,\theta_{\mathbf{k}};u)=&\hat{\mathbf{u}} \mathcal{F}_u^a(k,\theta_{\mathbf{k}};u)+\hat{\mathbf{k}} \mathcal{F}_k^a(k,\theta_{\mathbf{k}};u)\\
=&\hat{\mathbf{u}}\left[-\frac{1}{k}\sin2\varphi_{\mathbf{k}}(u)\partial_k\varphi_{\mathbf{k}}(u)\right]\\
&+\hat{\mathbf{k}}\left[\frac{1}{k}\sin2\varphi_{\mathbf{k}}(u)\partial_u\varphi_{\mathbf{k}}(u)\right].
\end{split}
\end{equation}
$\hat{\mathbf{u}}$ and $\hat{\mathbf{k}}$ are unit vectors along the renormalization direction and the
momentum direction, respectively.
This is an emergent Berry curvature due to the extra renormalization direction $\hat{\mathbf{u}}$.
By checking the other three phases with the same procedures, one can find that $\vec{\mathcal{F}}^{c}(k,\theta_{\mathbf{k}};u)$
for the relativistic insulators with $m>0$ has the same expression as that in Eq.\ (\ref{vector}).
For the other two phases with $i=b$ and $d$, one has
\begin{equation}\label{Fbd}
\begin{split}
\vec{\mathcal{F}}^{b(d)}(k,\theta_{\mathbf{k}};u)=&\hat{\mathbf{u}}\left[\frac{1}{k}\sin2\varphi_k^{b(d)}(u)\partial_k\varphi_k^{b(d)}(u)\right]\\
&+\hat{\mathbf{k}}\left[-\frac{1}{k}\sin2\varphi_k^{b(d)}(u)\partial_u\varphi_k^{b(d)}(u)\right].
\end{split}
\end{equation}
Again, the sign difference between cases $a(c)$ and $b(d)$ is caused by the sign change
of mass term. To check the validity of the formulas in Eqs.\ (\ref{vector}) and (\ref{Fbd}),
we compare the cMERA constructed $\mathbf{\mathcal{F}}^i_u(k,\theta_{\mathbf{k}};u=u_{\text{UV}})$
with the exact results in the low energy physics region.
As shown in Fig.\ \ref{BerryCompare}, the cMERA results agree with the
exact results  in an excellent way.

\begin{figure}
\includegraphics[width=3.15in]{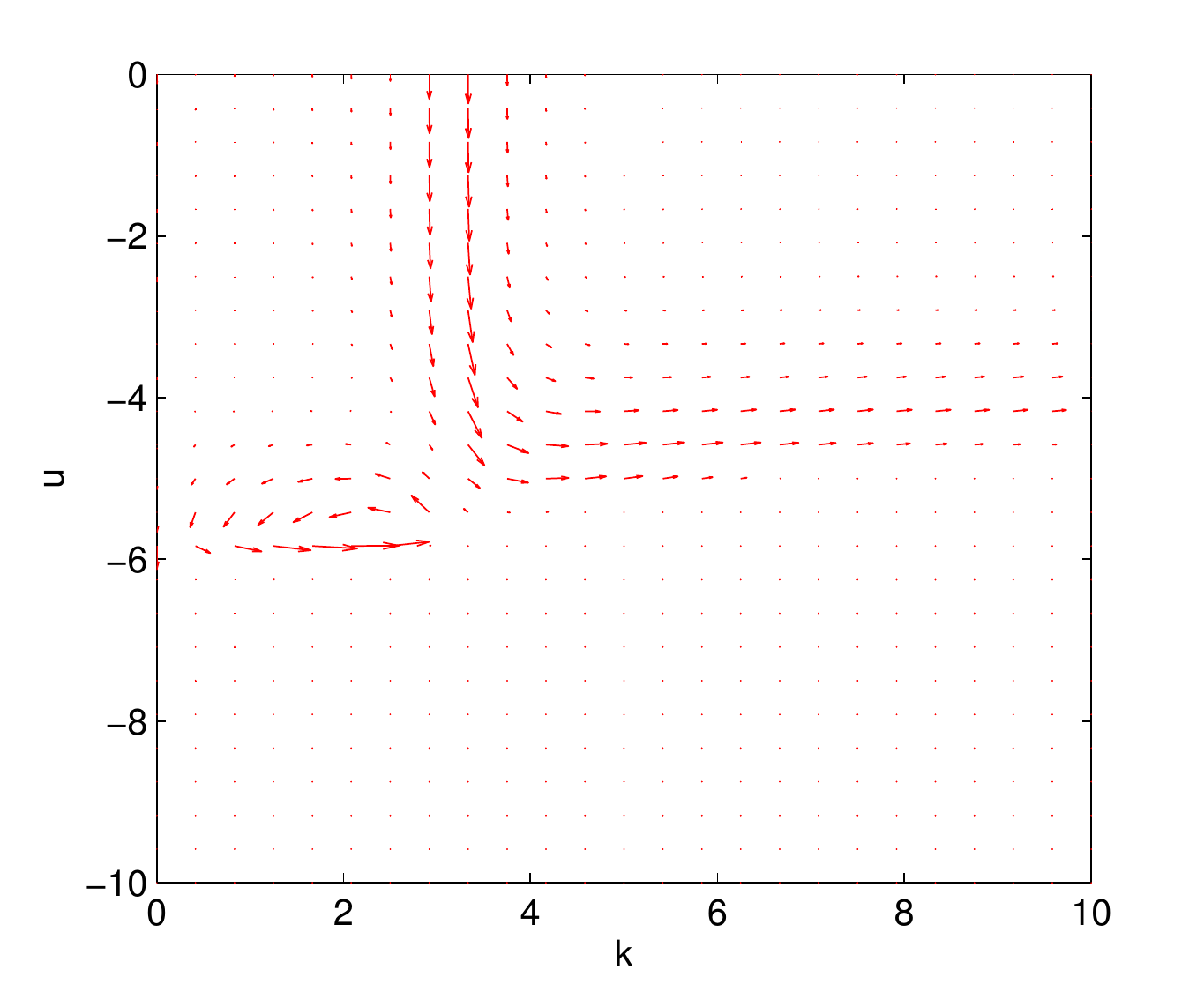}
\caption{ Berry curvature flow in cMERA of Chern insulators. The Berry curvature emanated from the
UV layer is bent backward along $k=\Lambda e^u$, before it reaches $u=u^{\ast}$.
In addition, a vortex feature develops near $u=u^{\ast}$. The parameters we use are $m=10$ and $\Lambda=1000$, based on which one has $u^{\ast}\simeq -5.76$. }\label{BerryFlowChern}
\end{figure}

Next, we will focus on the case of Chern insulators, and see what happens for
$\vec{\mathcal{F}}(k,\theta_{\mathbf{k}};u)$ if $u$ deviates from $u=u_{\text{UV}}$ and goes deeper
towards $u_{\text{IR}}$. In other words, we hope to study the Berry curvature flow in the
bulk of cMERA. As shown in Fig.\ \ref{BerryFlowChern},
according to Eqs.\ (\ref{DvarphiK}) and (\ref{vector}),  we plot $\vec{\mathcal{F}}(k,\theta_{\mathbf{k}};u)$ as
a function of momentum $k=|\mathbf{k}|$ and layer $u$.
It is found that the Berry curvature $\mathcal{\vec{F}}(k,\theta_{\mathbf{k}};u)$
emanated from the UV layer $u_{\text{UV}}$ flows towards the IR layer $u_{\text{IR}}$.
Before it reaches $u=u^{\ast}$, $\mathcal{\vec{F}}(k,\theta_{\mathbf{k}};u)$ is bent backwards
along $k=\Lambda e^u$. In addition, it can be observed that a vortex feature develops
near $u=u^{\ast}$.

Then we may ask two questions. (i) How does the vortex feature in $\mathcal{\vec{F}}(k,\theta_{\mathbf{k}};u)$
arise? (In the appendices, we also calculate the Berry curvature flow for the other three phases,
and there is no vortex feature for these three phases.) (ii) Now that the Berry curvature is
bent backwards along $k=\Lambda e^u$, where does the Berry curvature flow finally? For question (i),
as discussed in detail in Appendices \ref{vortex}, it is shown that the vortex feature in the Berry curvature
flow is mainly caused by the sign change of $g^a(u)$ in Fig.\ \ref{gu}(a). Now we are mainly interested in question (ii) as follows.

\begin{figure}
\includegraphics[width=3.50in]{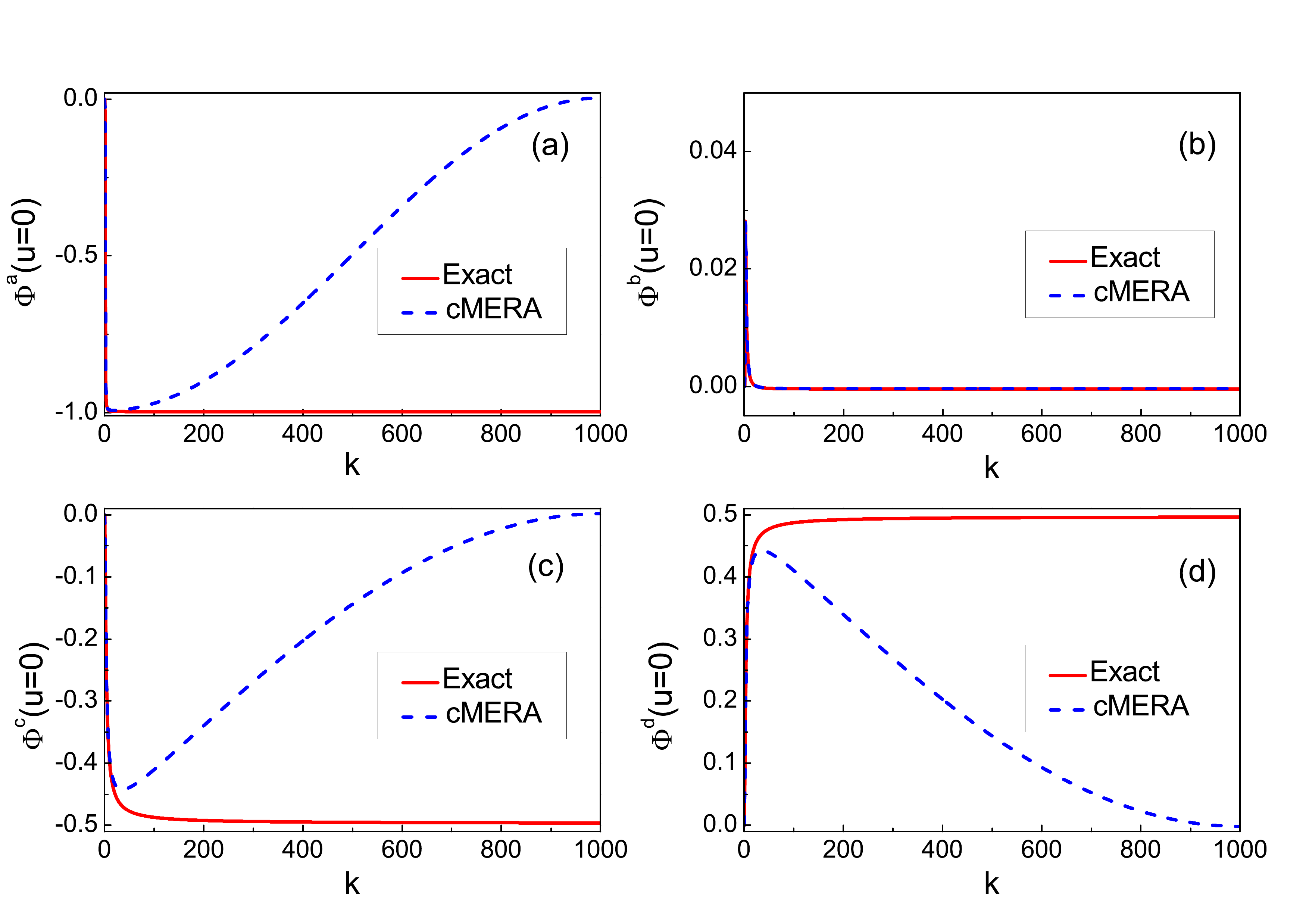}
\caption{Berry flux $\Phi^i(u=0)/2\pi$ for (a) a non-relativistic Chern insulator (b) a non-relativistic trivial insulator
(c) a relativistic Chern insulator with $m>0$ and (d) a relativistic Chern insulator with $m<0$. The parameters we used are (a) $m=2$ (b) $m=-2$ (c) $m=2$ and (d) $m=-2$. We use $\Lambda=1000$ for all cases. }\label{Flux}
\end{figure}

At the UV layer $u_{\text{UV}}$, we calculate the Berry flux in the region
$k'\le k$, {\it i.e.},
\begin{equation}\nonumber
\Phi(k,u=0)=\int d\theta_{\mathbf{k}} \int^k k' dk'\mathcal{F}_u(k',\theta_{\mathbf{k}};u=0),
\end{equation}
and compare it with the exact result,  as shown in Fig.\ \ref{Flux}.
For the case of Chern insulators in Fig.\ \ref{Flux} (a), one can find that
for $k\ll \Lambda$, the Berry flux $\Phi(k,u=0)/2\pi$ calculated from cMERA
agrees with the exact result very well, and it reaches $-1$ at certain $k$,
which is much smaller than $\Lambda$. However, as $k$ increases
further, the Berry flux deviates from the exact result, and decays from $-1$
to $0$ gradually as $k\to \Lambda$. This indicates that the cMERA result
is not exact for large $k$, which was also observed in Ref.\ \onlinecite{Haegeman}.
In addition, because we do not find any `source' or `drain' for the Berry curvature
in the bulk of cMERA, this total zero flux $\Phi(k=\Lambda,u=0)=0$ indicates that all
the Berry curvature emanated from the low energy physics region of the UV layer
flows back to the UV layer itself.

\begin{figure*}[htp]
\centering
\includegraphics[width=5.00in]{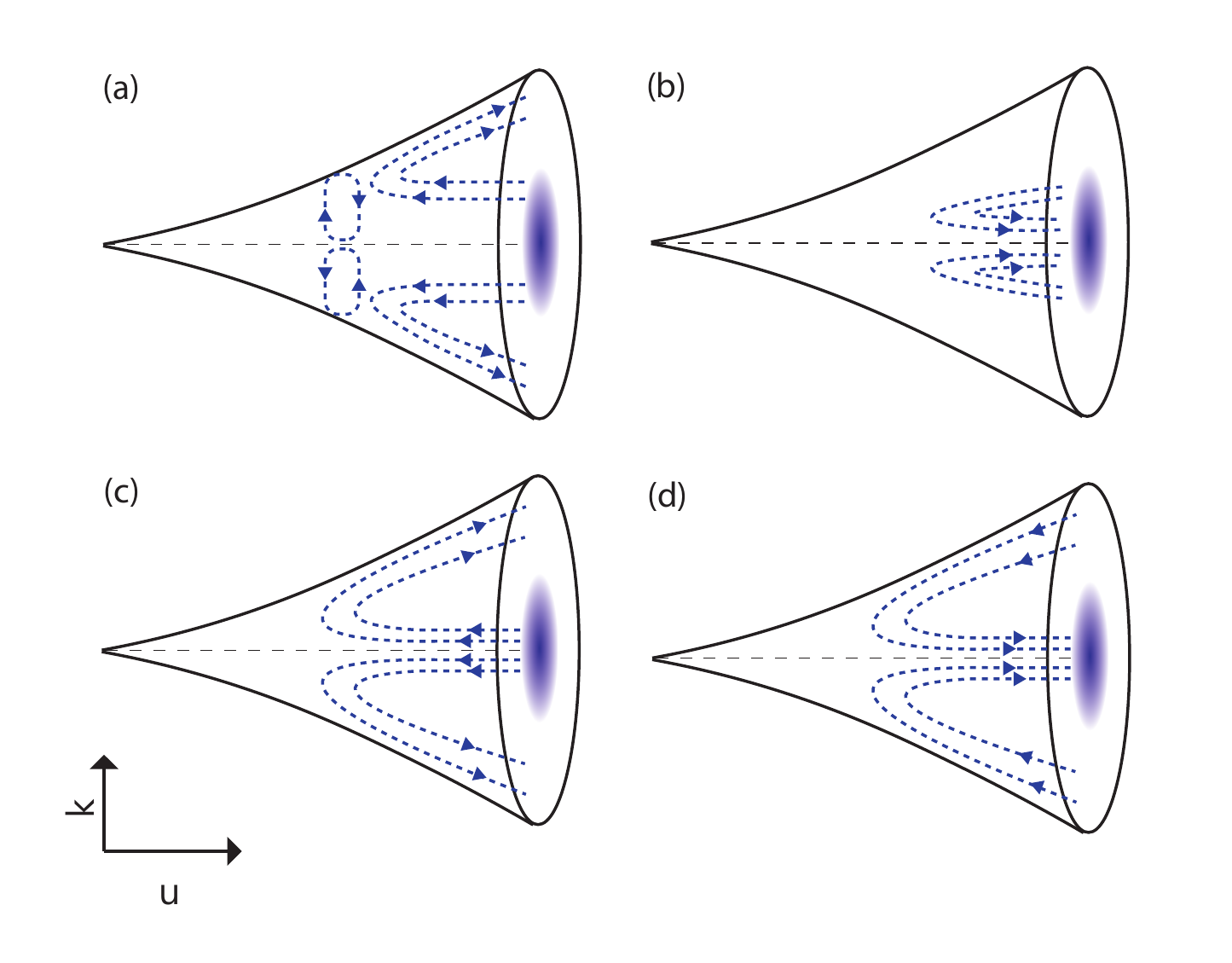}
\caption{ Schematic plot of Berry curvature flow for (a) a non-relativistic Chern insulator, (b) a non-relativistic trivial insulator, (c) a relativistic Chern insulator with $m>0$ and (d) a relativistic Chern insulator with $m<0$.
}\label{BerryFlow4}
\end{figure*}

In fact, the conclusion above can be more transparently understood by checking the cMERA constructed wavefunction.
It is noted there is no singularity in the cMERA constructed wavefunction
$|\Psi(\mathbf{k},u)\rangle$. Therefore, based on Eq.\ (\ref{connection}), the Berry flux can be expressed as
\begin{equation}\nonumber
\begin{split}
\frac{\Phi(k,u=0)}{2\pi}=&\frac{1}{2\pi}\int  \mathcal{A}(k,\theta_{\mathbf{k}};u=0)k d\theta_{\mathbf{k}}\\
=&-\sin^2\varphi_{\mathbf{k}}(u=0).
\end{split}
\end{equation}
Considering $\varphi_{\mathbf{k}}(u=0)=0$ for $k=\Lambda$, one immediately obtains
\begin{equation}\nonumber
\Phi(k=\Lambda,u=0)=0,
\end{equation}
which agrees with our numerical calculation. Similarly, at each layer $u$, one can find
$
\Phi(k=\Lambda,u)=0.
$
Therefore, the total Berry flux at each layer is conserved to be zero.

To conclude, in this part we study the Berry curvature flow in the bulk of cMERA for a Chern insulator.
In the low energy physics region $k\ll \Lambda$, cMERA can reproduce the exact
results on Berry curvature in the UV layer. However, it is found that
the Berry curvature, which is emanated from the low energy physics region, after
bent back near $u=u^{\ast}$, flows backwards to the large $k$ region in the layer
$u=u_{\text{UV}}$, as schematically shown in Fig.\ \ref{BerryFlow4} (a).
Therefore, the cMERA constructed wavefunction in the whole layer $u=u_{\text{UV}}$
is topologically trivial, although we can see the band inversion feature in
the low energy physics region. From this point of view, the `topological phase
transition' we found in the previous part is not a true phase transition.

It is interesting to compare the Berry curvature flow in cMERA for all the four phases.
By repeating the same procedures for Chern insulators (see Appendices \ref{cMERA4}), we obtain
the Berry curvature flow in the low energy physics region. (See Fig.\ \ref{BerryB},
Fig.\ \ref{BerryC} and Fig.\ \ref{BerryD}, respectively.) It is found that there is no
vortex feature in cMERA for the other three phases, because there is no sign change in the corresponding $g^i(u)$.

We also check the Berry flux distribution in the whole region $0\le |k|\le \Lambda$ for the four phases,
as shown in Fig.\ \ref{Flux}. For relativistic insulators with both $m>0$ and $m<0$,
one has similar conclusions as that of non-relativistic Chern insulators. The Berry
flux $\Phi(k,u)/2\pi$ reaches $\sim\mp\frac{1}{2}$ at certain $k$ which is much
smaller than $\Lambda$, and then decays to zero gradually as $k$ increases
to $\Lambda$. On the other hand, for the non-relativistic trivial insulators, the Berry
flux $\Phi(k,u)/2\pi$ obtained from cMERA agrees with the exact result in the whole
region $0\le |k|\le \Lambda$. In addition, one can find that the total Berry flux in the low energy
physics region $|k|\ll \Lambda$ is already zero, and the large $k$ region does not contribute
any Berry curvature.

Based on the analysis above, we summarize the features of Berry curvature
flow in cMERA for the four phases as follows, with the schematic plotting shown in Fig.\ \ref{BerryFlow4}:

\textit{(a) Non-relativistic Chern insulator:}

 A bundle of Berry curvature with a total Berry
flux $-2\pi$ is emanated from the low energy physics region in the UV layer
$u=u_{\text{UV}}$. These Berry curvature is bent backwards near $u=u^{\ast}$,
and flows back to the large $k$ region in the UV layer. In addition, a vortex feature
develops near $u=u^{\ast}$.

\textit{(b) Non-relativistic trivial insulator:}

 A bundle of Berry curvature is emanated from
the low energy physics region in the UV layer $u=u_{\text{UV}}$.
These Berry curvature is bent backwards near $u=u^{\ast}$, and flows back to
the low energy physics region itself in the UV layer. No Berry curvature is emanated
 or absorbed in the large $k$ region.

\textit{(c) Relativistic insulator with $m>0$:}

 A bundle of Berry curvature with a total
Berry flux $-\pi$ is emanated from the low energy physics region in the UV layer
$u=u_{\text{UV}}$. These Berry curvature is bent backwards near $u=u^{\ast}$,
and flows back to the large $k$ region in the UV layer.

\textit{(d) Relativistic insulator with $m<0$:}

 A bundle of Berry curvature with a total
Berry flux $-\pi$ is emanated from the large $k$ region in the UV layer
$u=u_{\text{UV}}$. These Berry curvature is bent backwards near $u=u^{\ast}$,
and flows back to the low energy physics region in the UV layer. In other words,
we simply reverse the direction of Berry curvature flow in (c).

\section{cMERA of topological insulators with a topologically nontrivial IR state}
\label{SecIV}

We show in the previous section that, with a topologically trivial IR state,
one cannot construct the exact ground state of a Chern insulator with a nonzero Chern number.
To recover the nontrivial topological
property of the exact ground state, we may have to consider a cMERA with a topologically
nontrivial IR state.

Before we move on to the cMERA with a topologically nontrivial IR state, it is helpful to
review the prior works on the lattice MERA construction of topological phases.
In Refs.\ \onlinecite{Vidal2008top} and \onlinecite{Vidal2009top}, the lattice MERA of Kitaev's toric code
model and Levin-Wen's string-net model have been constructed in an exact way.
It is found that the state at each layer of the lattice MERA has nontrivial topological
properties, and it will never flow to a topologically trivial IR state. Recently, the symmetry
protected entanglement renormalization was proposed\cite{Vidal2013}, which is applied
to the lattice MERA construction of a symmetry protected topological (SPT) phase.
In particular, for the AKLT state, it is found that as long as the $Z_2^{T}$ symmetry
is preserved in the process of RG flow, the state in each layer of the lattice MERA is
nontrivial in topology. In addition, in Ref.\ \onlinecite{Swingle14}, although a
concrete lattice MERA network for a Chern band insulator is still difficult to find,
procedures to construct the lattice MERA are proposed: Starting from a `top' tensor,
which represents the exact ground state of a small cluster of a Chern insulator,
by using disentangler and isometry operations on and on, one may be able
to construct the ground state of a Chern insulator in a very large size.
Apparently, the state at each layer inherits the topologically nontrivial property from
the `top' tensor.
In short, based on previous works, it suggests that in the lattice MERA, a topologically nontrivial UV state
always flows to a topologically nontrivial IR state. Therefore, we believe that in cMERA, a
continuous version of the lattice MERA, we may have a parallel story.

\subsection{cMERA of Chern insulators in (2+1)D with a topologically nontrivial IR state}

The main procedures are the same as those in Sec. III. The cMERA constructed many-body wavefunction
at each layer has the form
\begin{equation}\label{WFchernNontrivial}
\begin{split}
|\Psi(u)\rangle
=&
\prod_{|\mathbf{k}|\le \Lambda}\left(
P_{\mathbf{k}}(u)\psi^{\dag}_2(\mathbf{k})-Q_{\mathbf{k}}(u)\psi^{\dag}_1(\mathbf{k})
\right)|\text{vac}\rangle.
\end{split}
\end{equation}
where the expressions of $Q_{\mathbf{k}}(u)$ and $P_{\mathbf{k}}(u)$ can be found in Eq.\ (\ref{QP}).
For convenience, we rewrite them here
\begin{equation}\label{QP002}
\left\{
\begin{split}
Q_{\mathbf{k}}(u)=&-ie^{-i\theta_{\mathbf{k}}}\left(
A^{\ast}e^{-i\int^u g^r_{\mathbf{k}}(u')du'}-B^{\ast}e^{i\int^u g^r_{\mathbf{k}}(u')du'}
\right),\\
P_{\mathbf{k}}(u)=&Ae^{i\int^u g^r_{\mathbf{k}}(u')du'}+Be^{-i\int^u g^r_{\mathbf{k}}(u')du'}.
\end{split}
\right.
\end{equation}
where $g_{\mathbf{k}}^r(u)$ is defined through Eq.\ (\ref{gku001}).
Instead of choosing $A=B=1/2$, to have a topologically nontrivial IR state,
we choose $A=-B=-i/2$.
Then one can obtain
\begin{equation}\label{WFchern001}
\left\{
\begin{split}
Q_{\mathbf{k}}(u)=&e^{-i\theta_{\mathbf{k}}}\cos \int^u_{u_{\text{IR}}} g^r_{\mathbf{k}}(u')du'\\
P_{\mathbf{k}}(u)=&\sin \int^u_{u_{\text{IR}}} g^r_{\mathbf{k}}(u')du',
\end{split}
\right.
\end{equation}
In particular, in the IR limit, one has
\begin{equation}\label{IRnontrivial}
|\Omega^{\text{nontrivial}}\rangle:=
|\Psi(u\to u_{\text{IR}})\rangle=
\prod_{|\mathbf{k}|\le \Lambda}\left(-e^{-i\theta_{\mathbf{k}}}\psi_1^{\dag}(\mathbf{k})\right)|\text{vac}\rangle.
\end{equation}
Here we use `nontrivial' because the cMERA constructed wavefunction $|\Psi(u)\rangle$
at arbitrarily finite layer $u$  carries a nonzero Berry
flux $\Phi/2\pi=-1$, as discussed in detail later.
Next, by requiring
\begin{equation}
|\Psi(u=u_{\text{UV}})\rangle=|\Psi\rangle,
\end{equation}
where $|\Psi\rangle$ is the exact ground state of a Chern insulator in Eq.\ (\ref{target}), one can obtain
\begin{equation}\label{gNontrivial}
\begin{split}
g^{\text{nontrivial}}(u)
&=\frac{1}{2}\frac{\Lambda e^u(m+\Lambda^2 e^{2u})}{(m-\Lambda^2 e^{2u})^2+\Lambda^2 e^{2u}}\\
+&\text{arctan}\frac{\sqrt{(m-\Lambda^2 e^{2u})^2+\Lambda^2 e^{2u}}+(m-\Lambda^2 e^{2u})}{\Lambda e^u}.
\end{split}
\end{equation}
It is straightforward to check that
\begin{equation}
g^{\text{nontrivial}}(u)=g^{\text{trivial}}(u)+\frac{\pi}{2},
\end{equation}
where $g^{\text{trivial}}(u)$ is $g(u)$ obtained from the cMERA with a topologically
trivial IR state(see Eq.\ (\ref{gua})).

Following the same procedures in Sec.III, one can also define the emergent holographic
metric with the expression
\begin{equation}
g^{\text{nontrivial}}_{uu}(u)=\frac{1}{2}\left[g^{\text{nontrivial}}(u)\right]^2.
\end{equation}

\subsection{Band inversion in cMERA of Chern insulators}

\begin{figure}
\includegraphics[width=3.30in]{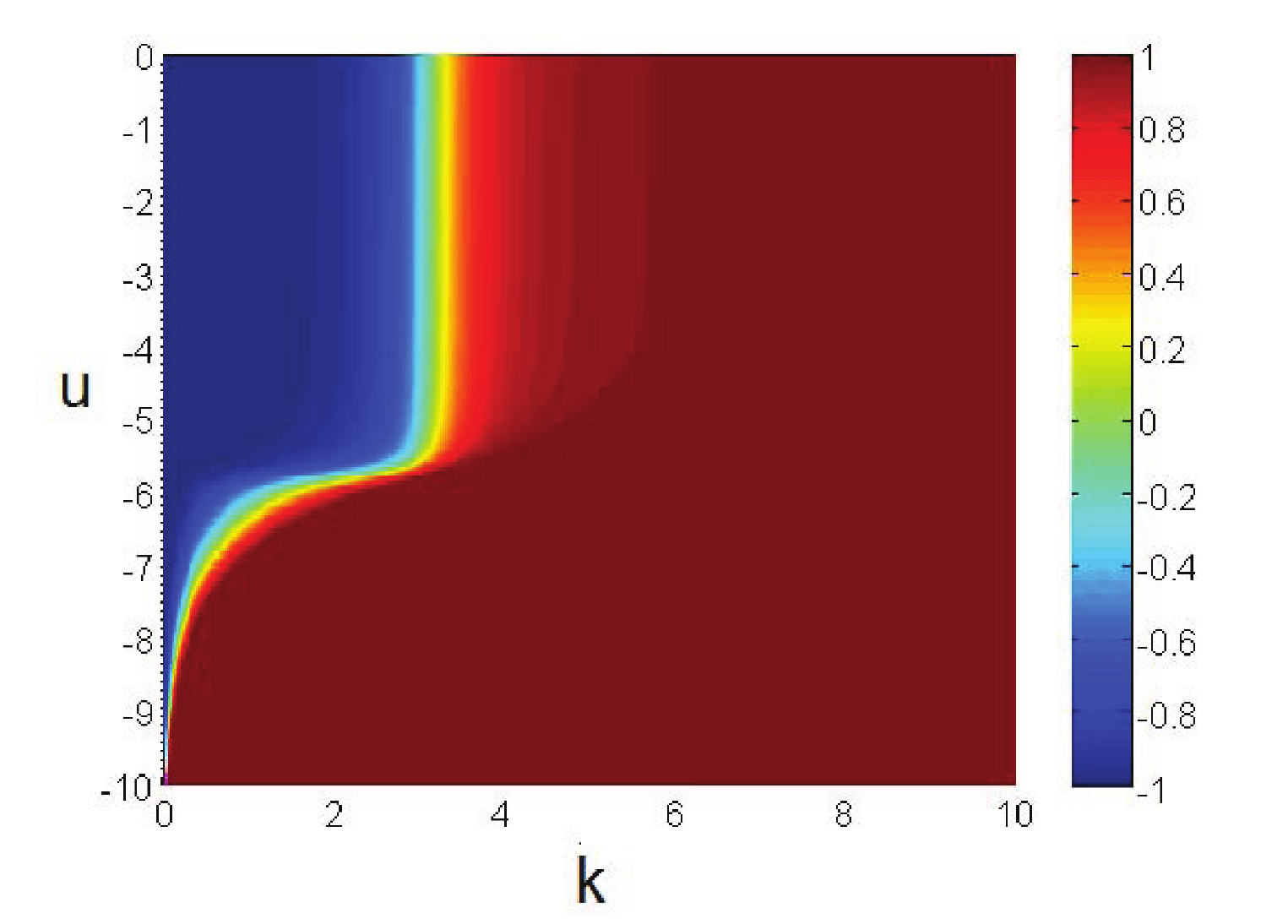}
\caption{ $d_z(\mathbf{k},u)$ in the bulk of cMERA for a non-relativistic Chern insulator with a nontrivial IR state. The parameters we use are $m=10$ and $\Lambda=1000$, based on which one has $u^{\ast}\simeq -5.76$. Note that band inversion happens at each layer, which indicates that the state at each layer is topologically nontrivial.
}\label{BandInversionNonTrivialIR}
\end{figure}

To understand the topological property of the state at each layer,
we study the band inversion behavior in the bulk of cMERA with a topologically nontrivial IR state.
As discussed in Sec.III, we use $d_z(\mathbf{k},u)=\langle \Psi(\mathbf{k},u)|\sigma^z|\Psi(\mathbf{k},u)\rangle$
to track the band inversion, where $|\Psi(\mathbf{k},u)\rangle
=\left(P_{\mathbf{k}}(u)\psi^{\dag}_2(\mathbf{k})-Q_{\mathbf{k}}(u)\psi^{\dag}_1(\mathbf{k})
\right)|\text{vac}\rangle$ is the single particle wave-function.
By using the expression of $P_{\mathbf{k}}(u)$ and $Q_{\mathbf{k}}(u)$ in Eq.\ (\ref{WFchern001}), one can obtain
\begin{equation}
d_z(\mathbf{k},u)=\cos 2\varphi_{\mathbf{k}}(u),
\end{equation}
where $\varphi_{\mathbf{k}}(u)$ is now expressed in terms of $g^{\text{nontrivial}}(u)$ (see Eq.\ (\ref{DvarphiK})).
The plot of $d_z(\mathbf{k},u)$ is shown in Fig.\ \ref{BandInversionNonTrivialIR}.
Different from the case with a topologically trivial IR state,
it is found that the band inversion happens at each layer $u$, which indicates that the state at each layer is topologically nontrivial.

To have a better understanding of the band inversion, it is helpful to see how
$d_z(\mathbf{k},u)$ is related with $g^{\text{nontrivial}}(u)$. In the following,
we discuss the behavior of $d_z(\mathbf{k},u)$ in separate regions:
\begin{equation}\nonumber
\left\{
\begin{split}
&\text{Region I}: \ \ \ u^{\ast}<u\le u_{\text{UV}}, \ \ k>k^{\ast}\\
&\text{Region II}: \ \ \ u_{\text{IR}}<u<u^{\ast}, \ \ k> \Lambda e^u\\
&\text{Region III}: \ \ \ u_{\text{IR}}<u<u_{\text{UV}}, \ \  k \ll \text{min}[\Lambda e^{u^{\ast}}, \Lambda e^u].
\end{split}
\right.
\end{equation}

(i) $\text{Region I}:$
This region corresponds to the upper right corner in Fig.\ \ref{BandInversionNonTrivialIR}.
In this region, one has $g^{\text{nontrivial}}(u)\simeq0$, and therefore $\varphi_{\mathbf{k}}(u)\simeq0$. Then we have
\begin{equation}
d_z(\mathbf{k},u)=\cos 2\varphi_{\mathbf{k}}(u)\simeq 1.
\end{equation}

(ii) $\text{Region II}:$
This region is trivial in the sense that the single-particle state $|\Psi(\mathbf{k},u)\rangle$ is the same as the IR state $|\Omega^{\text{nontrivial}}(\mathbf{k})\rangle$, because no entanglement is created/removed in this region.
Based on Eq.\ (\ref{IRnontrivial}), one has
\begin{equation}
d_z(\mathbf{k},u)=\langle \Omega^{\text{nontrivial}}(\mathbf{k})|\sigma^z| \Omega^{\text{nontrivial}}(\mathbf{k})\rangle=1.
\end{equation}

(iii) $\text{Region III}:$
In this region, to make an estimation of $d_z(\mathbf{k},u)$, we use the approximated
expression of $g^{\text{nontrivial}}(u)$, \textit{i.e.},
\begin{equation}\label{approximation}
g^{\text{nontrivial}}(u)\simeq\left\{
\begin{split}
&\frac{\pi}{2}, \ \ \ &u<u^{\ast}\\
&0, \ \ \ &u>u^{\ast}.
\end{split}
\right.
\end{equation}
Then $\varphi_{\mathbf{k}}(u)$ can be expressed as
\begin{equation}\label{apparoximationVarPhi}
\begin{split}
\varphi_{\mathbf{k}}(u)\simeq&\frac{\pi}{2}\int_{\log\frac{k}{\Lambda}}^{\text{min}[u^{\ast}, u]}ds \frac{k}{\Lambda e^s}=\frac{\pi}{2}\left(1-\frac{k}{\text{min}[\Lambda e^{u^{\ast}}, \Lambda e^u]}\right),
\end{split}
\end{equation}
based on which one has
\begin{equation}
d_z(\mathbf{k},u)=\cos\left[\pi\left(1-\frac{k}{\text{min} \left[ \Lambda e^{u^{\ast}},\Lambda e^u  \right] }\right)\right].
\end{equation}
By considering $k \ll \text{min}[\Lambda e^{u^{\ast}}, \Lambda e^u]$, then
$d_z(\mathbf{k},u)$ can be evaluated as
\begin{equation}
d_z(\mathbf{k},u)=\cos\left[\pi\left(1-0^+\right)\right]\simeq -1.
\end{equation}

Based on the discussions above, it is apparent that as we increase $k$ from $k=0$, $d_z(\mathbf{k},u)$
changes as follows:
\begin{equation}
d_z(\mathbf{k},u):\ \ \ -1\to +1,
\end{equation}
which happens in each layer $u$ from $u_{\text{IR}}$ to $u_{\text{UV}}$.
In other words, the band inversion happens in each layer $u$.
This is in agreement with the calculation in Fig.\ \ref{BandInversionNonTrivialIR}.
It is emphasized that the discussion above applies to the whole region with $0\le k \le \Lambda$
and $u_{\text{IR}}<u<u_{\text{UV}}$, which indicates that the state in each layer is topologically nontrivial.

\begin{figure}[htp]
\includegraphics[width=3.05in]{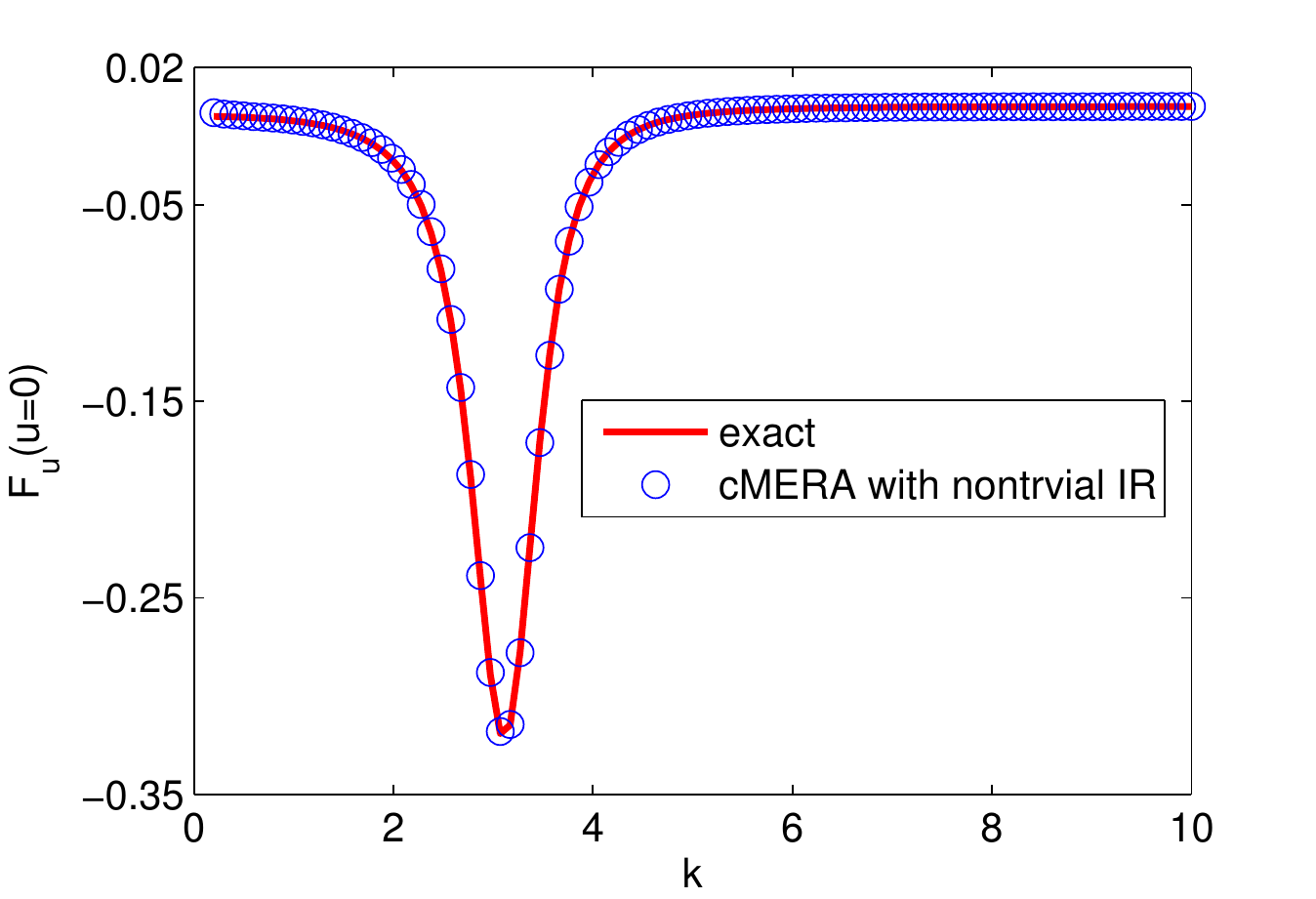}
\caption{Comparison of cMERA constructed Berry curvature $\mathcal{F}_u(k,\theta_{\mathbf{k}};u=0)$ with the exact result for a Chern insulator. The parameters we use are $m=10$ and $\Lambda=1000$.
 }\label{smallK001}
\end{figure}

\subsection{Berry curvature flow in cMERA of Chern insulators}

To further identify the topological property in each layer of cMERA, in this part, we will study the Berry curvature flow
in the bulk of cMERA.

Following the previous section, based on the wavefunction in Eqs.\ (\ref{WFchernNontrivial}) and (\ref{WFchern001}),
one can obtain the Berry connection as follows
\begin{small}
\begin{equation}\label{connection001}
\left\{
\begin{split}
\mathcal{A}_k(k,\theta_{\mathbf{k}};u)=&-i\langle \Psi(\mathbf{k},u)|\partial_{k} |\Psi(\mathbf{k},u)\rangle=0,\\
\mathcal{A}_{\theta_{\mathbf{k}}}(k,\theta_{\mathbf{k}};u)=&-\frac{i}{k} \langle \Psi(\mathbf{k},u)|\partial_{\theta_{\mathbf{k}}} |\Psi(\mathbf{k},u)\rangle
=-\frac{1}{k}\cos^2\varphi_{\mathbf{k}}(u),\\
\mathcal{A}_u(k,\theta_{\mathbf{k}};u)=&-i\langle \Psi(\mathbf{k},u)|\partial_{u} |\Psi(\mathbf{k},u)\rangle=0,\\
\end{split}
\right.
\end{equation}
\end{small}
where $\varphi_{\mathbf{k}}(u)$ is defined as
\begin{equation}\label{varphiNontrivial}
\varphi_{\mathbf{k}}(u)=\int_{\log k/\Lambda}^udsg^{\text{nontrivial}}(s)\frac{ke^{-s}}{\Lambda}.
\end{equation}
Therefore, the Berry curvature can be obtained by calculating
$\vec{\mathcal{F}}=\mathbf{\nabla}\times \vec{\mathcal{A}}$.
Then we have
\begin{equation}\label{vector001}
\begin{split}
\vec{\mathcal{F}}(k,\theta_{\mathbf{k}};u):=&\hat{\mathbf{u}} \mathcal{F}_u(k,\theta_{\mathbf{k}};u)+\hat{\mathbf{k}} \mathcal{F}_k(k,\theta_{\mathbf{k}};u)\\
=&\hat{\mathbf{u}}\left[\frac{1}{k}\sin2\varphi_k(u)\partial_k\varphi_k(u)\right]\\
&+\hat{\mathbf{k}}\left[-\frac{1}{k}\sin2\varphi_k(u)\partial_u\varphi_k(u)\right].
\end{split}
\end{equation}

\begin{figure}
\includegraphics[width=3.05in]{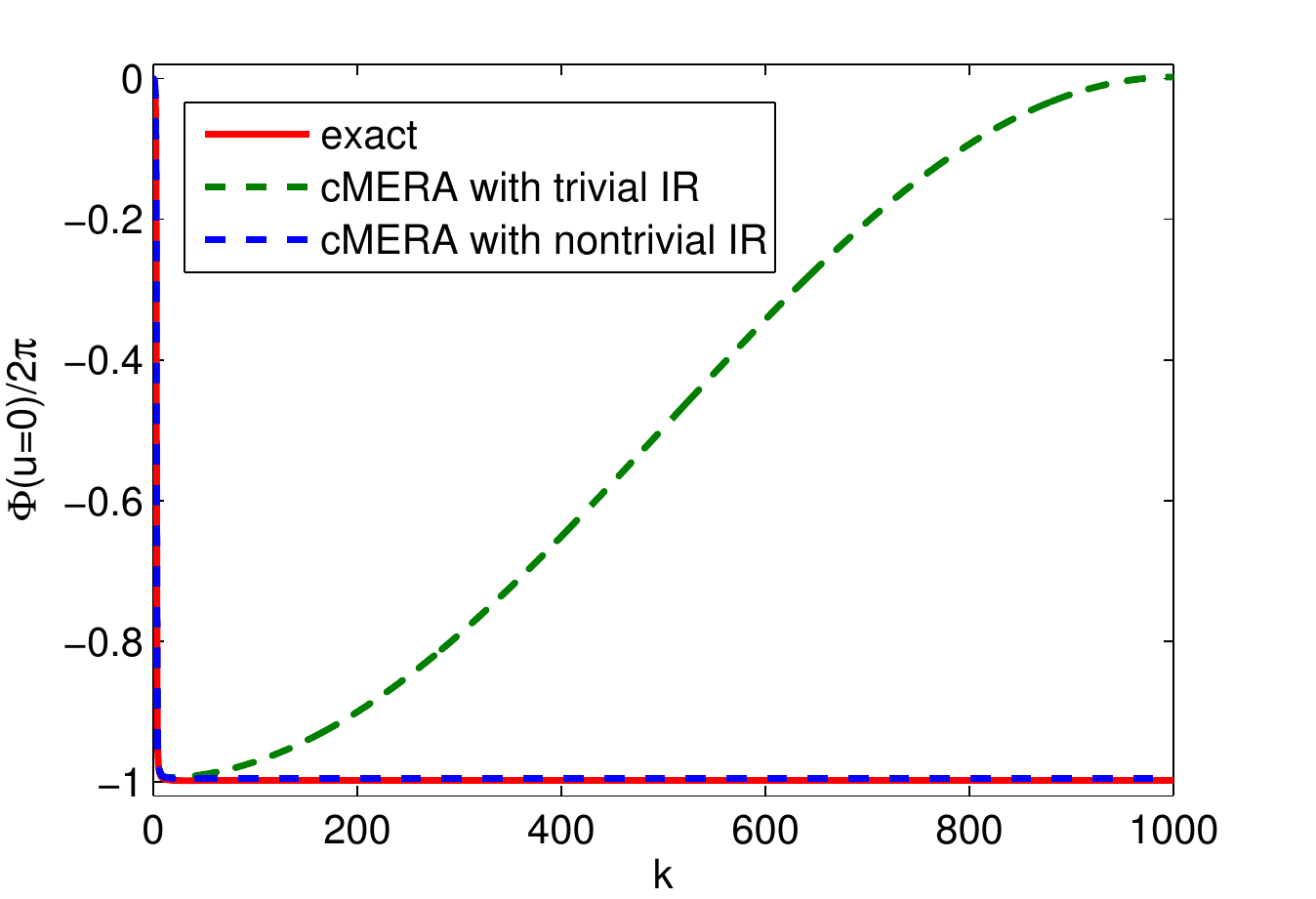}
\caption{Comparison of cMERA constructed Berry flux $\Phi(u=0)/2\pi$ in layer $u=u_{\text{UV}}=0$ with the exact result for a Chern insulator.  For cMERA with a topologically trivial IR state, the Berry flux deviates from the exact result from
certain momentum $k$, and decays to $0$ gradually as $k\to\Lambda$. For cMERA with a topologically
nontrivial IR state, however, it agrees with the exact result in the whole region. The parameters we use are $m=10$ and $\Lambda=1000$.
 }\label{largeKnontrivial}
\end{figure}

As the first step, we check if the Berry curvature and Berry flux obtained from cMERA agrees
the exact results in the UV layer $u=u_{\text{UV}}$.
As shown in Fig.\ \ref{smallK001}, we compare the cMERA constructed
$\mathcal{F}_u(k,\theta_{\mathbf{k}};u=0)$ and the exact result of Berry curvature
in Eq.\ (\ref{exactBerrycurvature}) in the low energy physics region
$k\ll \Lambda$, and they agree with each other very well.
Then in
Fig.\ \ref{largeKnontrivial}, we compare the Berry flux obtained from cMERA
and the exact result in the whole region $0\le k\le \Lambda $. For cMERA with a topologically trivial IR state, the Berry
flux agrees with the exact result only in the region $k\ll\Lambda$.
As $k$ increases, the Berry flux  deviates from the the exact result,
and decays to zero gradually. For cMERA with a topologically nontrivial
IR state, the Berry flux obtained from cMERA agrees with the exact result
in the whole region $0\le k\le\Lambda$,
which indicates that cMERA with a topologically nontrivial IR state respects the topological property
of the exact ground state.

\begin{figure}
\includegraphics[width=3.05in]{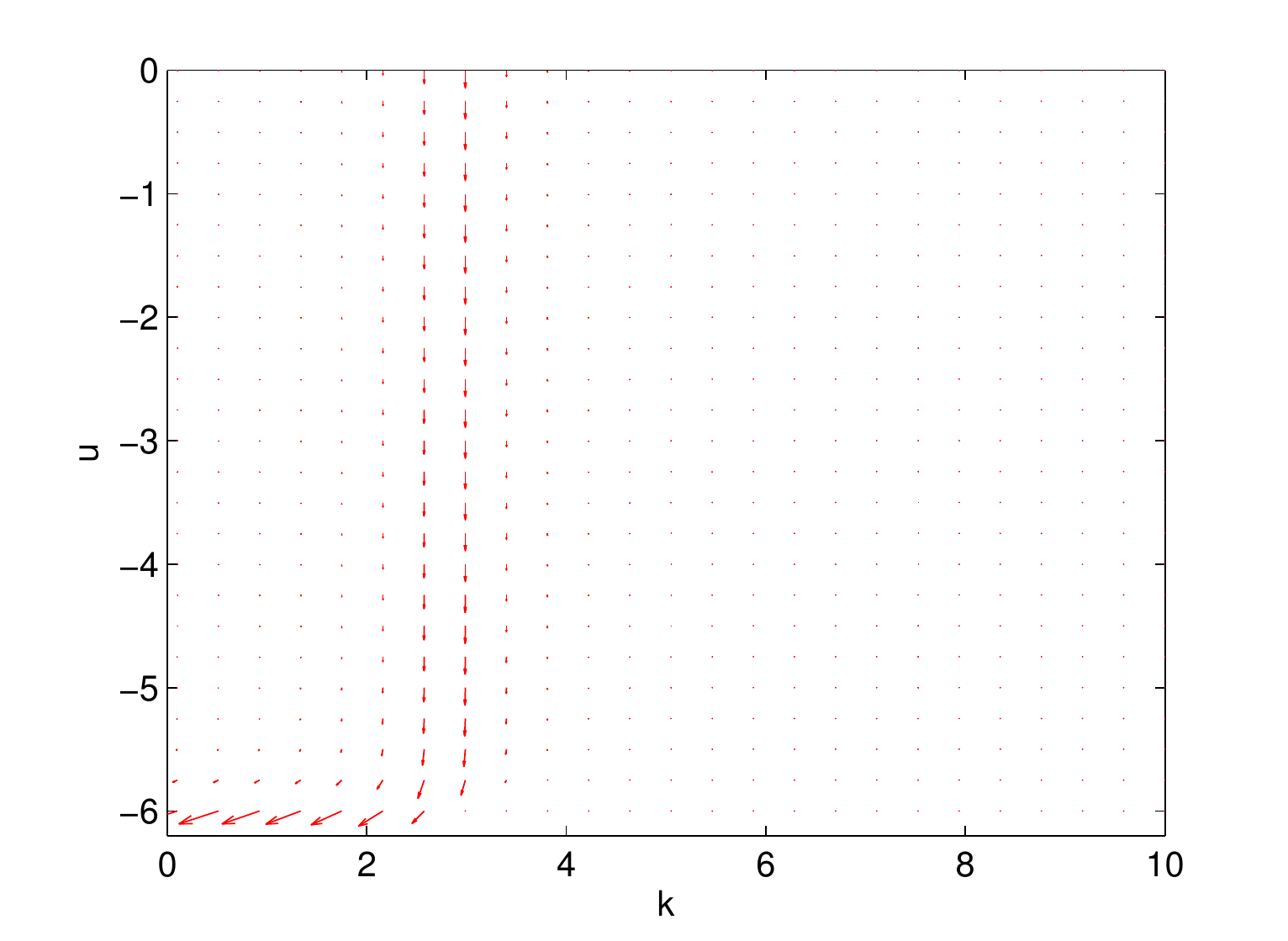}
\caption{Berry curvature flow in the bulk of cMERA for a Chern insulator with a topologically nontrivial IR state.
The parameters we use are $m=10$ and $\Lambda=1000$, based on which one has $u^{\ast}\simeq -5.76$.
}\label{BerryFlow001}
\end{figure}

Then we will study the Berry curvature flow in the bulk of cMERA in the following.
As shown in Fig.\ \ref{BerryFlow001}, we plot the vector
field $\vec{\mathcal{F}}(k,\theta_{\mathbf{k}};u)$ based on $\mathcal{F}_u(k,\theta_{\mathbf{k}};u)$
and $\mathcal{F}_k(k,\theta_{\mathbf{k}};u)$.
Quite different from the results in cMERA with a topologically trivial IR state, here the Berry curvature
is not bent backwards near $u^{\ast}$. On the contrary, the Berry curvature is
bent towards smaller $k$, and then flows towards the IR layers.
Note that the Berry curvature flow in IR layers ($u_{\text{IR}}<u<u^{\ast}$)
is not shown here, because the Berry curvature converges to smaller $k$
and the field strength is very strong (Therefore, to have a good contrast of display for the
Berry curvature flow near $u^{\ast}$, we only plot $\vec{F}(k,\theta_{\mathbf{k}};u)$ in the finite region.).
Nevertheless, the behavior of Berry curvature flow in the whole region is schematically shown in
Fig.\ \ref{scheme00B}. In Appendix \ref{BerryFlowNontrivialIR}, we give a detailed analysis
on how the Berry curvature flow $\vec{\mathcal{F}}(k,\theta_{\mathbf{k}};u)$ in the bulk of cMERA
is related with the behavior of $g^{\text{nontrivial}}(u)$.

In addition, we check the total Berry flux $\Phi(u)$ in different layers, and find that the Berry
flux in each layer is conserved to be $\Phi(u)=-2\pi$. This can be easily understood
by writing down the form of Berry flux  explicitly
\begin{equation}
\begin{split}
\Phi(u,k=\Lambda e^u)=&\oint \mathcal{A}(k,\theta_{\mathbf{k}};u)kd\theta_{\mathbf{k}}\big|_{k=\Lambda e^u}\\
=&-2\pi\cos^2\varphi_{\mathbf{k}}(u)\big|_{k=\Lambda e^u}.
\end{split}
\end{equation}
By noting that $\varphi_{\mathbf{k}}(u)|_{k=\Lambda e^u}=0$, one has
\begin{equation}
\begin{split}
\Phi(u,k=\Lambda e^u)=-2\pi,
\end{split}
\end{equation}
which is independent of the layer $u$.

As a short sum in this part, we find that \emph{all} the Berry curvature emanated from the
UV layer flows to the IR layer, and the total Berry flux at each layer $u$ is conserved to
be $-2\pi$. This verifies that the cMERA constructed wavefunction $|\Psi(u)\rangle$
at each layer $u$ is topologically nontrivial.  Our result parallels with the story in the lattice MERA\cite{Vidal2008top,Vidal2009top,Vidal2013}.

We give some remarks before ending this part.
It is noticed that if we focus on the IR state $|\Omega^{\text{nontrivial}}\rangle$ in Eq.\ (\ref{IRnontrivial}),
there is \textit{no} real space entanglement. However, for an arbitrary finite layer $u$, the state in Eq.\ (\ref{WFchernNontrivial}) carries finite real space entanglement,
because of its topologically nontrivial property.
This is as expected, because we cannot
remove all the entanglement of a Chern insulator by simply using a local unitary operation within finite depth.

In Appendices \ref{Hch}, we also discuss the cMERA construction for a Chern insulator with
higher Chern numbers. Both topologically trivial and nontrivial IR states are considered.
The physical pictures are basically the same as the case with $\text{Ch}_1=-1$  as discussed in
the main text.

\section{Discussions and conclusions}
\label{conclusion}

In this paper, we studied the entanglement renormalization group flows of topological band insulators
in (2+1)D with cMERA. In particular, we constructed the cMERA for a Chern band insulator with topologically
trivial and nontrivial IR states, respectively.

For the cMERA of a Chern insulator with a topologically trivial IR state, the UV state constructed
from cMERA agrees with the exact ground state in the low energy physics region $k\ll \Lambda$.
The topological properties in the bulk of cMERA were studied through band inversion and Berry curvature
flow. In the low energy physics region, it was found that band inversion happens in the region
$u^{\ast}<u<u_{\text{UV}}$, where $u^{\ast}$ is determined by the mass term. In the region
$u_{\text{IR}}<u<u^{\ast}$, however, there is no band inversion. This indicates a `topological
phase transition' in the renormalization direction. Then we studied the Berry curvature flow
in the bulk of cMERA. It was found that the Berry curvature, which is emanated from the low
energy physics region in the UV layer,  is bent backwards near $u=u^{\ast}$.
Finally, these Berry curvature flows to the large $k$ region in the UV layer, which results
in a total zero Berry flux in each layer of cMERA. Therefore, the cMERA constructed UV state cannot
recover the exact ground state of a Chern band insulator in the whole region
$0\le k\le \Lambda$. Besides the topological properties, we also studied the geometric
properties in the bulk of cMERA by calculating the holographic metric.

For the cMERA of a Chern insulator with a topologically nontrivial IR state, the UV state
constructed from cMERA agrees with the exact ground state in the whole region
$0\le k\le \Lambda$. It was found that band inversion happens in each layer of cMERA,
and the total Berry flux in each layer is conserved to be $-2\pi$. Furthermore, we studied
how the Berry curvature flows in the bulk of cMERA. We found that all the Berry curvature
emanated from the UV layer flows to the IR layer. This means
a topologically nontrivial UV state corresponds to topologically nontrivial states in the bulk of cMERA.
This parallels with the story in the lattice MERA, where it is found that if the UV state
is nontrivial in topology, then the state in each layer of the bulk is similarly nontrivial.

\begin{figure}
\centering
\includegraphics[width=3.05in]{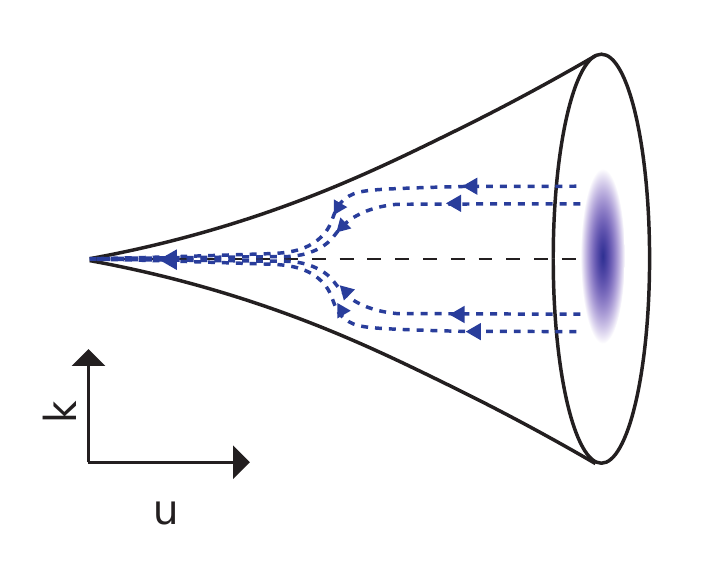}
\caption{ Schematic plot of Berry curvature flow in the bulk of cMERA for a Chern insulator with a topologically
nontrivial IR state. All the Berry curvature emanated from the UV layer flows to the IR layer. The total Berry flux in each layer is conserved to be $-2\pi$.
}\label{scheme00B}
\end{figure}

Finally, we mention some interesting future problems  as follows.

$\bullet$ \emph{Finite temperature effect on cMERA of topological insulators}

Our current work focuses on the cMERA construction of topological insulators at
zero temperature. Most recently, topological insulators at finite temperature
were studied by introducing two quantities: the Uhlmann phase in (1+1)D systems
and the Uhlmann number in (2+1)D systems\cite{Viyuela1d,Viyuela2d,Huang}, which
are used to characterize the topological invariant of the system at finite temperatures.
In particular, it is found that, for topological insulators, there exists a critical
temperature $T_c$ where thermal topological phase transitions may happen.
It may prove interesting to study how finite temperature $T$ affects the topological
property as well as the geometric property in the bulk of cMERA, and in particular,
how the thermal topological phase transition reveals itself in the bulk of cMERA.

$\bullet$ \emph{cMERA for interacting topological phases}

The topological band insulators we discussed here are noninteracting systems.
Generalizing our method to topological phases with interactions,
such as fractional quantum hall states or fractional Chern insulators, remains an open problem.
To obtain the ground state of fractional quantum hall systems or fractional
Chern insulators, one may project copies of free fermion states onto
a gauge invariant subspace \cite{Wen1999}. How such projections affect the bulk properties of cMERA
is unknown at this moment.

$\bullet$ \emph{Quench dynamics in cMERA}

Recently, quench dynamics in AdS/CFT correspondence has been discussed
intensively. In particular, the time evolution of cMERA after a global quantum
quench has been studied in free field theories \cite{Ryu2013}. It is found that
the behavior of the holographic metric  qualitatively agrees with
its gravity dual given by a half of the AdS Schwarzschild black hole spacetime \cite{Hartman}.
As studied in our current work, the geometric and topological properties are
closely related with each other through the disentangler in cMERA. Therefore, it will be of
great interest to study how the quantum quench affects the topological quantities, \emph{e.g.},
the Berry curvature flow in the bulk of cMERA.

$\bullet$ \emph{The relation of cMERA and exact holographic mapping}

In a companion paper\cite{Gu}, by using the exact holographic mapping (EHM), the holographic duality between a
(2+1)D Chern insulator and a (3+1)D topological insulator is studied. In the EHM approach, the Chern number of the boundary theory gets distributed to different positions of the bulk. Therefore the two different approaches lead to different bulk theories. It will be interesting to have more direct comparison of the dual geometry obtained in these two approaches in future works.

\section{Acknowledgement}

We are grateful to the KITP Program “Entanglement in Strongly-Correlated Quantum
Matter” (Apr 6 - Jul 2, 2015).
This work is supported by the NSF under
Grants  No. DMR-1455296 (X.W. and S.R.), 
No. DMR-1151786 (YG and XLQ),
the Alfred P. Sloan foundation
(SR), the Brain Korea 21 PLUS Project of Korea
Government (GYC), as well as by the David and Lucile
Packard Foundation (XLQ). 
 PLSL acknowledges support from FAPESP under grants 2009/18336-0 and 2012/03210-3.

\section{Appendices}

\subsection{Brief review of the lattice MERA}
\label{LatticeMERA}

Some nice  reviews of the lattice MERA can be found in Refs.\ \onlinecite{Vidalreview01,Vidalreview02}.
For the completeness of this paper, we give a brief introduction to the lattice MERA here. The construction
of the lattice MERA can be understood in the following two ways. First, it can be considered as a coarse
graining transformation (combined with disentangling operations) that maps the lattice $\mathcal{L}_{u}$ in
layer $u$ to a sequence of coarser lattices $\mathcal{L}_{u-1}, \mathcal{L}_{u-2}, \cdots$, and
therefore it leads to a real space renormalization group transformation. Secondly, the lattice MERA
can be viewed as quantum circuits with the output as the states living on the lattice at $u=0$
and the quantum gates as the disentanglers and `coarse grainers' (isometries). With appropriate
quantum gates, the lattice MERA can transform the input, which is an unentangled state at the IR layer,
into the target state $|\Psi\rangle$, which faithfully represents the ground state.

\begin{figure}
\includegraphics[width=2.85in]{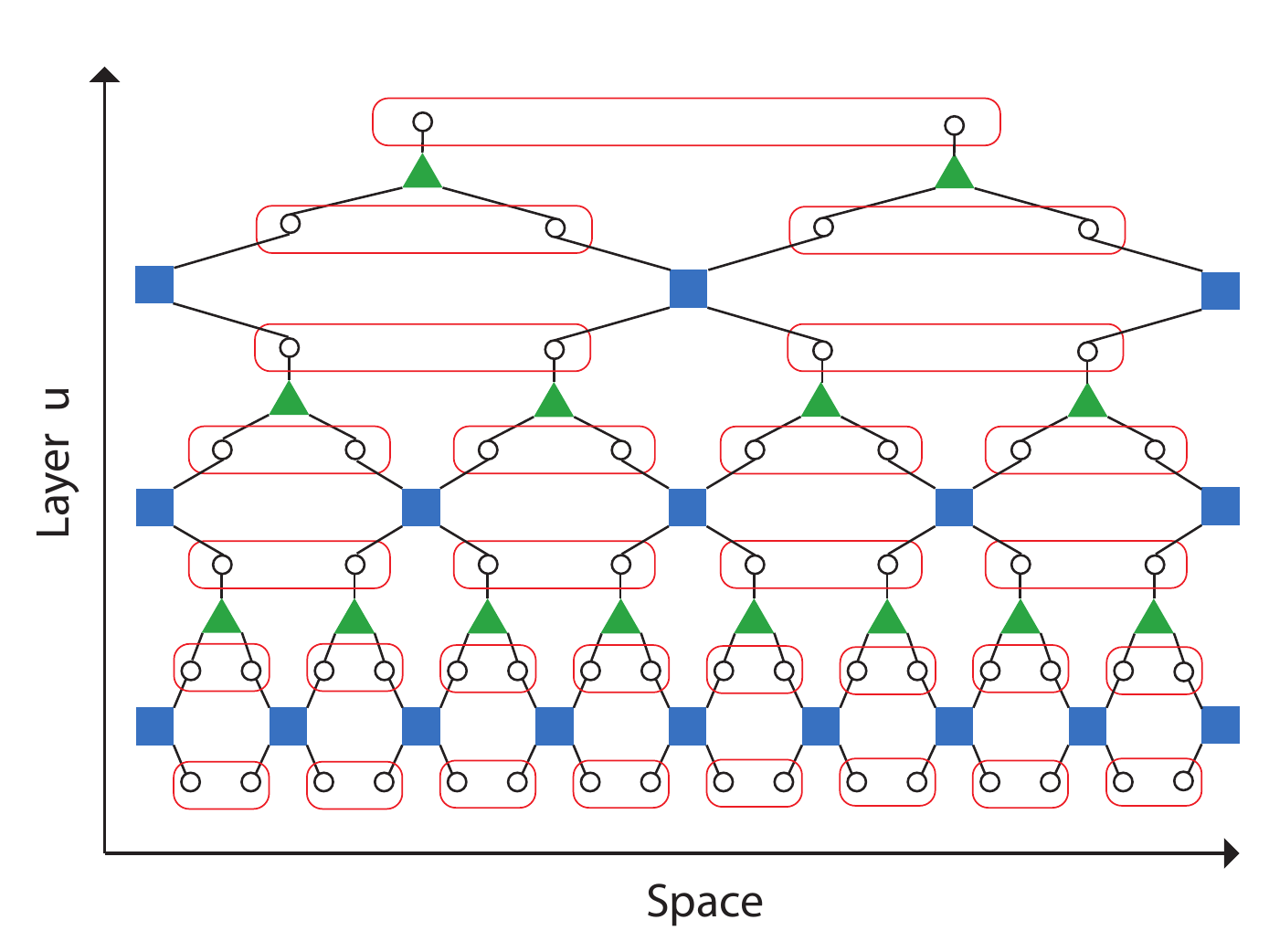}
\caption{Tensor network structure of the lattice MERA. Circles are lattice sites at different
coarse-graining scales. Blue squares are unitary disentaglers which are used to remove
short range entanglement between neighboring blocks, and green triangles are isometric
coarse graining transformations which map a block of sites into a single site in the next layer.
In different layers,  entanglements are removed in different length scales. For a deeper layer
$u$, we create/remove entanglement in a larger length scale $ae^{-u}$, where $a$ is
the lattice constant.}
\label{Fig1}
\end{figure}

Here we choose the language of renormalization group transformation for concreteness.
Denoting $\mathcal{L}$ as the lattice with $N$ sites living in $(d+1)$ dimensions in which
the bare lattice Hamiltonian and its ground state are written on, the lattice MERA is
composed of tensors living in $|T|\simeq \log N$ different layers, with each layer
containing a row of disentanglers $v$ and a row of isometries $w$. Let us take the lattice
MERA in (1+1)D for example, as shown in Fig.\ \ref{Fig1}.
We start from the original lattice $\mathcal{L}_0\equiv \mathcal{L}$.
By applying disentanglers $V_{-1}^{\dag}=\prod v^{\dag}$ to remove the short
range entanglement between neighboring blocks, and then applying isometries
$W_{-1}^{\dag}=\prod w^{\dag}$ to map blocks of sites in $\mathcal{L}_0$
into single sites in the next layer, we obtain $\mathcal{L}_{-1}$ which is the first
step coarse grained lattice of $\mathcal{L}_0$. Repeating this procedure, we get
a sequence of lattices in the lattice MERA:
\begin{equation}
\mathcal{L}_0\overset{U^{\dag}_{-1}}{\longrightarrow} \mathcal{L}_{-1}\overset{U^{\dag}_{-2}}\longrightarrow \cdots \mathcal{L}_{u}\overset{U^{\dag}_{u-1}}\longrightarrow \mathcal{L}_{u-1}\cdots \overset{U^{\dag}_T}{\longrightarrow}\mathcal{L}_{T},
\end{equation}
where $U^{\dag}_{u}=W_{u}^{\dag}V_{u}^{\dag}$. We get the increasingly
coarse grained states, corresponding to the increasingly coarser lattices
$\{\mathcal{L}_0, \mathcal{L}_{-1},\cdots, \mathcal{L}_T\}$, as follows
\begin{equation}
|\Psi_0\rangle\overset{U^{\dag}_{-1}}{\longrightarrow} |\Psi_{-1}\rangle\overset{U^{\dag}_{-2}}\longrightarrow \cdots |\Psi_{u}\rangle\overset{U^{\dag}_{u-1}}\longrightarrow |\Psi_{u-1}\rangle\cdots \overset{U^{\dag}_T}{\longrightarrow}|\Psi_{T}\rangle.
\end{equation}
To be more precise, $|\Psi_{u-1}\rangle=U_{u-1}^{\dag}|\Psi_{u}\rangle$,
from which one can obtain
\begin{figure}
\includegraphics[width=2.65in]{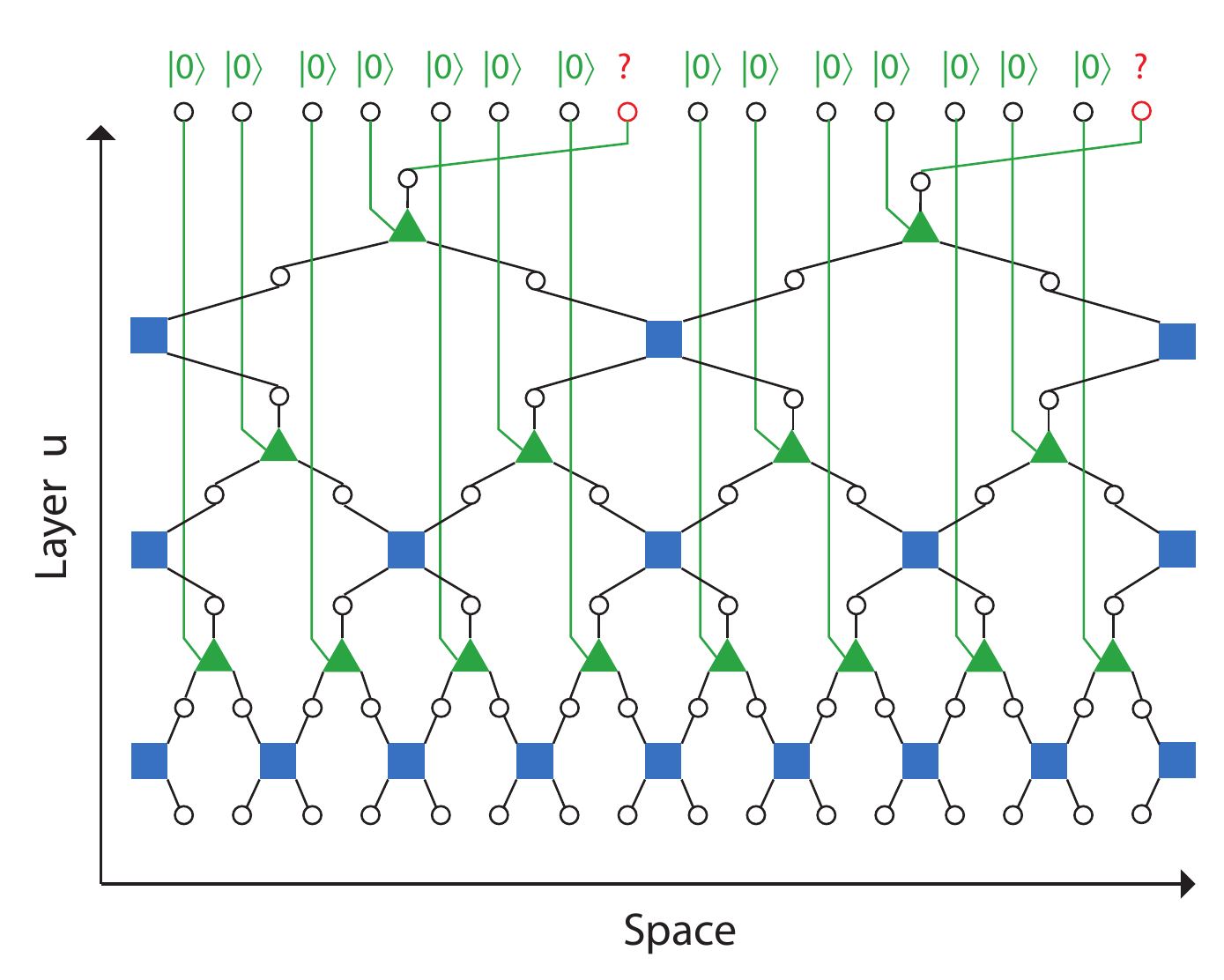}
\caption{`Interaction' picture of MERA. We add a dummy $|0\rangle$ at each
isometry operation $w$ to keep the Hilbert space conserved. The circles, squares, and triangles have the
same meaning as those in Fig.\ref{Fig1}.}
\label{InMERA}
\end{figure}
\begin{equation}\label{Psi0}
\begin{split}
|\Psi_0\rangle=&U_{-1}U_{-2}\cdots U_{T}|\Psi_T\rangle=\mathcal{P}\prod_{u}V_{u}W_{u}|\Psi_T\rangle,
\end{split}
\end{equation}
where the symbol $\mathcal{P}$ is a path-ordering which puts all operators with smaller
$u$ to the right.
Eq.\ (\ref{Psi0}) is very useful for the following reasons: (i) It is straightforward to generalize to
continuum MERA, as will be seen clearly later. (ii) It makes the construction of the lattice
MERA intuitive. Given $|\Psi_T\rangle$ which may be unentangled, by doing dilation (scaling) $W$ and
adding short-ranged entanglement $V$ repeatedly, we can obtain the target state $|\Psi_0\rangle$
which is the ground state for a given Hamiltonian.
(iii) It makes clear that at each layer the disentangler $W(u)$ acts in different length scales.
As $u$ goes deeper towards $T$, the quantum entanglement is created/removed in larger
length scales $a e^{-u}$, and thus smaller momentum scales $e^{u}/a$, where $a$ is the
lattice constant. For the algorithm to optimize the disentanglers $V$ and isometries $W$,
one can refer to the detailed descriptions in Refs.\ \onlinecite{Vidalreview01,Vidalreview02}.

Next, we introduce the `interaction' picture of MERA, as shown in Fig.\ \ref{InMERA}.
The convenience of the `interaction' picture of MERA is that as the layer $u$ varies, the size of the
Hilbert space is conserved. This is in contrast with the conventional picture of MERA in
Fig.\ \ref{Fig1}, where the size of the Hilbert space is reduced by a half as $u\to u-1$.
The strategy of constructing the `interaction' picture of MERA is simple; at each isometry
(scaling), we add a dummy state $|0\rangle$ replacing the state in the Hilbert space to be truncated
in the isometry process. Therefore, as $u$ goes deeper towards $u_{\text{T}}$, we get a lot
of extra $|0\rangle$'s which are un-entangled in $|\Psi_T\rangle$. This also supports an
intuitive picture of the MERA that as $u$ varies from $u_T$ to $u_0$, we are adding
entanglement on the un-entangled state $|0\rangle \otimes |0\rangle \otimes \cdots |0\rangle $
at different length scales depending on the layer $u$. As discussed in the main text, the `interaction'
picture of MERA is useful in the construction of cMERA.

\subsection{cMERA of different phases with topologically trivial IR states}
\label{cMERA4}

\subsubsection{cMERA of nonrelativistic Chern insulators}

A Chern insulator in (2+1) dimensions can be described by a simple
two-band model with the Hamiltonian\cite{QiPRB,QiRMP} (See also Eq.\ (\ref{H0}) for more details.)
\begin{equation}\label{Hchern}
\begin{split}
H=&\int d^2\mathbf{k} \psi^{\dag}(\mathbf{k}) \left[\mathbf{R}(\mathbf{k})\cdot\mathbf{\sigma}\right] \psi(\mathbf{k}),\\
\end{split}
\end{equation}
where
\begin{equation}
\mathbf{R}(\mathbf{k})=(k_x,\ k_y,\ m-\mathbf{k}\cdot \mathbf{k}),
\end{equation}
and  $m>0$.
The Hamiltonian in Eq. \eqref{Hchern} can be diagonalized by using a unitary transformation
\begin{equation}
\psi(\mathbf{k})=U(\mathbf{k})\widetilde{\psi}(\mathbf{k}),
\end{equation}
where
\begin{equation}\label{Umatrix}
\begin{split}
U(\mathbf{k})=&\frac{1}{\sqrt{2R\left(R+R_3\right)}}
\left(
\begin{matrix}
&R_3+R        &-(R_1-iR_2)  \\
&R_1+iR_2     &R_3+R
\end{matrix}
\right)\\
=:&
\left(
\begin{matrix}
&u_{\mathbf{k}}       &-v_{\mathbf{k}}   \\
&v^{\ast}_{\mathbf{k}}   &u_{\mathbf{k}}
\end{matrix}
\right),\\
\end{split}
\end{equation}
with $R(\mathbf{k})=|\mathbf{R}(\mathbf{k})|$. Then one can get
\begin{equation}
H=\int d^2\mathbf{k}\left[ R(\mathbf{k})\widetilde{\psi}_1^{\dag}(\mathbf{k})\widetilde{\psi}_1(\mathbf{k})-R(\mathbf{k})\widetilde{\psi}_2^{\dag}(\mathbf{k})\widetilde{\psi}_2(\mathbf{k})\right],
\end{equation}
and the ground state $|\Psi\rangle$ is defined by
\begin{equation}
\widetilde{\psi}_1(\mathbf{k})|\Psi\rangle=0,\ \ \widetilde{\psi}^\dag_2(\mathbf{k})|\Psi\rangle=0.
\end{equation}
This condition is met uniquely by the state
\begin{equation}\label{targetA0}
|\Psi\rangle=\prod_{|\mathbf{k}|\le \Lambda}\widetilde{\psi}^\dag_2(\mathbf{k})|\text{vac}\rangle=
\prod_{|\mathbf{k}|\le \Lambda}
\left(
u_{\mathbf{k}}\psi_2^{\dag}(\mathbf{k})-v_{\mathbf{k}}\psi_1^{\dag}(\mathbf{k})
\right)|\text{vac}\rangle,
\end{equation}
where
\begin{equation}
\left\{
\begin{split}
u_{\mathbf{k}}=&\frac{1}{\sqrt{N}}\left((m-k^2)+\sqrt{(m-k^2)^2+k^2}\right),\\
v_{\mathbf{k}}=&\frac{1}{\sqrt{N}}\left(ke^{-i\theta_{\mathbf{k}}}   \right),
\end{split}
\right.
\end{equation}
and $N$ is a normalization factor so that $|u_{\mathbf{k}}|^2+|v_{\mathbf{k}}|^2=1$,
$k\equiv|\mathbf{k}|$, and $\theta_{\mathbf{k}}$ is defined through
$k\cos\theta_{\mathbf{k}}=k_x$ and $k\sin\theta_{\mathbf{k}}=k_y$.
Based on the wavefunction, one can calculate the Berry curvature at momentum $\mathbf{k}$ as follows
\begin{equation}\label{exactBerrycurvature}
\begin{split}
\mathcal{F}(k,\theta_{\mathbf{k}})=&-\frac{1}{2}\frac{m+k^2}{\Big[k^2+(m-k^2)^2\Big]^{\frac{3}{2}}}.
\end{split}
\end{equation}
Next we will use cMERA to construct the ground state in Eq. (\ref{targetA0}).
In the main text, we have found that the wavefunction at layer $u$ is expressed as (See Eq.\ (\ref{GeneralForm001}))
\begin{equation}
\begin{split}
|\Psi(u)\rangle
=&
\prod_{|\mathbf{k}|\le \Lambda}\left(
P_{\mathbf{k}}(u)\psi^{\dag}_2(\mathbf{k})-Q_{\mathbf{k}}(u)\psi^{\dag}_1(\mathbf{k})
\right)|\text{vac}\rangle.
\end{split}
\end{equation}
where the expression of $P_{\mathbf{k}}(u)$ and $Q_{\mathbf{k}}(u)$ can be found in Eq.\ (\ref{WFchern}).
Then,
by requiring that
$
|\Psi(u=u_{\text{UV}})\rangle=|\Psi\rangle,
$
and defining
\begin{equation}\label{phik001}
\varphi_{\mathbf{k}}(u):=\int_{u_{\text{IR}} }^u du' g_{\mathbf{k}}^r(u')=
\int_{\log k/\Lambda}^u du' g(u')\frac{ke^{-u'}}{\Lambda},
\end{equation}
one has
\begin{equation}\label{varphi}
\varphi_{\mathbf{k}}(u_{\text{UV}})=\arctan\frac{-k}{m-k^2+\sqrt{(m-k^2)^2+k^2}}.
\end{equation}
After some straightforward algebra, one can obtain the form of $g(u)$ in the following,
\begin{equation}\nonumber
\begin{split}
g(u)=&-\frac{k^2}{\Lambda}\frac{\partial}{\partial k}\left(\frac{\Lambda}{k}\varphi_k(u_{\text{UV}})\right)\big|_{k=\Lambda e^u}\\
=&\varphi_k(u_{\text{UV}})-k\partial_k \varphi_k(u_{\text{UV}})\big|_{k=\Lambda e^u},
\end{split}
\end{equation}
which, after plugging Eq.\ (\ref{varphi}) in, results in
\begin{equation}\label{guChern001}
\begin{split}
g(u)
&=\frac{1}{2}\frac{\Lambda e^u(m+\Lambda^2 e^{2u})}{(m-\Lambda^2 e^{2u})^2+\Lambda^2 e^{2u}}\\
-&\text{arctan}\frac{\Lambda e^u}{\sqrt{(m-\Lambda^2 e^{2u})^2+\Lambda^2 e^{2u}}+(m-\Lambda^2 e^{2u})}.
\end{split}
\end{equation}
Then we obtain $g(u)$ in Eq.\ (\ref{gua}).

\subsubsection{cMERA of nonrelativistic trivial insulators}

The cMERA of non-relativistic trivial insulators is slightly different from Chern insulators because of
the sign change of mass term $m$, as discussed below.

\begin{figure}
\includegraphics[width=3.25in]{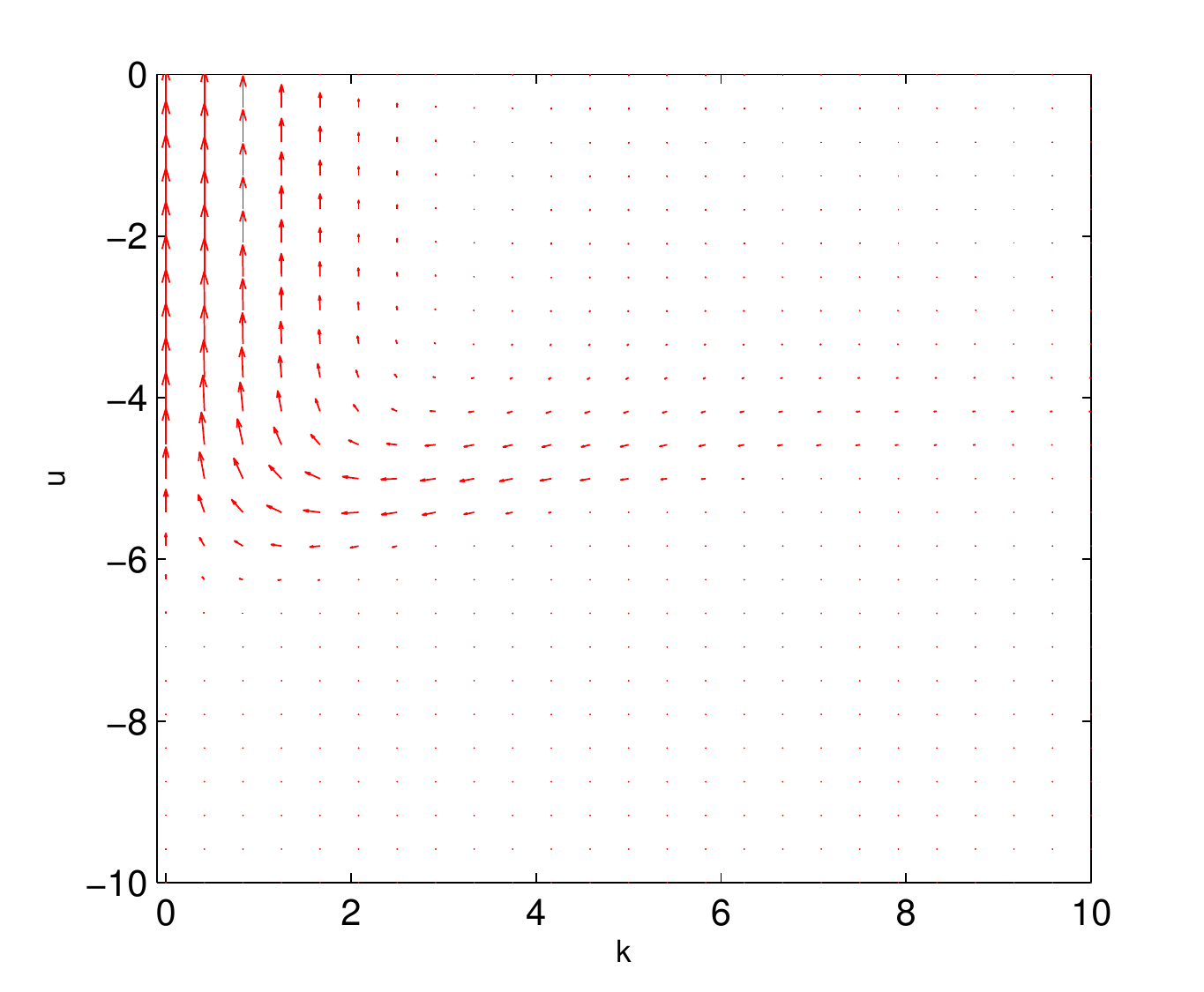}
\caption{
Berry curvature flow in cMERA of a  non-relativistic insulator with $m<0$.
The parameters we use are $m=-10$ and $\Lambda=1000$. }\label{BerryB}
\end{figure}

For non-relativistic trivial insulators, one has
\begin{equation}
\mathbf{R}(\mathbf{k})=(k_x,k_y,m-k^2),
\end{equation}
with $m<0$.
However, we should be careful when using the expression of ground state $|\Psi\rangle$ in Eq.\ (\ref{target}) (with
$u(\mathbf{k})$ and $v(\mathbf{k})$ expressed in Eq.\ (\ref{Umatrix})),
since one can find that $|\Psi\rangle$ is not well defined at $k=0$ when $m<0$.
Therefore, we need to change the gauge so that
\begin{equation}\label{UkVkN}
\left\{
\begin{split}
u_{\mathbf{k}}=&-\frac{R_1+iR_2}{\sqrt{2R\left(R-R_3\right)}},\\
v_{\mathbf{k}}=&-\frac{R-R_3}{\sqrt{2R\left(R-R_3\right)}}.
\end{split}
\right.
\end{equation}
which can be written explicitly as
\begin{equation}
\left\{
\begin{split}
u_{\mathbf{k}}=&-\frac{1}{\sqrt{N}}ke^{i\theta_{\mathbf{k}}},\\
v_{\mathbf{k}}=&-\frac{1}{\sqrt{N}}\left(\sqrt{(m-k^2)^2+k^2}-(m-k^2)\right).\\
\end{split}
\right.
\end{equation}
 The Berry curvature has the same form as that of Chern insulators, {\it i.e.},
\begin{equation}\label{exactBerrycurvatureTrivial}
\begin{split}
\mathcal{F}(k,\theta_{\mathbf{k}})=&-\frac{1}{2}\frac{m+k^2}{\Big[k^2+(m-k^2)^2\Big]^{\frac{3}{2}}},
\end{split}
\end{equation}
but with $m<0$.
By comparing the cMERA constructed wavefunction $|\Psi(u)\rangle$ in Eq.\ (\ref{GeneralForm001}) and the
exact ground state wavefunction, we can set
$A=-B=ie^{i\theta_{\mathbf{k}}}/2$. Then one has
\begin{equation}\label{QkPkN}
\left\{
\begin{split}
Q_{\mathbf{k}}=&-\cos \varphi_{\mathbf{k}}(u),\\
P_{\mathbf{k}}=&-e^{i\theta_{\mathbf{k}}}\sin \varphi_{\mathbf{k}}(u).
\end{split}
\right.
\end{equation}
For $u=u_{\text{IR}}$, one has
\begin{equation}
|\Omega\rangle=|\Psi(u\to u_{\text{IR}})\rangle=\prod_{\mathbf{k}\le \Lambda}\psi_1^{\dag}(\mathbf{k})|\text{vac}\rangle.
\end{equation}
Then, by requiring $|\Psi(u_{\text{UV}})\rangle=|\Psi\rangle$, where $|\Psi\rangle$ is the exact ground state of a nonrelativistic
insulator, one has
\begin{equation}\label{varphiTrivial}
\varphi_{\mathbf{k}}(u_{\text{UV}})=\arctan\frac{k}{\sqrt{(m-k^2)^2+k^2}-(m-k^2)},
\end{equation}
based on which one can find
\begin{equation}
\begin{split}
g(u)&=\varphi_{\mathbf{k}}(u_{\text{UV}})-k\partial_k \varphi_{\mathbf{k}}(u_{\text{UV}})\big|_{k=\Lambda e^u}\\
&=\frac{1}{2}\frac{\Lambda e^u(m+\Lambda^2 e^{2u})}{(m-\Lambda^2 e^{2u})^2+\Lambda^2 e^{2u}}\\
+&\text{arctan}\frac{\Lambda e^u}{\sqrt{(m-\Lambda^2 e^{2u})^2+\Lambda^2 e^{2u}}-(m-\Lambda^2 e^{2u})}.
\end{split}
\end{equation}
One can simply check that in the IR limit and UV limit,
\begin{equation}
g(u)=\left\{
\begin{split}
&0   & u=u_{\text{IR}},\\
&0   & u=u_{\text{UV}}.
\end{split}
\right.
\end{equation}
With the wavefunction $|\Psi(u)\rangle$ in layer $u$, one can get the Berry curvature
\begin{equation}
\begin{split}
\vec{\mathcal{F}}(k,\theta_{\mathbf{k}};u)=&\hat{\mathbf{u}}\left[\frac{1}{k}\sin2\varphi_{\mathbf{k}}(u)\partial_k\varphi_{\mathbf{k}}(u)\right]\\
&+\hat{\mathbf{k}}\left[-\frac{1}{k}\sin2\varphi_{\mathbf{k}}(u)\partial_u\varphi_{\mathbf{k}}(u)\right],
\end{split}
\end{equation}
based on which one can plot the Berry curvature flow in the bulk of cMERA as shown in Fig.\ \ref{BerryB}.

\subsubsection{cMERA of relativistic insulators with $m>0$}

The cMERA construction of relativistic insulators with $m>0$ in (2+1) dimensions is similar with that of non-relativistic Chern insulators as discussed in the main text. In this case, one has
\begin{equation}\label{RkR01}
\mathbf{R}(\mathbf{k})=(k_x,\ k_y,\ m).
\end{equation}
Then based on Eq.\ (\ref{Umatrix}), one can obtain
\begin{equation}\label{GstateDiracMP}
\left\{
\begin{split}
 u_{\mathbf{k}}=&\frac{1}{\sqrt{N}}\left(m+\sqrt{m^2+k^2}\right),\\
 v_{\mathbf{k}}=&\frac{1}{\sqrt{N}}\left(ke^{-i\theta_{\mathbf{k}}}\right), \\
\end{split}
\right.
\end{equation}
where $N$ is the normalization factor. Based on the ground state wavefunction, one can get the Berry curvature :
\begin{equation}\label{exactBerrycurvatureDirac}
\begin{split}
\mathcal{F}(k,\theta_{\mathbf{k}})=&-\frac{1}{2}\frac{m}{\left(m^2+k^2\right)^{\frac{3}{2}}}.
\end{split}
\end{equation}
\begin{figure}
\includegraphics[width=3.25in]{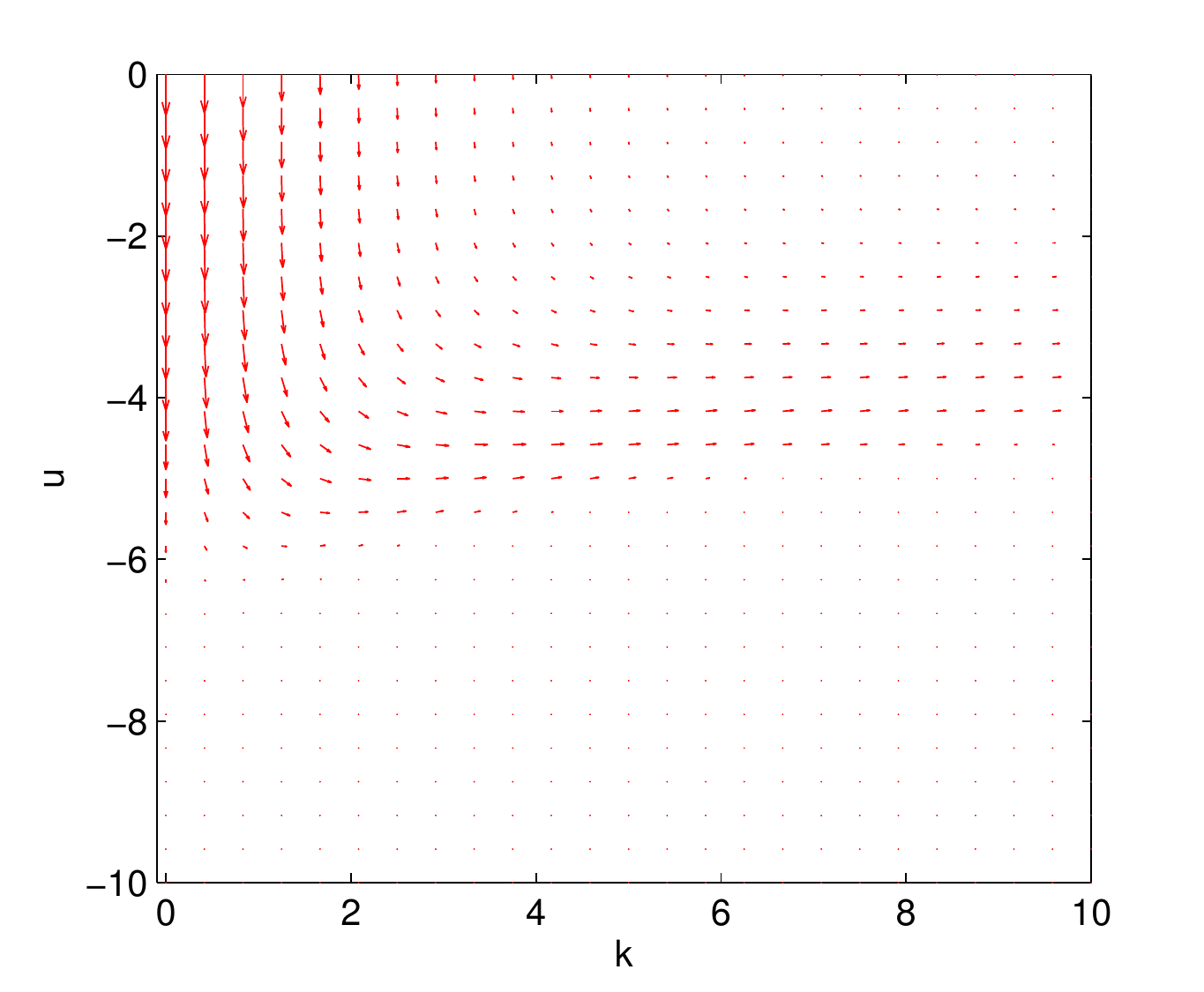}
\caption{Berry curvature flow in cMERA of a relativistic insulator with $m>0$. The parameters we used are $m=2$ and $\Lambda=1000$. }\label{BerryC}
\end{figure}
The cMERA constructed $P_{\mathbf{k}}(u)$ and $Q_{\mathbf{k}}(u)$ have the same expressions as those in
Eq.\ (\ref{WFchern}).
By requiring $|\Psi(u)\rangle=|\Psi\rangle$, one can obtain
\begin{equation}\label{varphiDiracMP}
\varphi_{\mathbf{k}}(u_{\text{UV}})=\arctan\frac{-k}{m+\sqrt{m^2+k^2}},
\end{equation}
based on which one can get the form of $g(u)$:
\begin{equation}
\begin{split}
g(u)=&\varphi_{\mathbf{k}}(u_{\text{UV}})-k\partial_k \varphi_{\mathbf{k}}(u_{\text{UV}})\big|_{k=\Lambda e^u}\\
=&\frac{1}{2}\frac{m\Lambda e^u}{m^2+\Lambda^2 e^{2u}}-\text{arctan}\frac{\Lambda e^u}{\sqrt{m^2+\Lambda^2 e^{2u}}+m}.
\end{split}
\end{equation}
It is straightforward to check that in the IR limit and UV limit, one has
\begin{equation}
g(u)=\left\{
\begin{split}
&0   & u=u_{\text{IR}},\\
&-\frac{\pi}{4}  & u=u_{\text{UV}}.
\end{split}
\right.
\end{equation}
With the wavefunction $|\Psi(u)\rangle$ in layer $u$, one can get the Berry curvature
\begin{equation}
\begin{split}
\vec{\mathcal{F}}(k,\theta_{\mathbf{k}};u)=&\hat{\mathbf{u}}\left[-\frac{1}{k}\sin2\varphi_k(u)\partial_k\varphi_k(u)\right]\\
&+\hat{\mathbf{k}}\left[\frac{1}{k}\sin2\varphi_k(u)\partial_u\varphi_k(u)\right],
\end{split}
\end{equation}
based on which one can obtain the Berry curvature flow in the bulk of cMERA as shown in Fig.\ \ref{BerryC}.

\begin{figure}
\includegraphics[width=3.25in]{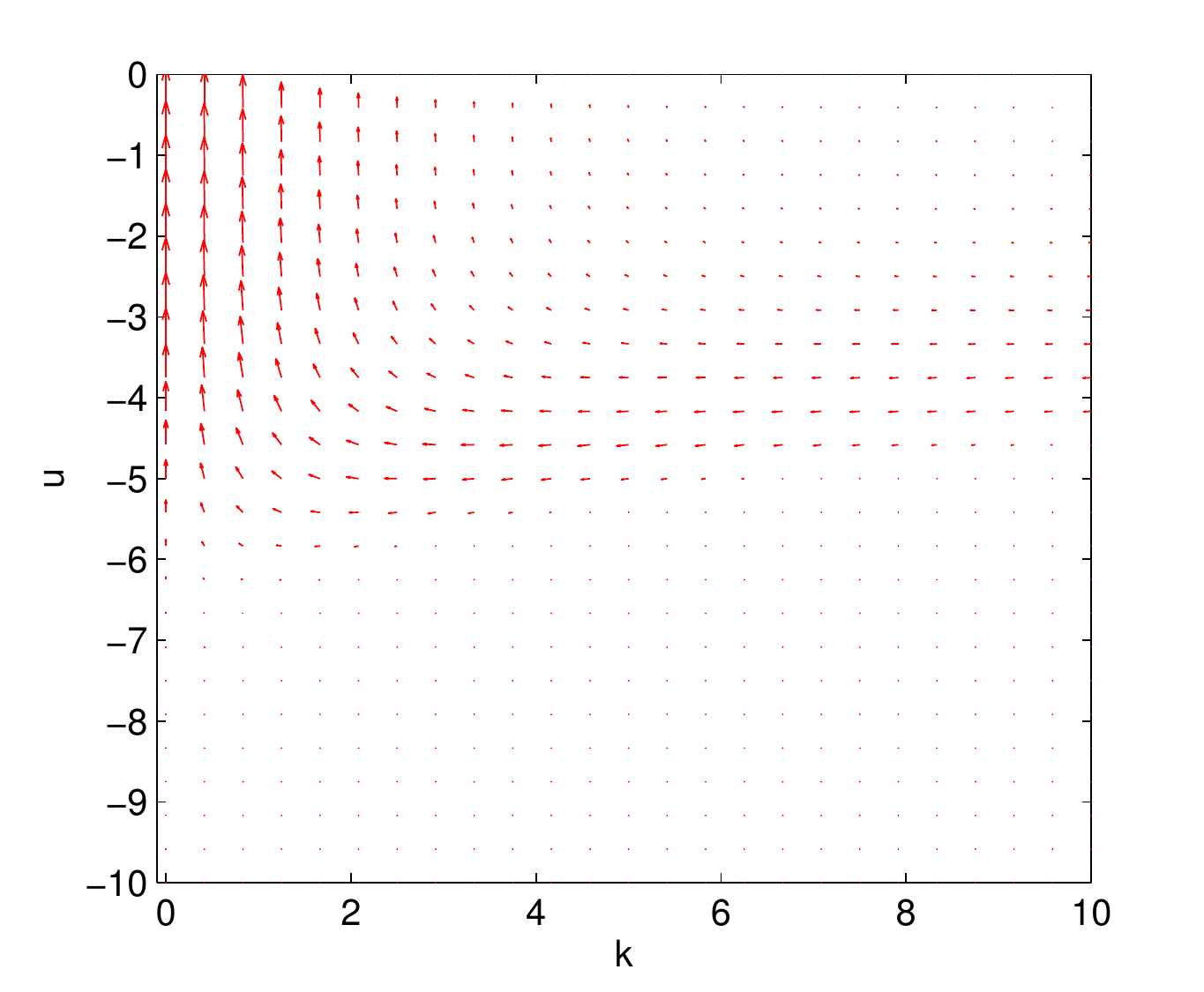}
\caption{ Berry curvature flow in cMERA of a relativistic insulator with $m<0$.  The parameters we use are $m=-2$ and $\Lambda=1000$. }\label{BerryD}
\end{figure}

\subsubsection{cMERA of relativistic insulators with $m<0$}

The cMERA construction of relativistic insulators with $m<0$ in (2+1) dimensions is similar with that of non-relativistic trivial insulators. $\mathbf{R}(\mathbf{k})$ has the same expression as Eq.\ (\ref{RkR01})
except that we use $m<0$ now. $u_{\mathbf{k}}$ and $v_{\mathbf{k}}$ can be obtained based on the
expression in Eq.\ (\ref{UkVkN}), and are expressed as
\begin{equation}\label{GstateDiracMN}
\left\{
\begin{split}
u_{\mathbf{k}}=&-\frac{1}{\sqrt{N}}ke^{i\theta_{\mathbf{k}}},\\
v_{\mathbf{k}}=&-\frac{1}{\sqrt{N}}\left(\sqrt{m^2+k^2}-m\right),
\end{split}
\right.
\end{equation}
where $N$ is the normalization factor. Based on the ground state wavefunction, one can get the Berry curvature with the same expression in Eq.\ (\ref{exactBerrycurvatureDirac}), {\it i.e.},
\begin{equation}\label{exactBerrycurvatureDiracN}
\begin{split}
\mathcal{F}(k,\theta_{\mathbf{k}})=&-\frac{1}{2}\frac{m}{\left(m^2+k^2\right)^{\frac{3}{2}}},
\end{split}
\end{equation}
with $m<0$.
In cMERA construction, similar with the case of non-relativistic trivial insulators, $P_{\mathbf{k}}$ and $Q_{\mathbf{k}}$
have the expressions
\begin{equation}
\left\{
\begin{split}
Q_{\mathbf{k}}=&-\cos \varphi_{\mathbf{k}}(u),\\
P_{\mathbf{k}}=&-e^{i\theta_{\mathbf{k}}}\sin \varphi_{\mathbf{k}}(u).
\end{split}
\right.
\end{equation}
For $u=u_{\text{IR}}$, one has
\begin{equation}
|\Omega\rangle=|\Psi(u\to u_{\text{IR}})\rangle=\prod_{\mathbf{k}\le \Lambda}\psi_1^{\dag}(\mathbf{k})|\text{vac}\rangle.
\end{equation}
By requiring that $|\Psi(u_{\text{UV}})\rangle=|\Psi\rangle$, where $|\Psi\rangle$ is the exact ground state, one has
\begin{equation}\label{varphiDiracMN}
\varphi_{\mathbf{k}}(u_{\text{UV}})=\arctan\frac{k}{\sqrt{m^2+k^2}-m},
\end{equation}
based on which one can obtain the form of $g(u)$:
\begin{equation}
\begin{split}
g(u)=&\varphi_{\mathbf{k}}(u_{\text{UV}})-k\partial_k \varphi_{\mathbf{k}}(u_{\text{UV}})\big|_{k=\Lambda e^u}\\
 =&\frac{1}{2}\frac{m\Lambda e^u}{m^2+\Lambda^2 e^{2u}}+\text{arctan}\frac{\Lambda e^u}{\sqrt{m^2+\Lambda^2 e^{2u}}-m}.
\end{split}
\end{equation}
It is straightforward to check that in the IR limit and UV limit, one has
\begin{equation}
g(u)=\left\{
\begin{split}
&0   & u=u_{\text{IR}},\\
&\frac{\pi}{4}  & u=u_{\text{UV}}.
\end{split}
\right.
\end{equation}
With the wavefunctions $|\Psi(u)\rangle$ in layer $u$, one can obtain the Berry curvature
\begin{equation}
\begin{split}
\vec{\mathcal{F}}(k,\theta_{\mathbf{k}};u)=&\hat{\mathbf{u}}\left[\frac{1}{k}\sin2\varphi_{\mathbf{k}}(u)\partial_{\mathbf{k}}\varphi_k(u)\right]\\
&+\hat{\mathbf{k}}\left[-\frac{1}{k}\sin2\varphi_{\mathbf{k}}(u)\partial_u\varphi_{\mathbf{k}}(u)\right],
\end{split}
\end{equation}
based on which one can obtain the Berry curvature flow in the bulk of cMERA as shown in Fig.\ \ref{BerryD}.

\subsubsection{Vortex feature in the Berry curvature flow}
\label{vortex}

In this part, we analyze the vortex feature near $u=u^{\ast}$ in Fig.\ref{BerryFlowChern}.
Acrossing the vortex core, there are sign changes in both $\mathcal{F}_u(k,\theta_{\mathbf{k}};u)$ and
$\mathcal{F}_k(k,\theta_{\mathbf{k}};u)$.
For convenience, let us rewrite the expression of $\mathcal{F}_u(k,\theta_{\mathbf{k}};u)$ and $\mathcal{F}_k(k,\theta_{\mathbf{k}};u)$ here
\begin{equation}\nonumber
\left\{
\begin{split}
\mathcal{F}_u(k,\theta_{\mathbf{k}};u)=&-\frac{1}{k}\sin2\varphi_{\mathbf{k}}(u)\partial_k\varphi_{\mathbf{k}}(u),\\
\mathcal{F}_k(k,\theta_{\mathbf{k}};u)=&\frac{1}{k}\sin2\varphi_{\mathbf{k}}(u)\partial_u\varphi_{\mathbf{k}}(u).
\end{split}
\right.
\end{equation}

First, let us discuss the sign change in $\mathcal{F}_u(k,\theta_{\mathbf{k}};u)$
as we change $k$ while keeping $u=u^{\ast}$ fixed.
In $\mathcal{F}_u(k,\theta_{\mathbf{k}};u)$, the explicit form of $\partial_k\varphi_{\mathbf{k}}(u)$ is
\begin{equation}\label{partialkphi}
\partial_{\mathbf{k}}\varphi_k(u)=-\frac{g(s)}{\Lambda e^s}\Big|_{s=\log k/\Lambda}+\int_{\log k/\Lambda}^u ds \frac{g(s)}{\Lambda e^s}.
\end{equation}
For $k\ll k^{\ast}$, the boundary term $-\frac{g(s)}{\Lambda e^s}\Big|_{s=\log k/\Lambda}$ equals zero and does not
play any role, and therefore one has positive $\partial_k\varphi_{\mathbf{k}}(u)$.
As $k\to k^{\ast}$, however, the boundary term dominates and one has negative $\partial_k\varphi_{\mathbf{k}}(u)$, which explains the sign change in
$\mathcal{F}_u(k,\theta_{\mathbf{k}};u)$.

Second, let us discuss the sign change
in $\mathcal{F}_k(k,\theta_{\mathbf{k}};u)$ as we change $u$ across $u^{\ast}$
while keeping $k$ fixed.
This is directly related with the sign change of $g^{(a)}(u)$(see Fig.\ \ref{gu}) by considering
\begin{equation}\label{partialuUvarphik001}
\partial_u\varphi_{\mathbf{k}}(u)=\frac{k}{\Lambda e^u}g(u).
\end{equation}
In a short sum, the vortex feature of $\vec{\mathcal{F}}(k,\theta_{\mathbf{k}};u)$ in
cMERA of Chern insulators is closely related with the sign change in $g(u)$.
Note that in the other three phases, {\it i.e.}, non-relativistic trivial insulators,
relativistic insulators with $m>0$ and relativistic insulators with $m<0$,
there is no sign change in the corresponding $g^i(u)$ (see Fig.\ \ref{gu}).
Therefore, the vortex feature in Berry curvature flow only exists in cMERA
of Chern insulators.

\subsection{Other components of metric in cMERA}
\label{MetricOther}

Given the wavefunction $|\Psi^i(u)\rangle$, we can also calculate other components of the metric at each layer $u$.
Similar with the method to define $g_{uu}(\mathbf{k},u)$, we consider the overlap of
wavefunctions $|\Psi^i(\mathbf{k},u)\rangle$ and $|\Psi^i(\mathbf{k}+d\mathbf{k},u)\rangle$,
where $|\Psi^i(\mathbf{k},u)\rangle$ is the single-particle wavefunction.
Then one can get
\begin{equation}
\begin{split}
g^i_{kk}(\mathbf{k},u)=&\text{Re}\langle \partial_k\Psi^i(\mathbf{k},u)|\partial_k \Psi^i(\mathbf{k},u)\rangle\\
&-\langle \partial_k\Psi^i(\mathbf{k},u)|\Psi^i(\mathbf{k},u)\rangle\langle \Psi^i(\mathbf{k},u)|\partial_k\Psi^i(\mathbf{k},u)\rangle.
\end{split}
\end{equation}
which can be further simplified as
\begin{equation}\label{gkk}
g^i_{kk}(\mathbf{k},u)=\left[\partial_k\varphi_{\mathbf{k}}^i(u)\right]^2.
\end{equation}
Following similar procedures, one can calculate $g_{ku}^i(\mathbf{k},u)$, and the result is
\begin{equation}\label{gku}
g_{ku}^i(\mathbf{k},u)=\partial_k\varphi^i_{\mathbf{k}}(u)\partial_u \varphi^i_{\mathbf{k}}(u),
\end{equation}
where $\partial_k\varphi^i_{\mathbf{k}}(u)$ is expressed in Eq.\ (\ref{partialkphi}) and
$\partial_u\varphi^i_{\mathbf{k}}(u)$ is expressed as in Eq.\ (\ref{partialuUvarphik001}).

\subsection{Berry curvature flow in cMERA of a Chern insulator with a topologically nontrival IR state}
\label{BerryFlowNontrivialIR}

In this part, we study how the feature of Berry curvature flow in Fig.\ \ref{BerryFlow001}
is related with the behavior of $g^{\text{nontrivial}}(u)$.

\emph{(i) Behavior of $F_{k}(k,\theta_{\mathbf{k}};u)$}:

Based on the expression of $\vec{\mathcal{F}}(k,\theta_{\mathbf{k}};u)$ in Eq.\ (\ref{vector001}), one has
\begin{equation}\nonumber
\mathcal{F}_k(k,\theta_{\mathbf{k}};u)=-\frac{1}{k}\sin2\varphi_{\mathbf{k}}(u)\partial_u\varphi_{\mathbf{k}}(u).
\end{equation}
Then by using the expression of $\partial_u\varphi_{\mathbf{k}}(u)$ in Eq.\ (\ref{partialuUvarphik001}),
$\mathcal{F}_k(k,\theta_{\mathbf{k}};u)$ can be expressed as
\begin{equation}\label{fk001}
\mathcal{F}_k(k,\theta_{\mathbf{k}};u)=-\frac{\sin 2\varphi_{\mathbf{k}}(u)}{\Lambda e^u}g^{\text{nontrivial}}(u).
\end{equation}
To make an estimation of $\mathcal{F}_k(k,\theta_{\mathbf{k}};u)$, we simply use the approximated form of $g^{\text{nontrivial}}(u)$ as follows
\begin{equation}\label{approximation001}
g^{\text{nontrivial}}(u)\simeq\left\{
\begin{split}
&\frac{\pi}{2}, \ \ \ &u<u^{\ast}\\
&0, \ \ \ &u>u^{\ast}.
\end{split}
\right.
\end{equation}
Based on Eqs.\ (\ref{fk001}) and (\ref{approximation001}), one can get
\begin{equation}\nonumber
\mathcal{F}_k(k,\theta_{\mathbf{k}};u)=0, \ \ \ \forall u>u^{\ast},
\end{equation}
which indicates that there is on Berry curvature flow in $k$ direction.
On the contrary, for $u<u^{\ast}$, one has
\begin{equation}\nonumber
\mathcal{F}_k(k,\theta_{\mathbf{k}};u)=-\frac{\pi}{2}\frac{\sin2\varphi_{\mathbf{k}}(u)}{\Lambda e^u},
\end{equation}
where
\begin{equation}\nonumber
\begin{split}
\varphi_{\mathbf{k}}(u)=\frac{\pi}{2}\int_{\log\frac{k}{\Lambda}}^{u}ds \frac{k}{\Lambda e^s}=\frac{\pi}{2}\left(1-\frac{k}{\Lambda e^{u}}\right).
\end{split}
\end{equation}
Considering that $k<\Lambda e^u$, one always has $\sin2\varphi_k(u)>0$. Therefore, one has
\begin{equation}\nonumber
\mathcal{F}_k(k,\theta_{\mathbf{k}};u)<0, \ \ \ \forall u<u^{\ast}.
\end{equation}
From the analysis above, it is found that the Berry curvature component $\mathcal{F}_k(k,\theta_{\mathbf{k}};u)$
is finite only in the IR layers and it points towards the smaller $k$ direction.

\emph{(ii) Behavior of $F_{u}(k,\theta_{\mathbf{k}};u)$}:

To study the bending of the Berry curvature flow, we are interested in the region $k<\text{min}\left[k^{\ast},\Lambda e^u\right]$. From Eq.\ (\ref{vector001}), one has
\begin{equation}
\mathcal{F}_u(k,\theta_{\mathbf{k}};u)=\frac{1}{k}\sin2\varphi_{\mathbf{k}}(u)\partial_{\mathbf{k}}\varphi_k(u).
\end{equation}
By using the expression of $\partial_k\varphi_{\mathbf{k}}(u)$ in Eq.\ (\ref{partialkphi}) and the approximation in Eq.\ (\ref{approximation001}) one has
\begin{equation}
\mathcal{F}_u(k,\theta_{\mathbf{k}};u)=
\left\{
\begin{split}
&0, \ \ \ \ &u>u^{\ast}\\
&-\frac{\pi}{2}\frac{1}{\Lambda e^u}\frac{\sin2\varphi_{\mathbf{k}}(u)}{k}, \ \ \ \ &u<u^{\ast}.
\end{split}
\right.
\end{equation}
Considering that $\sin2\varphi_{\mathbf{k}}(u)>0$, one always has $\mathcal{F}_u(k,\theta_{\mathbf{k}};u)<0$ for
$u<u^{\ast}$. This means that $\mathcal{F}_u(k,\theta_{\mathbf{k}};u)$ always flows towards the IR layer.

Based on the above analysis on $\mathcal{F}_k(k,\theta_{\mathbf{k}};u)$ and $\mathcal{F}_u(k,\theta_{\mathbf{k}};u)$,
we understand that the Berry curvature emanated from the UV layer is bent towards smaller
$k$ near $u^{\ast}$, and then flows towards the IR layer (see Fig.\ \ref{BerryFlow001} and
Fig.\ \ref{scheme00B}).

\subsection{Discussion on trivial and non-trivial IR states for cMERA of Chern insulators}

In the main text, we have studied the cMERA construction of a Chern insulator with  topologically trivial and nontrivial IR states, respectively. In both cases, we require that
\begin{equation}
|\Psi(\mathbf{k},u=u_{\text{UV}})\rangle=|\Psi\rangle,
\end{equation}
where $|\Psi\rangle$ is the exact ground state of a Chern insulator. One may ask why the cMERA
with a topologically nontrivial IR state can recover $|\Psi\rangle$ in the whole region ($0\le k\le \Lambda$) while
the cMERA with a topologically trivial IR state cannot fulfill this?
Here we will discuss this problem mainly from the mathematical point of view.

For cMERA of Chern insulators with a topologically trivial IR state, the wavefunction at each layer
is
$
|\Psi(u)\rangle=
\prod_{|\mathbf{k}|\le \Lambda}\left(
P_{\mathbf{k}}(u)\psi^{\dag}_2(\mathbf{k})-Q_{\mathbf{k}}(u)\psi^{\dag}_1(\mathbf{k})
\right)|\text{vac}\rangle,
$
where
\begin{equation}\label{wf0001}
\left\{
\begin{split}
Q_{\mathbf{k}}(u)=&-e^{-i\theta_{\mathbf{k}}}\sin \varphi_{\mathbf{k}}(u)\\
P_{\mathbf{k}}(u)=&\cos \varphi_{\mathbf{k}}(u).
\end{split}
\right.
\end{equation}
By requiring $|\Psi(\mathbf{k},u=u_{\text{UV}})\rangle=|\Psi\rangle$, we can obtain the form of $\varphi_{\mathbf{k}}(u)$ by solving
differential equations. However, It is found that the solution does not match the boundary condition
at $u=u_{\text{UV}}$ for large $k$.
Let us check this problem explicitly as follows.

In the large $k$ limit $k\to\Lambda$, one has
\begin{equation}
\varphi_{\mathbf{k}}(k\to\Lambda,u_{\text{UV}})=\lim_{k\to\Lambda}\int_{\log k/\Lambda}^{u_{\text{UV}}=0}ds g(s)\frac{k}{\Lambda e^s}= 0.
\end{equation}
Therefore, the cMERA constructed single-particle wavefunction for $|\mathbf{k}|\to \Lambda$ at $u_{\text{UV}}$ reads
\begin{equation}
|\Psi(\mathbf{k},u)\rangle=
\psi^{\dag}_2(\mathbf{k})
|\text{vac}\rangle.
\end{equation}
On the other hand, the exact  single-particle wavefunction at large momentum  $|\mathbf{k}|\to \Lambda$ reads
\begin{equation}
|\Psi(\mathbf{k})\rangle=
-e^{-i\theta_{\mathbf{k}}} \psi^{\dag}_1(\mathbf{k})|\text{vac}\rangle.
\end{equation}
Apparently, $|\Psi(\mathbf{k},u_{\text{UV}})\rangle \neq |\Psi(\mathbf{k})\rangle$ for $|\mathbf{k}|\to\Lambda$, \emph{i.e.}, the boundary condition does not match. To solve this problem, one
needs to modify $Q_{\mathbf{k}}(u)$ and $P_{\mathbf{k}}(u)$
 in Eq.\ (\ref{wf0001}) as
\begin{equation}\label{wf0002}
\left\{
\begin{split}
Q_{\mathbf{k}}(u)=&-e^{-i\theta_{\mathbf{k}}}\sin \left(\varphi_{\mathbf{k}}(u)-\frac{\pi}{2}\right)\\
P_{\mathbf{k}}(u)=&\cos \left(\varphi_{\mathbf{k}}(u)-\frac{\pi}{2}\right).
\end{split}
\right.
\end{equation}
In this way, one can find that in the large $k$ limit,
\begin{equation}
|\Psi'(\mathbf{k})\rangle\simeq
-e^{-i\theta_{\mathbf{k}}} \psi^{\dag}_1(\mathbf{k})|\text{vac}\rangle,
\end{equation}
which satisfies the boundary condition at large momentum $k$. One may be worried whether
 $|\Psi'(\mathbf{k},u_{\text{UV}})\rangle$ satisfies the boundary condition for small $k$. This can also be explicitly checked as follows. In the small $k$ limit, one can simply use the approximation in Eq.\ (\ref{apparoximationVarPhi}), and then one can obtain
\begin{equation}
\varphi_{\mathbf{k}\to0}(u)\simeq \frac{\pi}{2}.
\end{equation}
Therefore, the cMERA constructed single-particle wavefunction $|\Psi'(\mathbf{k},u)\rangle$ in small $k$ limit reads
\begin{equation}
|\Psi'(\mathbf{k}\to 0,u_{\text{UV}})\rangle\simeq
\psi^{\dag}_2(\mathbf{k})|\text{vac}\rangle,
\end{equation}
which agrees with the exact boundary condition for $\mathbf{k}\to 0$.

\subsection{Generalization to higher Chern number cases}
\label{Hch}

In this part, we generalize our cMERA method to construct the ground state of the Hamiltonian
\begin{equation}\label{H000H}
H=\int d^2\mathbf{k}\psi^{\dag}(\mathbf{k})h(\mathbf{k})\psi(\mathbf{k}),
\end{equation}
with
\begin{equation}
h(\mathbf{k})=\left(
\begin{matrix}
&m-(\mathbf{k}\cdot\mathbf{k})^{\gamma}        &(k_x-ik_y)^{\gamma}  \\
&(k_x+ik_y)^{\gamma}     &-(m-(\mathbf{k}\cdot\mathbf{k})^{\gamma})
\end{matrix}
\right),
\end{equation}
where $\gamma$ is an integer and $m>0$.
Alternatively, $h(\mathbf{k})$ can be rewritten as
\begin{equation}\label{H1h}
h(\mathbf{k})
=
\left(
\begin{matrix}
&m-k^{2\gamma}        &k^{\gamma}e^{-i\gamma\theta_{\mathbf{k}}}  \\
&k^{\gamma}e^{i\gamma\theta_{\mathbf{k}}}     &-(m-k^{2\gamma}  )
\end{matrix}
\right).
\end{equation}
The first Chern number corresponding to the
ground state of the Hamiltonian in Eq.\ (\ref{H000H}) is $\text{Ch}_1=-\gamma$.
The disentangler $\hat{K}(u)$ has the expression
\begin{equation}
\hat{K}(u)=i\int d^2\mathbf{k} \left[g_{\mathbf{k}}(u)\psi_1^{\dag}(\mathbf{k})\psi_2(\mathbf{k})+g_{\mathbf{k}}^{\ast}(u)\psi_1(\mathbf{k})\psi_2^{\dag}(\mathbf{k})\right]
\end{equation}
where
\begin{equation}
g_{\mathbf{k}}(u):=g(u)\Gamma\left(\frac{k}{\Lambda e^u}\right)\frac{k}{\Lambda e^u}e^{-i\gamma\theta_{\mathbf{k}}}.
\end{equation}
Following the procedures of cMERA construction in the main text, we can obtain
the cMERA constructed wavefunction
$
|\Psi(u)\rangle=
\prod_{|\mathbf{k}|\le \Lambda}\left(
P_{\mathbf{k}}(u)\psi^{\dag}_2(\mathbf{k})-Q_{\mathbf{k}}(u)\psi^{\dag}_1(\mathbf{k})
\right)|\text{vac}\rangle,
$
where
\begin{equation}\label{trivialIR2}
\left\{
\begin{split}
P_{\mathbf{k}}(u)=&\cos\varphi_k(u),\\
Q_{\mathbf{k}}(u)=&-e^{-i\gamma \theta_{\mathbf{k}}}\sin\varphi_k(u),\\
\end{split}
\right.
\end{equation}
and the corresponding IR state is $|\Omega\rangle=\prod_{|\mathbf{k}|\le \Lambda} \psi_2^{\dag}(\mathbf{k})|\text{vac}\rangle$,
which is topologically trivial.
By requiring that the cMERA constructed wavefunction at the UV layer $u=u_{\text{UV}}$ matches the exact ground state,
one can obtain
\begin{equation}\label{AppendixGuabcd}
\begin{split}
&g_{\text{trivial}}(u)
=\frac{\gamma}{2}\frac{(\Lambda e^u)^{\gamma}\left[m+(\Lambda e^{u})^{2\gamma}\right]}{\left[m-(\Lambda e^{u})^{2\gamma}\right]^2+(\Lambda e^{u})^{2\gamma}}\\
&-\text{arctan}\frac{(\Lambda e^u)^{\gamma}}{\sqrt{\left[m-(\Lambda e^{u})^{2\gamma}\right]^2+(\Lambda e^{u})^{2\gamma}}+m-(\Lambda e^{u})^{2\gamma}}.
\end{split}
\end{equation}
Note that when $\gamma=1$, we recover the results in Eq.\ (\ref{gua}).

Similar with our conclusion in the main text, to recover the exact ground state in the whole region $0\le k\le \Lambda$,
one should consider the cMERA with a nontrivial IR state. Then $P_{\mathbf{k}}(u)$ and $Q_{\mathbf{k}}(u)$
have the following expressions
\begin{equation}
\left\{
\begin{split}
P_{\mathbf{k}}(u)=&\sin\varphi_{\mathbf{k}}(u),\\
Q_{\mathbf{k}}(u)=&e^{-i\gamma \theta_{\mathbf{k}}}\cos\varphi_{\mathbf{k}}(u).\\
\end{split}
\right.
\end{equation}
The corresponding IR state is
\begin{equation}
|\Omega\rangle=\prod_{|\mathbf{k}|\le \Lambda}\left(- e^{-i\gamma\theta_{\mathbf{k}}} \psi_1^{\dag}(\mathbf{k})
\right)|\text{vac}\rangle.
\end{equation}
Compared to $g_{\text{trivial}}$ in Eq.\ (\ref{AppendixGuabcd}), $g_{\text{nontrivial}}$ can be expressed as
\begin{equation}
g_{\text{nontrivial}}(u)=g_{\text{trivial}}(u)+\frac{\pi}{2}.
\end{equation}

\bibliography{cMERAref}

\begin{thebibliography}{52}%
\makeatletter
\providecommand \@ifxundefined [1]{%
 \@ifx{#1\undefined}
}%
\providecommand \@ifnum [1]{%
 \ifnum #1\expandafter \@firstoftwo
 \else \expandafter \@secondoftwo
 \fi
}%
\providecommand \@ifx [1]{%
 \ifx #1\expandafter \@firstoftwo
 \else \expandafter \@secondoftwo
 \fi
}%
\providecommand \natexlab [1]{#1}%
\providecommand \enquote  [1]{``#1''}%
\providecommand \bibnamefont  [1]{#1}%
\providecommand \bibfnamefont [1]{#1}%
\providecommand \citenamefont [1]{#1}%
\providecommand \href@noop [0]{\@secondoftwo}%
\providecommand \href [0]{\begingroup \@sanitize@url \@href}%
\providecommand \@href[1]{\@@startlink{#1}\@@href}%
\providecommand \@@href[1]{\endgroup#1\@@endlink}%
\providecommand \@sanitize@url [0]{\catcode `\\12\catcode `\$12\catcode
  `\&12\catcode `\#12\catcode `\^12\catcode `\_12\catcode `\%12\relax}%
\providecommand \@@startlink[1]{}%
\providecommand \@@endlink[0]{}%
\providecommand \url  [0]{\begingroup\@sanitize@url \@url }%
\providecommand \@url [1]{\endgroup\@href {#1}{\urlprefix }}%
\providecommand \urlprefix  [0]{URL }%
\providecommand \Eprint [0]{\href }%
\providecommand \doibase [0]{http://dx.doi.org/}%
\providecommand \selectlanguage [0]{\@gobble}%
\providecommand \bibinfo  [0]{\@secondoftwo}%
\providecommand \bibfield  [0]{\@secondoftwo}%
\providecommand \translation [1]{[#1]}%
\providecommand \BibitemOpen [0]{}%
\providecommand \bibitemStop [0]{}%
\providecommand \bibitemNoStop [0]{.\EOS\space}%
\providecommand \EOS [0]{\spacefactor3000\relax}%
\providecommand \BibitemShut  [1]{\csname bibitem#1\endcsname}%
\let\auto@bib@innerbib\@empty
\bibitem [{\citenamefont {Vidal}(2007)}]{Vidal2005}%
  \BibitemOpen
  \bibfield  {author} {\bibinfo {author} {\bibfnamefont {G.}~\bibnamefont
  {Vidal}},\ }\bibfield  {title} {\enquote {\bibinfo {title} {Entanglement
  renormalization},}\ }\href {\doibase 10.1103/PhysRevLett.99.220405}
  {\bibfield  {journal} {\bibinfo  {journal} {Phys. Rev. Lett.}\ }\textbf
  {\bibinfo {volume} {99}},\ \bibinfo {pages} {220405} (\bibinfo {year}
  {2007})}\BibitemShut {NoStop}%
\bibitem [{\citenamefont {Vidal}(2008)}]{VidalPRL2008}%
  \BibitemOpen
  \bibfield  {author} {\bibinfo {author} {\bibfnamefont {G.}~\bibnamefont
  {Vidal}},\ }\bibfield  {title} {\enquote {\bibinfo {title} {Class of quantum
  many-body states that can be efficiently simulated},}\ }\href {\doibase
  10.1103/PhysRevLett.101.110501} {\bibfield  {journal} {\bibinfo  {journal}
  {Phys. Rev. Lett.}\ }\textbf {\bibinfo {volume} {101}},\ \bibinfo {pages}
  {110501} (\bibinfo {year} {2008})}\BibitemShut {NoStop}%
\bibitem [{\citenamefont {Evenbly}\ and\ \citenamefont
  {Vidal}(2010)}]{VidalPRB2008}%
  \BibitemOpen
  \bibfield  {author} {\bibinfo {author} {\bibfnamefont {G.}~\bibnamefont
  {Evenbly}}\ and\ \bibinfo {author} {\bibfnamefont {G.}~\bibnamefont
  {Vidal}},\ }\bibfield  {title} {\enquote {\bibinfo {title} {Entanglement
  renormalization in noninteracting fermionic systems},}\ }\href {\doibase
  10.1103/PhysRevB.81.235102} {\bibfield  {journal} {\bibinfo  {journal} {Phys.
  Rev. B}\ }\textbf {\bibinfo {volume} {81}},\ \bibinfo {pages} {235102}
  (\bibinfo {year} {2010})}\BibitemShut {NoStop}%
\bibitem [{\citenamefont {Cincio}\ \emph {et~al.}(2008)\citenamefont {Cincio},
  \citenamefont {Dziarmaga},\ and\ \citenamefont {Rams}}]{Rams2008}%
  \BibitemOpen
  \bibfield  {author} {\bibinfo {author} {\bibfnamefont {Lukasz}\ \bibnamefont
  {Cincio}}, \bibinfo {author} {\bibfnamefont {Jacek}\ \bibnamefont
  {Dziarmaga}}, \ and\ \bibinfo {author} {\bibfnamefont {Marek~M.}\
  \bibnamefont {Rams}},\ }\bibfield  {title} {\enquote {\bibinfo {title}
  {Multiscale entanglement renormalization ansatz in two dimensions: Quantum
  ising model},}\ }\href {\doibase 10.1103/PhysRevLett.100.240603} {\bibfield
  {journal} {\bibinfo  {journal} {Phys. Rev. Lett.}\ }\textbf {\bibinfo
  {volume} {100}},\ \bibinfo {pages} {240603} (\bibinfo {year}
  {2008})}\BibitemShut {NoStop}%
\bibitem [{\citenamefont {Evenbly}\ and\ \citenamefont
  {Vidal}(2009{\natexlab{a}})}]{VidalPRL2009}%
  \BibitemOpen
  \bibfield  {author} {\bibinfo {author} {\bibfnamefont {G.}~\bibnamefont
  {Evenbly}}\ and\ \bibinfo {author} {\bibfnamefont {G.}~\bibnamefont
  {Vidal}},\ }\bibfield  {title} {\enquote {\bibinfo {title} {Entanglement
  renormalization in two spatial dimensions},}\ }\href {\doibase
  10.1103/PhysRevLett.102.180406} {\bibfield  {journal} {\bibinfo  {journal}
  {Phys. Rev. Lett.}\ }\textbf {\bibinfo {volume} {102}},\ \bibinfo {pages}
  {180406} (\bibinfo {year} {2009}{\natexlab{a}})}\BibitemShut {NoStop}%
\bibitem [{\citenamefont {Pfeifer}\ \emph {et~al.}(2009)\citenamefont
  {Pfeifer}, \citenamefont {Evenbly},\ and\ \citenamefont
  {Vidal}}]{VidalPRA2009}%
  \BibitemOpen
  \bibfield  {author} {\bibinfo {author} {\bibfnamefont {Robert N.~C.}\
  \bibnamefont {Pfeifer}}, \bibinfo {author} {\bibfnamefont {Glen}\
  \bibnamefont {Evenbly}}, \ and\ \bibinfo {author} {\bibfnamefont {Guifr\'e}\
  \bibnamefont {Vidal}},\ }\bibfield  {title} {\enquote {\bibinfo {title}
  {Entanglement renormalization, scale invariance, and quantum criticality},}\
  }\href {\doibase 10.1103/PhysRevA.79.040301} {\bibfield  {journal} {\bibinfo
  {journal} {Phys. Rev. A}\ }\textbf {\bibinfo {volume} {79}},\ \bibinfo
  {pages} {040301} (\bibinfo {year} {2009})}\BibitemShut {NoStop}%
\bibitem [{\citenamefont {Aguado}\ and\ \citenamefont
  {Vidal}(2008)}]{Vidal2008top}%
  \BibitemOpen
  \bibfield  {author} {\bibinfo {author} {\bibfnamefont {Miguel}\ \bibnamefont
  {Aguado}}\ and\ \bibinfo {author} {\bibfnamefont {Guifr\'e}\ \bibnamefont
  {Vidal}},\ }\bibfield  {title} {\enquote {\bibinfo {title} {Entanglement
  renormalization and topological order},}\ }\href {\doibase
  10.1103/PhysRevLett.100.070404} {\bibfield  {journal} {\bibinfo  {journal}
  {Phys. Rev. Lett.}\ }\textbf {\bibinfo {volume} {100}},\ \bibinfo {pages}
  {070404} (\bibinfo {year} {2008})}\BibitemShut {NoStop}%
\bibitem [{\citenamefont {K\"onig}\ \emph {et~al.}(2009)\citenamefont
  {K\"onig}, \citenamefont {Reichardt},\ and\ \citenamefont
  {Vidal}}]{Vidal2009top}%
  \BibitemOpen
  \bibfield  {author} {\bibinfo {author} {\bibfnamefont {Robert}\ \bibnamefont
  {K\"onig}}, \bibinfo {author} {\bibfnamefont {Ben~W.}\ \bibnamefont
  {Reichardt}}, \ and\ \bibinfo {author} {\bibfnamefont {Guifr\'e}\
  \bibnamefont {Vidal}},\ }\bibfield  {title} {\enquote {\bibinfo {title}
  {Exact entanglement renormalization for string-net models},}\ }\href
  {\doibase 10.1103/PhysRevB.79.195123} {\bibfield  {journal} {\bibinfo
  {journal} {Phys. Rev. B}\ }\textbf {\bibinfo {volume} {79}},\ \bibinfo
  {pages} {195123} (\bibinfo {year} {2009})}\BibitemShut {NoStop}%
\bibitem [{\citenamefont {Haegeman}\ \emph {et~al.}(2013)\citenamefont
  {Haegeman}, \citenamefont {Osborne}, \citenamefont {Verschelde},\ and\
  \citenamefont {Verstraete}}]{Haegeman}%
  \BibitemOpen
  \bibfield  {author} {\bibinfo {author} {\bibfnamefont {Jutho}\ \bibnamefont
  {Haegeman}}, \bibinfo {author} {\bibfnamefont {Tobias~J.}\ \bibnamefont
  {Osborne}}, \bibinfo {author} {\bibfnamefont {Henri}\ \bibnamefont
  {Verschelde}}, \ and\ \bibinfo {author} {\bibfnamefont {Frank}\ \bibnamefont
  {Verstraete}},\ }\bibfield  {title} {\enquote {\bibinfo {title} {Entanglement
  renormalization for quantum fields in real space},}\ }\href {\doibase
  10.1103/PhysRevLett.110.100402} {\bibfield  {journal} {\bibinfo  {journal}
  {Phys. Rev. Lett.}\ }\textbf {\bibinfo {volume} {110}},\ \bibinfo {pages}
  {100402} (\bibinfo {year} {2013})}\BibitemShut {NoStop}%
\bibitem [{\citenamefont {Nozaki}\ \emph {et~al.}(2012)\citenamefont {Nozaki},
  \citenamefont {Ryu},\ and\ \citenamefont {Takayanagi}}]{Ryu2012}%
  \BibitemOpen
  \bibfield  {author} {\bibinfo {author} {\bibfnamefont {Masahiro}\
  \bibnamefont {Nozaki}}, \bibinfo {author} {\bibfnamefont {Shinsei}\
  \bibnamefont {Ryu}}, \ and\ \bibinfo {author} {\bibfnamefont {Tadashi}\
  \bibnamefont {Takayanagi}},\ }\bibfield  {title} {\enquote {\bibinfo {title}
  {Holographic geometry of entanglement renormalization in quantum field
  theories},}\ }\href {\doibase 10.1007/JHEP10(2012)193} {\bibfield  {journal}
  {\bibinfo  {journal} {Journal of High Energy Physics}\ }\textbf {\bibinfo
  {volume} {2012}},\ \bibinfo {eid} {193} (\bibinfo {year} {2012}),\
  10.1007/JHEP10(2012)193}\BibitemShut {NoStop}%
\bibitem [{\citenamefont {Swingle}(2012)}]{Swingle2012}%
  \BibitemOpen
  \bibfield  {author} {\bibinfo {author} {\bibfnamefont {Brian}\ \bibnamefont
  {Swingle}},\ }\bibfield  {title} {\enquote {\bibinfo {title} {Entanglement
  renormalization and holography},}\ }\href {\doibase
  10.1103/PhysRevD.86.065007} {\bibfield  {journal} {\bibinfo  {journal} {Phys.
  Rev. D}\ }\textbf {\bibinfo {volume} {86}},\ \bibinfo {pages} {065007}
  (\bibinfo {year} {2012})}\BibitemShut {NoStop}%
\bibitem [{\citenamefont {Van~Raamsdonk}(2010{\natexlab{a}})}]{Van2010}%
  \BibitemOpen
  \bibfield  {author} {\bibinfo {author} {\bibfnamefont {Mark}\ \bibnamefont
  {Van~Raamsdonk}},\ }\bibfield  {title} {\enquote {\bibinfo {title} {Building
  up space–time with quantum entanglement},}\ }\href {\doibase
  10.1142/S0218271810018529} {\bibfield  {journal} {\bibinfo  {journal}
  {International Journal of Modern Physics D}\ }\textbf {\bibinfo {volume}
  {19}},\ \bibinfo {pages} {2429--2435} (\bibinfo {year}
  {2010}{\natexlab{a}})}\BibitemShut {NoStop}%
\bibitem [{\citenamefont {Van~Raamsdonk}(2010{\natexlab{b}})}]{Van2010b}%
  \BibitemOpen
  \bibfield  {author} {\bibinfo {author} {\bibfnamefont {Mark}\ \bibnamefont
  {Van~Raamsdonk}},\ }\bibfield  {title} {\enquote {\bibinfo {title} {Building
  up spacetime with quantum entanglement},}\ }\href {\doibase
  10.1007/s10714-010-1034-0} {\bibfield  {journal} {\bibinfo  {journal}
  {General Relativity and Gravitation}\ }\textbf {\bibinfo {volume} {42}},\
  \bibinfo {pages} {2323--2329} (\bibinfo {year}
  {2010}{\natexlab{b}})}\BibitemShut {NoStop}%
\bibitem [{\citenamefont {Molina-Vilaplana}\ and\ \citenamefont
  {Sodano}(2011)}]{Vilaplana2011}%
  \BibitemOpen
  \bibfield  {author} {\bibinfo {author} {\bibfnamefont {Javier}\ \bibnamefont
  {Molina-Vilaplana}}\ and\ \bibinfo {author} {\bibfnamefont {Pasquale}\
  \bibnamefont {Sodano}},\ }\bibfield  {title} {\enquote {\bibinfo {title}
  {Holographic view on quantum correlations and mutual information between
  disjoint blocks of a quantum critical system},}\ }\href {\doibase
  10.1007/JHEP10(2011)011} {\bibfield  {journal} {\bibinfo  {journal} {Journal
  of High Energy Physics}\ }\textbf {\bibinfo {volume} {2011}},\ \bibinfo {eid}
  {11} (\bibinfo {year} {2011}),\ 10.1007/JHEP10(2011)011}\BibitemShut
  {NoStop}%
\bibitem [{\citenamefont {Swingle}(2010)}]{Swingle2}%
  \BibitemOpen
  \bibfield  {author} {\bibinfo {author} {\bibfnamefont {Brian}\ \bibnamefont
  {Swingle}},\ }\bibfield  {title} {\enquote {\bibinfo {title} {Mutual
  information and the structure of entanglement in quantum field theory},}\
  }\href@noop {} {\bibfield  {journal} {\bibinfo  {journal} {arXiv:1010.4038}\
  } (\bibinfo {year} {2010})}\BibitemShut {NoStop}%
\bibitem [{\citenamefont {Molina~Vilaplana}(2011)}]{Vilaplana2}%
  \BibitemOpen
  \bibfield  {author} {\bibinfo {author} {\bibfnamefont {Javier}\ \bibnamefont
  {Molina~Vilaplana}},\ }\bibfield  {title} {\enquote {\bibinfo {title}
  {Connecting entanglement renormalization and gauge/gravity dualities},}\
  }\href@noop {} {\bibfield  {journal} {\bibinfo  {journal} {arXiv:1109.5592}\
  } (\bibinfo {year} {2011})}\BibitemShut {NoStop}%
\bibitem [{\citenamefont {Matsueda}(2012)}]{Matsueda}%
  \BibitemOpen
  \bibfield  {author} {\bibinfo {author} {\bibfnamefont {Hiroaki}\ \bibnamefont
  {Matsueda}},\ }\bibfield  {title} {\enquote {\bibinfo {title} {Scaling of
  entanglement entropy and hyperbolic geometry},}\ }\href@noop {} {\bibfield
  {journal} {\bibinfo  {journal} {arXiv:1208.0206}\ } (\bibinfo {year}
  {2012})}\BibitemShut {NoStop}%
\bibitem [{\citenamefont {Okunishi}(2012)}]{Okunishi}%
  \BibitemOpen
  \bibfield  {author} {\bibinfo {author} {\bibfnamefont {Kouichi}\ \bibnamefont
  {Okunishi}},\ }\bibfield  {title} {\enquote {\bibinfo {title} {Wilson's
  numerical renormalization group and ads$_3$ geometry},}\ }\href@noop {}
  {\bibfield  {journal} {\bibinfo  {journal} {arXiv:1208.1645}\ } (\bibinfo
  {year} {2012})}\BibitemShut {NoStop}%
\bibitem [{\citenamefont {Bao}\ \emph {et~al.}(2015)\citenamefont {Bao},
  \citenamefont {Cao}, \citenamefont {Carroll}, \citenamefont {Chatwin-Davies},
  \citenamefont {Hunter-Jones}, \citenamefont {Pollack},\ and\ \citenamefont
  {Remmen}}]{Bao}%
  \BibitemOpen
  \bibfield  {author} {\bibinfo {author} {\bibfnamefont {Ning}\ \bibnamefont
  {Bao}}, \bibinfo {author} {\bibfnamefont {ChunJun}\ \bibnamefont {Cao}},
  \bibinfo {author} {\bibfnamefont {Sean~M.}\ \bibnamefont {Carroll}}, \bibinfo
  {author} {\bibfnamefont {Aidan}\ \bibnamefont {Chatwin-Davies}}, \bibinfo
  {author} {\bibfnamefont {Nicholas}\ \bibnamefont {Hunter-Jones}}, \bibinfo
  {author} {\bibfnamefont {Jason}\ \bibnamefont {Pollack}}, \ and\ \bibinfo
  {author} {\bibfnamefont {Grant~N.}\ \bibnamefont {Remmen}},\ }\bibfield
  {title} {\enquote {\bibinfo {title} {Consistency conditions for an ads/mera
  correspondence},}\ }\href@noop {} {\bibfield  {journal} {\bibinfo  {journal}
  {arXiv:1504.06632}\ } (\bibinfo {year} {2015})}\BibitemShut {NoStop}%
\bibitem [{\citenamefont {Miyaji}\ \emph
  {et~al.}(2015{\natexlab{a}})\citenamefont {Miyaji}, \citenamefont {Ryu},
  \citenamefont {Takayanagi},\ and\ \citenamefont {Wen}}]{Miyaji2015A}%
  \BibitemOpen
  \bibfield  {author} {\bibinfo {author} {\bibfnamefont {Masamichi}\
  \bibnamefont {Miyaji}}, \bibinfo {author} {\bibfnamefont {Shinsei}\
  \bibnamefont {Ryu}}, \bibinfo {author} {\bibfnamefont {Tadashi}\ \bibnamefont
  {Takayanagi}}, \ and\ \bibinfo {author} {\bibfnamefont {Xueda}\ \bibnamefont
  {Wen}},\ }\bibfield  {title} {\enquote {\bibinfo {title} {Boundary states as
  holographic duals of trivial spacetimes},}\ }\href {\doibase
  10.1007/JHEP05(2015)152} {\bibfield  {journal} {\bibinfo  {journal} {Journal
  of High Energy Physics}\ }\textbf {\bibinfo {volume} {2015}},\ \bibinfo {eid}
  {14} (\bibinfo {year} {2015}{\natexlab{a}}),\
  10.1007/JHEP05(2015)152}\BibitemShut {NoStop}%
\bibitem [{\citenamefont {Miyaji}\ and\ \citenamefont
  {Takayanagi}(2015)}]{Miyaji1503}%
  \BibitemOpen
  \bibfield  {author} {\bibinfo {author} {\bibfnamefont {Masamichi}\
  \bibnamefont {Miyaji}}\ and\ \bibinfo {author} {\bibfnamefont {Tadashi}\
  \bibnamefont {Takayanagi}},\ }\bibfield  {title} {\enquote {\bibinfo {title}
  {Surface/state correspondence as a generalized holography},}\ }\href@noop {}
  {\bibfield  {journal} {\bibinfo  {journal} {arXiv:1503.03542}\ } (\bibinfo
  {year} {2015})}\BibitemShut {NoStop}%
\bibitem [{\citenamefont {Miyaji}\ \emph
  {et~al.}(2015{\natexlab{b}})\citenamefont {Miyaji}, \citenamefont {Numasawa},
  \citenamefont {Shiba}, \citenamefont {Takayanagi},\ and\ \citenamefont
  {Watanabe}}]{Miyaji2015B}%
  \BibitemOpen
  \bibfield  {author} {\bibinfo {author} {\bibfnamefont {Masamichi}\
  \bibnamefont {Miyaji}}, \bibinfo {author} {\bibfnamefont {Tokiro}\
  \bibnamefont {Numasawa}}, \bibinfo {author} {\bibfnamefont {Noburo}\
  \bibnamefont {Shiba}}, \bibinfo {author} {\bibfnamefont {Tadashi}\
  \bibnamefont {Takayanagi}}, \ and\ \bibinfo {author} {\bibfnamefont {Kento}\
  \bibnamefont {Watanabe}},\ }\bibfield  {title} {\enquote {\bibinfo {title}
  {Continuous multiscale entanglement renormalization ansatz as holographic
  surface-state correspondence},}\ }\href {\doibase
  10.1103/PhysRevLett.115.171602} {\bibfield  {journal} {\bibinfo  {journal}
  {Phys. Rev. Lett.}\ }\textbf {\bibinfo {volume} {115}},\ \bibinfo {pages}
  {171602} (\bibinfo {year} {2015}{\natexlab{b}})}\BibitemShut {NoStop}%
\bibitem [{\citenamefont {Qi}(2013)}]{EHM1}%
  \BibitemOpen
  \bibfield  {author} {\bibinfo {author} {\bibfnamefont {Xiao-Liang}\
  \bibnamefont {Qi}},\ }\bibfield  {title} {\enquote {\bibinfo {title} {Exact
  holographic mapping and emergent space-time geometry},}\ }\href@noop {}
  {\bibfield  {journal} {\bibinfo  {journal} {arXiv: 1309.6282}\ } (\bibinfo
  {year} {2013})}\BibitemShut {NoStop}%
\bibitem [{\citenamefont {Lee}\ and\ \citenamefont {Qi}(2016)}]{EHM2}%
  \BibitemOpen
  \bibfield  {author} {\bibinfo {author} {\bibfnamefont {Ching~Hua}\
  \bibnamefont {Lee}}\ and\ \bibinfo {author} {\bibfnamefont {Xiao-Liang}\
  \bibnamefont {Qi}},\ }\bibfield  {title} {\enquote {\bibinfo {title} {Exact
  holographic mapping in free fermion systems},}\ }\href {\doibase
  10.1103/PhysRevB.93.035112} {\bibfield  {journal} {\bibinfo  {journal} {Phys.
  Rev. B}\ }\textbf {\bibinfo {volume} {93}},\ \bibinfo {pages} {035112}
  (\bibinfo {year} {2016})}\BibitemShut {NoStop}%
\bibitem [{\citenamefont {Ryu}\ and\ \citenamefont
  {Takayanagi}(2006)}]{RyuPRL2006}%
  \BibitemOpen
  \bibfield  {author} {\bibinfo {author} {\bibfnamefont {Shinsei}\ \bibnamefont
  {Ryu}}\ and\ \bibinfo {author} {\bibfnamefont {Tadashi}\ \bibnamefont
  {Takayanagi}},\ }\bibfield  {title} {\enquote {\bibinfo {title} {Holographic
  derivation of entanglement entropy from the anti\char21{}de sitter
  space/conformal field theory correspondence},}\ }\href {\doibase
  10.1103/PhysRevLett.96.181602} {\bibfield  {journal} {\bibinfo  {journal}
  {Phys. Rev. Lett.}\ }\textbf {\bibinfo {volume} {96}},\ \bibinfo {pages}
  {181602} (\bibinfo {year} {2006})}\BibitemShut {NoStop}%
\bibitem [{\citenamefont {Mollabashi}\ \emph {et~al.}(2014)\citenamefont
  {Mollabashi}, \citenamefont {Naozaki}, \citenamefont {Ryu},\ and\
  \citenamefont {Takayanagi}}]{Ryu2013}%
  \BibitemOpen
  \bibfield  {author} {\bibinfo {author} {\bibfnamefont {Ali}\ \bibnamefont
  {Mollabashi}}, \bibinfo {author} {\bibfnamefont {Masahiro}\ \bibnamefont
  {Naozaki}}, \bibinfo {author} {\bibfnamefont {Shinsei}\ \bibnamefont {Ryu}},
  \ and\ \bibinfo {author} {\bibfnamefont {Tadashi}\ \bibnamefont
  {Takayanagi}},\ }\bibfield  {title} {\enquote {\bibinfo {title} {Holographic
  geometry of cmera for quantum quenches and finite temperature},}\ }\href
  {\doibase 10.1007/JHEP03(2014)098} {\bibfield  {journal} {\bibinfo  {journal}
  {Journal of High Energy Physics}\ }\textbf {\bibinfo {volume} {2014}},\
  \bibinfo {eid} {98} (\bibinfo {year} {2014}),\
  10.1007/JHEP03(2014)098}\BibitemShut {NoStop}%
\bibitem [{\citenamefont {Verlinde}(2015)}]{Verlinde}%
  \BibitemOpen
  \bibfield  {author} {\bibinfo {author} {\bibfnamefont {Herman}\ \bibnamefont
  {Verlinde}},\ }\bibfield  {title} {\enquote {\bibinfo {title} {Poking holes
  in ads/cft: Bulk fields from boundary states},}\ }\href@noop {} {\bibfield
  {journal} {\bibinfo  {journal} {arXiv:1505.05069}\ } (\bibinfo {year}
  {2015})}\BibitemShut {NoStop}%
\bibitem [{\citenamefont {Nakayama}\ and\ \citenamefont
  {Ooguri}(2015)}]{Nakayama2015}%
  \BibitemOpen
  \bibfield  {author} {\bibinfo {author} {\bibfnamefont {Yu}~\bibnamefont
  {Nakayama}}\ and\ \bibinfo {author} {\bibfnamefont {Hirosi}\ \bibnamefont
  {Ooguri}},\ }\bibfield  {title} {\enquote {\bibinfo {title} {Bulk locality
  and boundary creating operators},}\ }\href {\doibase 10.1007/JHEP10(2015)114}
  {\bibfield  {journal} {\bibinfo  {journal} {Journal of High Energy Physics}\
  }\textbf {\bibinfo {volume} {2015}},\ \bibinfo {pages} {1--8} (\bibinfo
  {year} {2015})}\BibitemShut {NoStop}%
\bibitem [{\citenamefont {Czech}\ \emph
  {et~al.}(2015{\natexlab{a}})\citenamefont {Czech}, \citenamefont {Lamprou},
  \citenamefont {McCandlish},\ and\ \citenamefont {Sully}}]{Czech1505}%
  \BibitemOpen
  \bibfield  {author} {\bibinfo {author} {\bibfnamefont {Bartlomiej}\
  \bibnamefont {Czech}}, \bibinfo {author} {\bibfnamefont {Lampros}\
  \bibnamefont {Lamprou}}, \bibinfo {author} {\bibfnamefont {Samuel}\
  \bibnamefont {McCandlish}}, \ and\ \bibinfo {author} {\bibfnamefont {James}\
  \bibnamefont {Sully}},\ }\bibfield  {title} {\enquote {\bibinfo {title}
  {Integral geometry and holography},}\ }\href@noop {} {\bibfield  {journal}
  {\bibinfo  {journal} {arXiv: 1505.05515}\ } (\bibinfo {year}
  {2015}{\natexlab{a}})}\BibitemShut {NoStop}%
\bibitem [{\citenamefont {Czech}\ \emph
  {et~al.}(2015{\natexlab{b}})\citenamefont {Czech}, \citenamefont {Lamprou},
  \citenamefont {McCandlish},\ and\ \citenamefont {Sully}}]{Czech1512}%
  \BibitemOpen
  \bibfield  {author} {\bibinfo {author} {\bibfnamefont {Bartlomiej}\
  \bibnamefont {Czech}}, \bibinfo {author} {\bibfnamefont {Lampros}\
  \bibnamefont {Lamprou}}, \bibinfo {author} {\bibfnamefont {Samuel}\
  \bibnamefont {McCandlish}}, \ and\ \bibinfo {author} {\bibfnamefont {James}\
  \bibnamefont {Sully}},\ }\bibfield  {title} {\enquote {\bibinfo {title}
  {Tensor networks from kinematic space},}\ }\href@noop {} {\bibfield
  {journal} {\bibinfo  {journal} {arXiv: 1512.01548}\ } (\bibinfo {year}
  {2015}{\natexlab{b}})}\BibitemShut {NoStop}%
\bibitem [{\citenamefont {Fujita}\ \emph {et~al.}(2009)\citenamefont {Fujita},
  \citenamefont {Li}, \citenamefont {Ryu},\ and\ \citenamefont
  {Takayanagi}}]{Ryu_CS}%
  \BibitemOpen
  \bibfield  {author} {\bibinfo {author} {\bibfnamefont {Mitsutoshi}\
  \bibnamefont {Fujita}}, \bibinfo {author} {\bibfnamefont {Wei}\ \bibnamefont
  {Li}}, \bibinfo {author} {\bibfnamefont {Shinsei}\ \bibnamefont {Ryu}}, \
  and\ \bibinfo {author} {\bibfnamefont {Tadashi}\ \bibnamefont {Takayanagi}},\
  }\bibfield  {title} {\enquote {\bibinfo {title} {Fractional quantum hall
  effect via holography: Chern-simons, edge states and hierarchy},}\
  }\href@noop {} {\bibfield  {journal} {\bibinfo  {journal} {Journal of High
  Energy Physics}\ }\textbf {\bibinfo {volume} {2009}},\ \bibinfo {pages} {066}
  (\bibinfo {year} {2009})}\BibitemShut {NoStop}%
\bibitem [{\citenamefont {Dubail}\ and\ \citenamefont {Read}(2013)}]{Read13}%
  \BibitemOpen
  \bibfield  {author} {\bibinfo {author} {\bibfnamefont {J}~\bibnamefont
  {Dubail}}\ and\ \bibinfo {author} {\bibfnamefont {N}~\bibnamefont {Read}},\
  }\bibfield  {title} {\enquote {\bibinfo {title} {Tensor network trial states
  for chiral topological phases in two dimensions},}\ }\href@noop {} {\bibfield
   {journal} {\bibinfo  {journal} {arXiv:1307.7726}\ } (\bibinfo {year}
  {2013})}\BibitemShut {NoStop}%
\bibitem [{\citenamefont {Wahl}\ \emph {et~al.}(2013)\citenamefont {Wahl},
  \citenamefont {Tu}, \citenamefont {Schuch},\ and\ \citenamefont
  {Cirac}}]{Cirac13}%
  \BibitemOpen
  \bibfield  {author} {\bibinfo {author} {\bibfnamefont {T.~B.}\ \bibnamefont
  {Wahl}}, \bibinfo {author} {\bibfnamefont {H.-H.}\ \bibnamefont {Tu}},
  \bibinfo {author} {\bibfnamefont {N.}~\bibnamefont {Schuch}}, \ and\ \bibinfo
  {author} {\bibfnamefont {J.~I.}\ \bibnamefont {Cirac}},\ }\bibfield  {title}
  {\enquote {\bibinfo {title} {Projected entangled-pair states can describe
  chiral topological states},}\ }\href {\doibase
  10.1103/PhysRevLett.111.236805} {\bibfield  {journal} {\bibinfo  {journal}
  {Phys. Rev. Lett.}\ }\textbf {\bibinfo {volume} {111}},\ \bibinfo {pages}
  {236805} (\bibinfo {year} {2013})}\BibitemShut {NoStop}%
\bibitem [{\citenamefont {Wahl}\ \emph {et~al.}(2014)\citenamefont {Wahl},
  \citenamefont {Ha\ss{}ler}, \citenamefont {Tu}, \citenamefont {Cirac},\ and\
  \citenamefont {Schuch}}]{Cirac13b}%
  \BibitemOpen
  \bibfield  {author} {\bibinfo {author} {\bibfnamefont {Thorsten~B.}\
  \bibnamefont {Wahl}}, \bibinfo {author} {\bibfnamefont {Stefan~T.}\
  \bibnamefont {Ha\ss{}ler}}, \bibinfo {author} {\bibfnamefont {Hong-Hao}\
  \bibnamefont {Tu}}, \bibinfo {author} {\bibfnamefont {J.~Ignacio}\
  \bibnamefont {Cirac}}, \ and\ \bibinfo {author} {\bibfnamefont {Norbert}\
  \bibnamefont {Schuch}},\ }\bibfield  {title} {\enquote {\bibinfo {title}
  {Symmetries and boundary theories for chiral projected entangled pair
  states},}\ }\href {\doibase 10.1103/PhysRevB.90.115133} {\bibfield  {journal}
  {\bibinfo  {journal} {Phys. Rev. B}\ }\textbf {\bibinfo {volume} {90}},\
  \bibinfo {pages} {115133} (\bibinfo {year} {2014})}\BibitemShut {NoStop}%
\bibitem [{\citenamefont {Singh}\ and\ \citenamefont
  {Vidal}(2013)}]{Vidal2013}%
  \BibitemOpen
  \bibfield  {author} {\bibinfo {author} {\bibfnamefont {Sukhwinder}\
  \bibnamefont {Singh}}\ and\ \bibinfo {author} {\bibfnamefont {Guifre}\
  \bibnamefont {Vidal}},\ }\bibfield  {title} {\enquote {\bibinfo {title}
  {Symmetry-protected entanglement renormalization},}\ }\href {\doibase
  10.1103/PhysRevB.88.121108} {\bibfield  {journal} {\bibinfo  {journal} {Phys.
  Rev. B}\ }\textbf {\bibinfo {volume} {88}},\ \bibinfo {pages} {121108}
  (\bibinfo {year} {2013})}\BibitemShut {NoStop}%
\bibitem [{\citenamefont {Hasan}\ and\ \citenamefont {Kane}(2010)}]{Kane}%
  \BibitemOpen
  \bibfield  {author} {\bibinfo {author} {\bibfnamefont {M.~Z.}\ \bibnamefont
  {Hasan}}\ and\ \bibinfo {author} {\bibfnamefont {C.~L.}\ \bibnamefont
  {Kane}},\ }\bibfield  {title} {\enquote {\bibinfo {title}
  {\textit{Colloquium} : Topological insulators},}\ }\href {\doibase
  10.1103/RevModPhys.82.3045} {\bibfield  {journal} {\bibinfo  {journal} {Rev.
  Mod. Phys.}\ }\textbf {\bibinfo {volume} {82}},\ \bibinfo {pages}
  {3045--3067} (\bibinfo {year} {2010})}\BibitemShut {NoStop}%
\bibitem [{\citenamefont {Qi}\ and\ \citenamefont {Zhang}(2011)}]{QiRMP}%
  \BibitemOpen
  \bibfield  {author} {\bibinfo {author} {\bibfnamefont {Xiao-Liang}\
  \bibnamefont {Qi}}\ and\ \bibinfo {author} {\bibfnamefont {Shou-Cheng}\
  \bibnamefont {Zhang}},\ }\bibfield  {title} {\enquote {\bibinfo {title}
  {Topological insulators and superconductors},}\ }\href {\doibase
  10.1103/RevModPhys.83.1057} {\bibfield  {journal} {\bibinfo  {journal} {Rev.
  Mod. Phys.}\ }\textbf {\bibinfo {volume} {83}},\ \bibinfo {pages}
  {1057--1110} (\bibinfo {year} {2011})}\BibitemShut {NoStop}%
\bibitem [{\citenamefont {Qi}\ \emph {et~al.}(2008)\citenamefont {Qi},
  \citenamefont {Hughes},\ and\ \citenamefont {Zhang}}]{QiPRB}%
  \BibitemOpen
  \bibfield  {author} {\bibinfo {author} {\bibfnamefont {Xiao-Liang}\
  \bibnamefont {Qi}}, \bibinfo {author} {\bibfnamefont {Taylor~L.}\
  \bibnamefont {Hughes}}, \ and\ \bibinfo {author} {\bibfnamefont {Shou-Cheng}\
  \bibnamefont {Zhang}},\ }\bibfield  {title} {\enquote {\bibinfo {title}
  {Topological field theory of time-reversal invariant insulators},}\ }\href
  {\doibase 10.1103/PhysRevB.78.195424} {\bibfield  {journal} {\bibinfo
  {journal} {Phys. Rev. B}\ }\textbf {\bibinfo {volume} {78}},\ \bibinfo
  {pages} {195424} (\bibinfo {year} {2008})}\BibitemShut {NoStop}%
\bibitem [{\citenamefont {Swingle}\ and\ \citenamefont
  {McGreevy}(2014)}]{Swingle14}%
  \BibitemOpen
  \bibfield  {author} {\bibinfo {author} {\bibfnamefont {Brian}\ \bibnamefont
  {Swingle}}\ and\ \bibinfo {author} {\bibfnamefont {John}\ \bibnamefont
  {McGreevy}},\ }\bibfield  {title} {\enquote {\bibinfo {title}
  {Renormalization group constructions of topological quantum liquids and
  beyond},}\ }\href@noop {} {\bibfield  {journal} {\bibinfo  {journal} {arXiv:
  1407.8203}\ } (\bibinfo {year} {2014})}\BibitemShut {NoStop}%
\bibitem [{\citenamefont {Matsuura}\ and\ \citenamefont {Ryu}(2010)}]{Shunji}%
  \BibitemOpen
  \bibfield  {author} {\bibinfo {author} {\bibfnamefont {Shunji}\ \bibnamefont
  {Matsuura}}\ and\ \bibinfo {author} {\bibfnamefont {Shinsei}\ \bibnamefont
  {Ryu}},\ }\bibfield  {title} {\enquote {\bibinfo {title} {Momentum space
  metric, nonlocal operator, and topological insulators},}\ }\href {\doibase
  10.1103/PhysRevB.82.245113} {\bibfield  {journal} {\bibinfo  {journal} {Phys.
  Rev. B}\ }\textbf {\bibinfo {volume} {82}},\ \bibinfo {pages} {245113}
  (\bibinfo {year} {2010})}\BibitemShut {NoStop}%
\bibitem [{\citenamefont {Kadanoff}\ \emph {et~al.}(1967)\citenamefont
  {Kadanoff}, \citenamefont {Gotze}, \citenamefont {Hamblen}, \citenamefont
  {Hecht}, \citenamefont {Lewis}, \citenamefont {Palciauskas}, \citenamefont
  {Rayl}, \citenamefont {Swift},\ and\ \citenamefont {Aspnes}}]{Kadanoff}%
  \BibitemOpen
  \bibfield  {author} {\bibinfo {author} {\bibfnamefont {L.~P.}\ \bibnamefont
  {Kadanoff}}, \bibinfo {author} {\bibfnamefont {W.}~\bibnamefont {Gotze}},
  \bibinfo {author} {\bibfnamefont {D.}~\bibnamefont {Hamblen}}, \bibinfo
  {author} {\bibfnamefont {R.}~\bibnamefont {Hecht}}, \bibinfo {author}
  {\bibfnamefont {E.~A.~S.}\ \bibnamefont {Lewis}}, \bibinfo {author}
  {\bibfnamefont {V.~V.}\ \bibnamefont {Palciauskas}}, \bibinfo {author}
  {\bibfnamefont {M.}~\bibnamefont {Rayl}}, \bibinfo {author} {\bibfnamefont
  {J.}~\bibnamefont {Swift}}, \ and\ \bibinfo {author} {\bibfnamefont
  {J.}~\bibnamefont {Aspnes}, \bibfnamefont {D.and~Kane}},\ }\bibfield  {title}
  {\enquote {\bibinfo {title} {Static phenomena near critical points: Theory
  and experiment},}\ }\href {\doibase 10.1103/RevModPhys.39.395} {\bibfield
  {journal} {\bibinfo  {journal} {Rev. Mod. Phys.}\ }\textbf {\bibinfo {volume}
  {39}},\ \bibinfo {pages} {395--431} (\bibinfo {year} {1967})}\BibitemShut
  {NoStop}%
\bibitem [{\citenamefont {Wilson}(1975)}]{Wilson}%
  \BibitemOpen
  \bibfield  {author} {\bibinfo {author} {\bibfnamefont {Kenneth~G.}\
  \bibnamefont {Wilson}},\ }\bibfield  {title} {\enquote {\bibinfo {title} {The
  renormalization group: Critical phenomena and the kondo problem},}\ }\href
  {\doibase 10.1103/RevModPhys.47.773} {\bibfield  {journal} {\bibinfo
  {journal} {Rev. Mod. Phys.}\ }\textbf {\bibinfo {volume} {47}},\ \bibinfo
  {pages} {773--840} (\bibinfo {year} {1975})}\BibitemShut {NoStop}%
\bibitem [{\citenamefont {Fisher}(1998)}]{Fisher}%
  \BibitemOpen
  \bibfield  {author} {\bibinfo {author} {\bibfnamefont {Michael~E.}\
  \bibnamefont {Fisher}},\ }\bibfield  {title} {\enquote {\bibinfo {title}
  {Renormalization group theory: Its basis and formulation in statistical
  physics},}\ }\href {\doibase 10.1103/RevModPhys.70.653} {\bibfield  {journal}
  {\bibinfo  {journal} {Rev. Mod. Phys.}\ }\textbf {\bibinfo {volume} {70}},\
  \bibinfo {pages} {653--681} (\bibinfo {year} {1998})}\BibitemShut {NoStop}%
\bibitem [{Note1()}]{Note1}%
  \BibitemOpen
  \bibinfo {note} {Here we use the terminology `non-relativistic'
  (`relativistic') simply because the dispersion relation is $\sim k^2$ ($\sim
  k$) at UV limit $k\to \infty $. Alternatively, one can refer to these phases
  as insulators with (without) regularization at UV limit.}\BibitemShut {Stop}%
\bibitem [{\citenamefont {Evenbly}\ and\ \citenamefont
  {Vidal}(2009{\natexlab{b}})}]{Vidalreview01}%
  \BibitemOpen
  \bibfield  {author} {\bibinfo {author} {\bibfnamefont {G.}~\bibnamefont
  {Evenbly}}\ and\ \bibinfo {author} {\bibfnamefont {G.}~\bibnamefont
  {Vidal}},\ }\bibfield  {title} {\enquote {\bibinfo {title} {Algorithms for
  entanglement renormalization},}\ }\href {\doibase 10.1103/PhysRevB.79.144108}
  {\bibfield  {journal} {\bibinfo  {journal} {Phys. Rev. B}\ }\textbf {\bibinfo
  {volume} {79}},\ \bibinfo {pages} {144108} (\bibinfo {year}
  {2009}{\natexlab{b}})}\BibitemShut {NoStop}%
\bibitem [{\citenamefont {Evenbly}\ and\ \citenamefont
  {Vidal}(2011)}]{Vidalreview02}%
  \BibitemOpen
  \bibfield  {author} {\bibinfo {author} {\bibfnamefont {Glen}\ \bibnamefont
  {Evenbly}}\ and\ \bibinfo {author} {\bibfnamefont {Guifre}\ \bibnamefont
  {Vidal}},\ }\bibfield  {title} {\enquote {\bibinfo {title} {Quantum
  criticality with the multi-scale entanglement renormalization ansatz},}\
  }\href@noop {} {\bibfield  {journal} {\bibinfo  {journal} {arXiv:1109.5334}\
  } (\bibinfo {year} {2011})}\BibitemShut {NoStop}%
\bibitem [{\citenamefont {Viyuela}\ \emph
  {et~al.}(2014{\natexlab{a}})\citenamefont {Viyuela}, \citenamefont {Rivas},\
  and\ \citenamefont {Martin-Delgado}}]{Viyuela1d}%
  \BibitemOpen
  \bibfield  {author} {\bibinfo {author} {\bibfnamefont {O.}~\bibnamefont
  {Viyuela}}, \bibinfo {author} {\bibfnamefont {A.}~\bibnamefont {Rivas}}, \
  and\ \bibinfo {author} {\bibfnamefont {M.~A.}\ \bibnamefont
  {Martin-Delgado}},\ }\bibfield  {title} {\enquote {\bibinfo {title} {Uhlmann
  phase as a topological measure for one-dimensional fermion systems},}\ }\href
  {\doibase 10.1103/PhysRevLett.112.130401} {\bibfield  {journal} {\bibinfo
  {journal} {Phys. Rev. Lett.}\ }\textbf {\bibinfo {volume} {112}},\ \bibinfo
  {pages} {130401} (\bibinfo {year} {2014}{\natexlab{a}})}\BibitemShut
  {NoStop}%
\bibitem [{\citenamefont {Viyuela}\ \emph
  {et~al.}(2014{\natexlab{b}})\citenamefont {Viyuela}, \citenamefont {Rivas},\
  and\ \citenamefont {Martin-Delgado}}]{Viyuela2d}%
  \BibitemOpen
  \bibfield  {author} {\bibinfo {author} {\bibfnamefont {O.}~\bibnamefont
  {Viyuela}}, \bibinfo {author} {\bibfnamefont {A.}~\bibnamefont {Rivas}}, \
  and\ \bibinfo {author} {\bibfnamefont {M.~A.}\ \bibnamefont
  {Martin-Delgado}},\ }\bibfield  {title} {\enquote {\bibinfo {title}
  {Two-dimensional density-matrix topological fermionic phases: Topological
  uhlmann numbers},}\ }\href {\doibase 10.1103/PhysRevLett.113.076408}
  {\bibfield  {journal} {\bibinfo  {journal} {Phys. Rev. Lett.}\ }\textbf
  {\bibinfo {volume} {113}},\ \bibinfo {pages} {076408} (\bibinfo {year}
  {2014}{\natexlab{b}})}\BibitemShut {NoStop}%
\bibitem [{\citenamefont {Huang}\ and\ \citenamefont {Arovas}(2014)}]{Huang}%
  \BibitemOpen
  \bibfield  {author} {\bibinfo {author} {\bibfnamefont {Zhoushen}\
  \bibnamefont {Huang}}\ and\ \bibinfo {author} {\bibfnamefont {Daniel~P.}\
  \bibnamefont {Arovas}},\ }\bibfield  {title} {\enquote {\bibinfo {title}
  {Topological indices for open and thermal systems via uhlmann's phase},}\
  }\href {\doibase 10.1103/PhysRevLett.113.076407} {\bibfield  {journal}
  {\bibinfo  {journal} {Phys. Rev. Lett.}\ }\textbf {\bibinfo {volume} {113}},\
  \bibinfo {pages} {076407} (\bibinfo {year} {2014})}\BibitemShut {NoStop}%
\bibitem [{\citenamefont {Wen}(1999)}]{Wen1999}%
  \BibitemOpen
  \bibfield  {author} {\bibinfo {author} {\bibfnamefont {Xiao-Gang}\
  \bibnamefont {Wen}},\ }\bibfield  {title} {\enquote {\bibinfo {title}
  {Projective construction of non-abelian quantum hall liquids},}\ }\href
  {\doibase 10.1103/PhysRevB.60.8827} {\bibfield  {journal} {\bibinfo
  {journal} {Phys. Rev. B}\ }\textbf {\bibinfo {volume} {60}},\ \bibinfo
  {pages} {8827--8838} (\bibinfo {year} {1999})}\BibitemShut {NoStop}%
\bibitem [{\citenamefont {Hartman}\ and\ \citenamefont
  {Maldacena}(2013)}]{Hartman}%
  \BibitemOpen
  \bibfield  {author} {\bibinfo {author} {\bibfnamefont {Thomas}\ \bibnamefont
  {Hartman}}\ and\ \bibinfo {author} {\bibfnamefont {Juan}\ \bibnamefont
  {Maldacena}},\ }\bibfield  {title} {\enquote {\bibinfo {title} {Time
  evolution of entanglement entropy from black hole interiors},}\ }\href
  {\doibase 10.1007/JHEP05(2013)014} {\bibfield  {journal} {\bibinfo  {journal}
  {Journal of High Energy Physics}\ }\textbf {\bibinfo {volume} {2013}},\
  \bibinfo {eid} {14} (\bibinfo {year} {2013}),\
  10.1007/JHEP05(2013)014}\BibitemShut {NoStop}%
\bibitem [{\citenamefont {Gu}\ \emph {et~al.}(2016)\citenamefont {Gu},
  \citenamefont {Lee}, \citenamefont {Wen}, \citenamefont {Cho}, \citenamefont
  {Ryu},\ and\ \citenamefont {Qi}}]{Gu}%
  \BibitemOpen
  \bibfield  {author} {\bibinfo {author} {\bibfnamefont {Yingfei}\ \bibnamefont
  {Gu}}, \bibinfo {author} {\bibfnamefont {Ching~Hua}\ \bibnamefont {Lee}},
  \bibinfo {author} {\bibfnamefont {Xueda}\ \bibnamefont {Wen}}, \bibinfo
  {author} {\bibfnamefont {Gil~Young}\ \bibnamefont {Cho}}, \bibinfo {author}
  {\bibfnamefont {Shinsei}\ \bibnamefont {Ryu}}, \ and\ \bibinfo {author}
  {\bibfnamefont {Xiao-Liang}\ \bibnamefont {Qi}},\ }\bibfield  {title}
  {\enquote {\bibinfo {title} {Holographic duality between (2+1)-d quantum
  anomalous hall state and (3+1)-d topological insulators},}\ }\href@noop {}
  {\bibfield  {journal} {\bibinfo  {journal} {arXiv:1605.00570}\ } (\bibinfo
  {year} {2016})}\BibitemShut {NoStop}%
\end{thebibliography}%

\end{document}